%% file: ms.tex
\documentclass[preprint2]{aastex}
\usepackage{graphicx}
\usepackage{psfig}
\usepackage{epsfig}
\usepackage{amsmath}
\usepackage{natbib}

\newcommand{\msun}{\,\hbox{$M_{\odot}$}}
\newcommand{\lsun}{\,\hbox{$L_{\odot}$}}

\newcommand{\reff}{\mbox{$R_{\rm eff}$}}
\newcommand{\lir}{\,\hbox{$L_{\rm IR}$}}

\newcommand{\spi}{{\it Spitzer}}

\newcommand{\psix}{{PAH 6.2}}
\newcommand{\psev}{{PAH 7.7}}
\newcommand{\peig}{{PAH 8.6}}
\newcommand{\pele}{{PAH 11.3}}
\newcommand{\ptwe}{{PAH 12.7}}
\newcommand{\pts}{{PAH 17.0}}
\newcommand{\neiitab}{\,\hbox{[NeII]}}
\newcommand{\neiiitab}{\,\hbox{[NeIII]}}
\newcommand{\neii}{\,\hbox{[\ion{Ne}{2}]}}
\newcommand{\neiii}{\,\hbox{[\ion{Ne}{3}]}}
\newcommand{\nev}{\,\hbox{[\ion{Ne}{5}]}}
\newcommand{\clii}{\,\hbox{[\ion{Cl}{2}]}}

\shorttitle{A spectroscopic catalog of 0.3$\le$z$\le$3.5 infrared$-$luminous, 24 micron selected galaxies} 
\shortauthors{Dasyra et al.}

\begin{document}

\title{The $\sim$0.9 mJy sample: A mid-infrared spectroscopic catalog of 150 infrared$-$luminous, 24 micron selected galaxies at 0.3$\le$z$\le$3.5}

\author{Kalliopi M. Dasyra\altaffilmark{1}, Lin Yan\altaffilmark{2}, George Helou\altaffilmark{1,2}, Anna Sajina\altaffilmark{3},  Dario Fadda\altaffilmark{2}, Michel Zamojski\altaffilmark{1}, Lee Armus\altaffilmark{1}, Bruce Draine\altaffilmark{4}, David Frayer\altaffilmark{2} }

\altaffiltext{1}{Spitzer Science Center, California Institute of Technology,
Mail Code 220-6, 1200 East California Blvd, Pasadena, CA 91125}
\altaffiltext{2}{Infrared Processing and Analysis Center, California
Institute of Technology, Mail Code 100-22, Pasadena, CA 91125}
\altaffiltext{3}{Department of Astronomy, Haverford College, Haverford, PA 19041}
\altaffiltext{4}{Department of Astrophysical Sciences, Princeton University, Princeton, NJ 08544-1001}

\begin{abstract}
We present a catalog of mid-infrared (MIR) spectra of 150 infrared (IR) luminous galaxies in the \spi\ extragalactic first look survey 
obtained with the IR spectrograph on board the \spi\ Space Telescope. The sample is selected to be brighter than $\sim$0.9 mJy 
at 24 \micron\ and it has a redshift distribution in the range [0.3,3.5], with a peak at $z$$=$1.  It primarily comprises  ultraluminous 
IR galaxies at z$\gtrsim$1 and luminous IR galaxies at z$<$1, as estimated from their monochromatic rest-frame 14 \micron\ 
luminosities. The number of sources with spectra that are dominated by an active galactic nucleus (AGN) continuum is 49, while 
39 sources have strong,  star-formation related features. For this classification,  we used the equivalent width (EW) of the 11.3 
\micron\ polycyclic aromatic hydrocarbon (PAH) feature. Several intermediate and high $z$ starbursts have higher PAH EW than 
local ULIRGs, which is indicative of an elevated star formation activity. Moreover, an increase in the AGN activity is observed with 
increasing $z$ and luminosity, based on the decreasing EW of PAHs and the increasing \neiii /\neii\ ratio. Spectral stacking leads 
to the detection of weak features such as the 3.3 \micron\ PAH, the H$_2$ 0-0 S(1) and S(3) lines, and the \nev\ line. We observe 
differences in the flux ratios of PAHs in the stacked spectra of IR-luminous galaxies with redshift or luminosity, which cannot be 
attributed to extinction effects since both the depth and the profile of the silicate absorption feature at 9.7 \micron\ remain the same 
at $z$$<$1 and $z$$\ge$1. When placing the observed galaxies on IR color-color diagrams, we find that the wedge defining AGN 
comprises most sources with continuum-dominated spectra, but also contains many starbursts and sources with strong silicate 
absorption at 9.7 \micron. The comparison of the 11.3 \micron\ PAH EW and the $H$-band effective radius, measured from 
{\it Hubble} Space Telescope data, indicates that sources with EWs$\ge$2 \micron , are typically more extended than $\sim$3 kpc. 
However, there is no strong correlation between the MIR spectral type and the near-IR extent of the sources. 
\end{abstract}


\section{Introduction}
\label{sec:intro}
Infrared (IR) bright galaxies at intermediate and high redshift, i.e., 0.5$\lesssim$$z$$\lesssim$3, play 
an important role in driving galaxy evolution since they are in the process of forming new stars   
(\citealt{carilli05}; \citealt{solomon05}; \citealt{tacconi06}) or growing black holes (e.g. \citealt{alexander08}, 
\citealt{sajina08}), and since they are often associated with galactic interactions (\citealt{zheng04}; 
\citealt{dasyra08a}). The number density of both IR-bright galaxies and galactic interactions evolves strongly 
with redshift, increasing to at least $z$$\sim$1 (\citealt{chary_elbaz}; \citealt{lefloch05}; \citealt{perez05};
\citealt{conselice06}; \citealt{jeyhan07}). This cosmologically significant population of galaxies 
with IR-excess, which often comprises galaxies that are selected in sub-millimeter wavelengths 
(\citealt{valiante07}; Men\'endez-Delmestre et al. 2007; 2009; \citealt{pope08}) or that have extremely red 
colors in the optical and near-infrared (NIR) wavelengths (\citealt{yan04}; \citealt{daddi05}; \citealt{papovich06}), 
is believed to form many of the early-type galaxies (\citealt{daddi04} \citealt{hyan04}; \citealt{swinbank06}) and 
possibly some of the late-type galaxies (\citealt{hammer05}) that we observe in the local Universe. 

Thousands of new IR-bright galaxies were discovered with the \spi\  Space Telescope in the 24 \micron\ 
catalog of various surveys (\citealt{fadda04}; \citealt{papovich04}; \citealt{rigby04}; \citealt{houck05}; 
\citealt{surace05}) thanks to the high sensitivity of the Multiband Imaging Photometer for \spi\  (MIPS).
Many of these systems were spectroscopically followed up with the IR spectrograph (IRS) on board 
\spi\ to determine their redshifts and to reveal the origin of their IR emission. The $z$ determination
from mid-infrared (MIR) spectra was very efficient because of the bright and broad polycyclic aromatic 
hydrocarbon (PAH) features that can be observed within the IRS spectral range, $\sim$5-35 \micron , for 
sources at $z$$\lesssim$3.5. This led to the discovery of a large number of galaxies at 1$<$$z$$<$3, 
complementing previous studies that aimed to populate the so-called  'redshift desert' of optical spectroscopy 
(\citealt{adelberger04}; \citealt{steidel04}). 

The IRS spectra often revealed different powering mechanisms for the MIR emission of these distant IR-bright
sources because of the different selection techniques that were applied for each program. \cite{dey08} imposed 
a high 24 \micron\ flux,  f$_{24}$, to $R$-band flux ratio to select systems more obscured than local ultraluminous 
IR galaxies (ULIRGs). They detected a large number of sources with continuum-dominated spectra, especially for 
f$_{24}$$\gtrsim$1 mJy (\citealt{dey09}).  Other authors used a radio flux cutoff that also led to the identification of 
several active galactic nuclei (AGN; \citealt{martinez08}; \citealt{weedman06}). Star-forming galaxies, as determined 
by their PAH emission, were found in samples with different selection criteria.  Yan et al. (2005; 2007) identified new 
1$<$$z$$<$3 starbursts by imposing that the ratio of their 24 \micron\ flux to their 8 \micron\  flux and the ratio of their 24 
\micron\ flux to their $R$-band flux are as low as those of local starbursts. \cite{farrah08}, \cite{huang09}, and \cite{desai09} 
collected IRS spectra for sources with a flux excess in Infrared Array Camera (IRAC) channels that corresponded 
to the 1.6 \micron\ stellar bump. Their samples comprised a large number of $z$$\sim$2 star-forming galaxies.  
Sub-millimeter galaxies were also followed-up with IRS and were found to be primarily star-forming, but to also 
have a nonnegligible AGN contribution to their MIR luminosity (\citealt{valiante07}; Men\'endez-Delmestre et al. 
2007; 2009; \citealt{pope08}). \cite{hernan09} observed with IRS a sample of $z$$>$1 galaxies that have detections 
in several optical bands. Their sample comprised several composite systems, i.e., systems with weak PAH or silicate 
absorption feature(s) and an AGN continuum.  

The goal of this project is to use a purely flux-limited sample of 150 IR-bright galaxies spanning a wide 
$z$ range to investigate what are the processes that lead to an excess of IR activity at different epochs.  
We have been awarded time with \spi\ and with the {\it Hubble} Space Telescope ({\it HST}) to collect IR 
imaging and spectroscopic datasets to address what  type of galaxies undergo IR-luminous phases
at different $z$. Our program aims to investigate whether there are changes in the interstellar medium 
properties of these galaxies with look-back time, and whether there is any relation between the dominant MIR 
emission mechanism and the morphologies of these galaxies. In this paper, we present the catalog of the 
IRS spectra and we primarily study the behavior of MIR line and feature properties as a function of $z$.

The rest of this paper is organized as follows. In \S~\ref{sec:sample} we describe the sample 
selection technique, followed by the presentation of ancillary and complementary datasets for
the sample in \S~\ref{sec:ancillary}. The methods that we used to reduce and analyze the spectra
are presented in \S~\ref{sec:reduction}. Results are presented in \S~\ref{sec:results}, followed by 
a discussion on MIR spectral properties with $z$ and a summary of our conclusions in  
\S~\ref{sec:discussion} and \S~\ref{sec:conclusions}, respectively. Throughout this paper we 
use a $\Lambda$CDM cosmology with $H_0$=70~km~s$^{-1}$~Mpc$^{-1}$, $\Omega_{m}$=0.3, 
and $\Omega_{\Lambda}$=0.7.


\section{Sample selection} 
\label{sec:sample}
\begin{figure*}
\centering
\includegraphics[width=17cm]{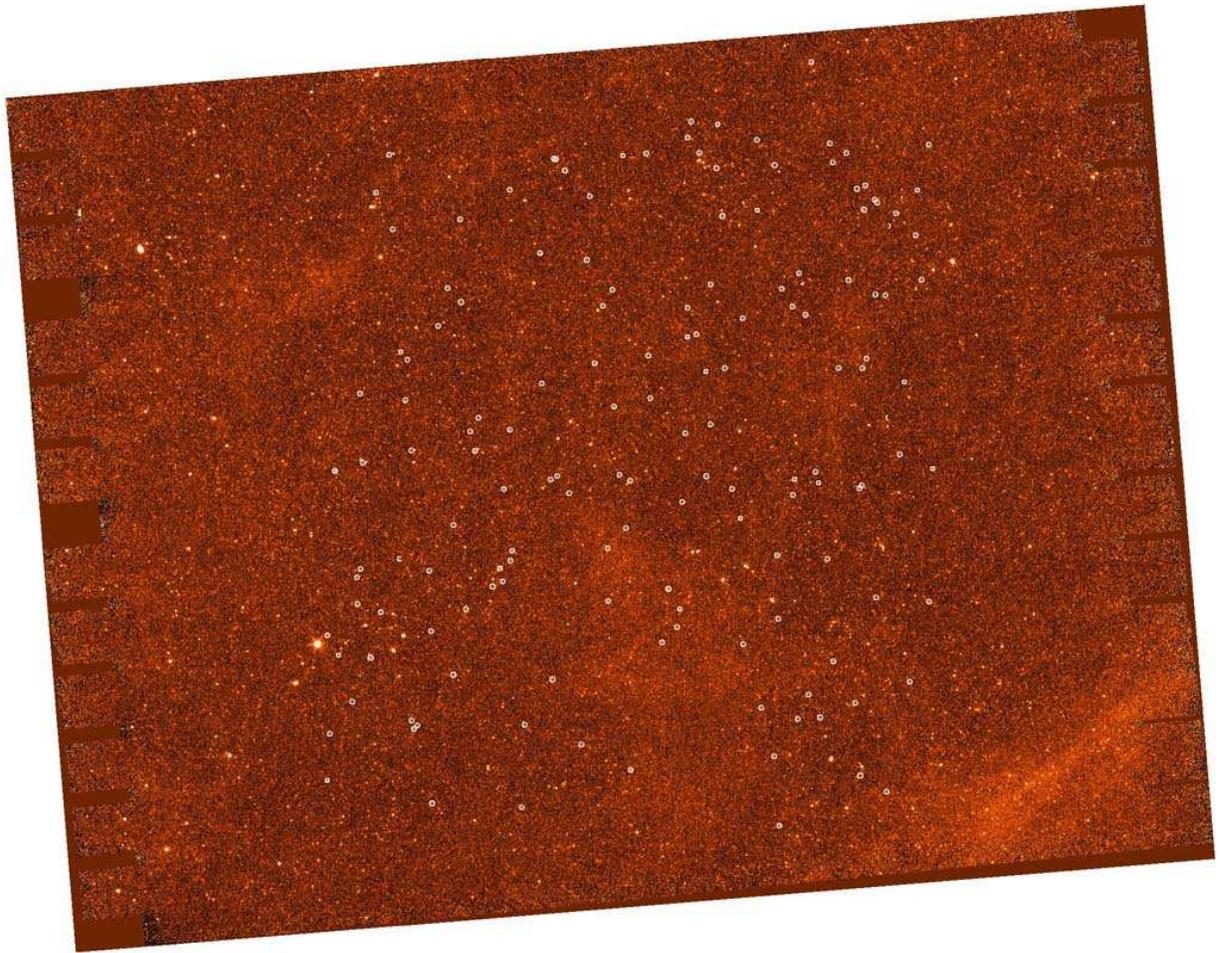}
\caption{\label{fig:region} Mosaic of the \spi\ extragalactic first look survey at 24 \micron\ 
(\citealt{fadda06}). The sources observed with IRS as part of the program with ID  20629 are shown 
as circles. }
\end{figure*}

The sample comprises 150 sources that were selected from the 24 \micron\ mosaic of the 3.7 deg$^2$ 
\spi\  extragalactic first look survey (XFLS; \citealt{fadda06}).  It was selected to be flux limited with a 
lower flux threshold of  $\sim$0.9 mJy at 24 \micron . To facilitate the selection of high-$z$ objects, we 
observed only sources that are fainter than 19 Vega magnitudes in the $R$ band. We also restricted the 
selection to an area that covers approximately three quarters of the total XFLS field of view and that is located 
at the center of the field (Fig.~\ref{fig:region}). An advantage of selecting sources from the central XFLS 
region is that it comprises a 0.25 deg$^2$ field with deep IRAC datasets (\citealt{lacy05}), the so-called 
'verification zone'.  Moreover, IRS spectra have also been acquired for several sources with 
f$_{24}$$\gtrsim$0.9 mJy in the same region as part of other programs (PIs Borys, Fazio, Lacy, Lagache, 
Martinez-Sansigre, Weedman, Yan).  These archival data can be merged with data from our program, with 
ID 20629, to create an extended sample that is highly complete at faint optical magnitudes and that 
will be used for luminosity function studies (Yan et al. 2009, in preparation).  

\begin{figure*}
\centering
\includegraphics[width=17cm]{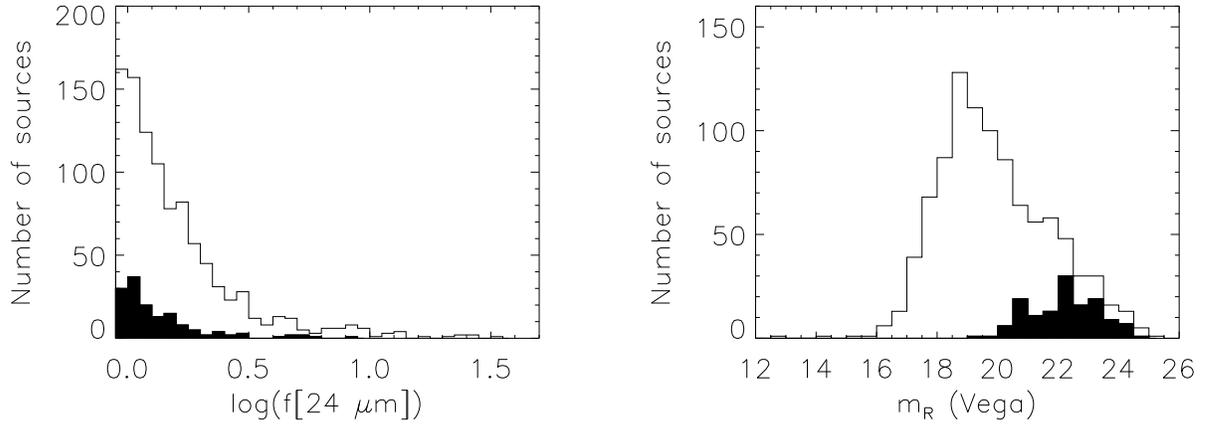}
\caption{\label{fig:selection}
Selection function of the sources in this sample. The histogram of the logarithm of the observed-frame 24 \micron\ fluxes 
is shown on the left panel, and the histogram of the observed-frame $R$-band magnitudes is shown 
on the right panel. In both panels, the outlined histogram comprises all galaxies in the central XFLS region
that are brighter than 0.9 mJy at 24 \micron. The filled histogram corresponds to the sources that were targeted 
by this program.
}
\centering
\end{figure*}

In total, new IRS observations were executed for 150 XFLS sources as part of the program with ID 20629. This
led to the acquisition of MIR spectra for one out of every three extragalactic sources in the central XFLS field 
with f$_{24}$$\ge$$0.9$mJy and m$_{\rm R}$$\ge$20 Vega mags (Fig.~\ref{fig:selection}).  Similarly, this sample is 57\% 
complete for  m$_{\rm R}$$\ge$22 Vega magnitudes within the selected area (Fig.~\ref{fig:selection}). The datasets 
were collected during the second  \spi\ cycle. The coordinates of the observed sources and their integration 
times per mode are presented in Table~\ref{tab:obs}. The total integration time per target varied from  25 mins 
to 2 hours on-source. The integration times were longer at wavelengths $\gtrsim$14 \micron\  than at wavelengths
$\lesssim$14 \micron\ to compensate for the lower sensitivity of the instrument at long wavelengths.

\section{Ancillary data collection and reduction}
\label{sec:ancillary}
The XFLS  region has a plethora of ancillary datasets. \cite{fadda04} presented an $R$-band mosaic 
with a corresponding catalog that has a depth of 25.5 Vega magnitudes.  Imaging datasets for all
IRAC channels, at 3.6, 4.5, 5.8, and 8.0 \micron ,   are available in the \cite{lacy05}
catalogs. Each IRAC flux is measured within a fixed aperture of either 6$\farcs$0, 9$\farcs$3, 
14$\farcs$9, or 24$\farcs$4 that is optimally determined for each galaxy and band (\citealt{lacy05}). 
When possible, we used the deep 'verification-zone' IRAC data. The ancillary datasets that we 
compiled, together with the MIPS 24 \micron\  fluxes of the sources (\citealt{fadda06}) are presented 
in Table~\ref{tab:obs}. 

To enhance the accuracy of the analysis presented in this paper, we examined the proper identification 
of the counterpart of each 24 \micron\ source in all IRAC channels. We remeasured IRAC fluxes or 
limits for 15\% of the galaxies because of counterpart misidentification or 
overlap of nearby sources. We deblended IRAC overlapping sources when one of them was a star or a 
galaxy that did not contribute to the MIPS 24 \micron\ flux. We also deblended overlapping sources 
when the 24 \micron\ emission peaked near one of the sources, unless their {\it HST} image showed clear 
evidence of interactions between them. Vice versa, we typically merged  the fluxes of two 
IRAC detections within the 24 \micron\ beam  if the 24  \micron\ flux peaked between them. Moreover, all 
IRAC limits were remeasured, since \cite{lacy05} used statistical completeness limits for their catalogs.

To perform aperture photometry, we used the Sextractor package (\citealt{sextractor}) with the same 
parameters as those used by \cite{lacy05}, except for the background and the minimum deblending  parameter.
We computed the background locally (within 1 square arc minute) and we set the deblending parameter to 0.001,
to easily deblend overlapping sources.  We used the optimal extraction aperture of each source, as chosen
by \cite{lacy05},  with the corresponding aperture flux correction for the specific channel. To measure the flux 
limit of a source that is undetected at a given channel, we used its aperture at the nearest channel(s) in which 
the source was detected. If two such apertures existed, we chose the largest of the two. To deblend overlapping
sources, we decomposed them into two or more point sources using the package GALFIT (\citealt{galfit}). 
After determining the intensity of each individual point-spread-function (PSF), we subtracted all PSF
models that were unrelated to the 24 \micron\ source. We then measured the flux of the deblended counterpart 
of the 24 $\micron$ source using Sextractor (with the same parameters as above), to avoid systematic effects 
in our measurements.

A cleaning analysis similar to that performed for the IRAC catalogs was also performed for the $R$-band 
catalog fluxes of \cite{fadda06}.  We remeasured the $R$-band fluxes for sources that were erroneously
identified as the counterpart of the 24 \micron\ source. Whenever we merged the IRAC fluxes of two overlapping 
sources, we also merged their $R$-band fluxes. To measure $R$-band fluxes, we used Sextractor with the 
zeropoints of the original catalog. We kept most parameters identical to those selected by \cite{fadda06}, including 
the automatic determination of the optimal aperture.  We set the area that is used to compute the background flux to
1 square arc minute. We used deblending parameter values in the range [0.1,0.001] to ensure the proper deblending 
of nearby sources.  In Table~\ref{tab:obs}, we summarize all remeasured IRAC and  $R$-band fluxes or limits.

Shallow MIPS 70 and 160 \micron\ catalogs are also available for the XFLS (\citealt{frayer06}).  We were
 awarded time to complement the 70 \micron\ catalog with deeper MIPS observations for two thirds 
of the sources in our sample. The MIPS datasets will be presented in Sajina et al. (2009, in preparation), 
together a multi-wavelength spectral energy distribution (SED) fitting that aims on the computation of the 
bolometric luminosities of the sources.  

As part of the same large, joint \spi\  
and {\it HST} program, we also acquired  {\it HST}  $H$-band images of several sources with the Near 
Infrared Camera and Multi-Object Spectrometer (NICMOS). In total, NICMOS successfully observed 
102 sources. We processed the NICMOS images similar to
those in \cite{dasyra08a}, and we used them to measure effective radii \reff\ using Sextractor 
(Table~\ref{tab:obs};  Zamojski et al. 2009, in preparation). The NICMOS data and their reduction 
techniques will be presented in detail in Zamojski et al. (2009, in preparation).

\section{IRS data reduction and analysis}
\label{sec:reduction}

\subsection{Data reduction procedures} 
The IRS data reduction began with the processing of the basic calibration datasets (BCDs) by the 
IRS pipeline version 15.3. The pipeline converts the up-the-ramp exposures to a single image and then 
performs dark subtraction, linearity correction, flat division and other instrument-related corrections to the image. 
Details on the pipeline products can be found at 
the IRS data and pipeline handbooks that are available at the \spi\ Science Center (SSC) webpage\footnote{
http://ssc.spitzer.caltech.edu/irs/dh/dh32.pdf  and http://ssc.spitzer.caltech.edu/irs/dh/irsPDDmar30.pdf
}. The pipeline provides the two-dimensional spectral images together with their uncertainties and
mask files for each of nod position and wavelength range. 

There are two wavelength ranges
for the low-resolution mode of the IRS spectrograph: the short-low (SL) and long-low (LL) modes that can obtain 
datasets in the wavelength ranges [5.2,14.7] \micron\ and [14.3,35.0] \micron, respectively. For LL data, we 
removed the sky background for each order and nod position using IDL routines that compute the median sky 
image for all other orders and nod positions of the same source. Because in SL we only obtained order
1 data, we simply subtracted one nod position from the other to remove the sky background.  The sky
subtraction was performed with the code used by \cite{yan05}. We then performed 
an automated first-order rejection of negative bad pixels, and an inspection and manual removal (when necessary) 
of positive bad pixels that deviated by more than 3 standard deviations $\sigma$ from the median value of the 
spectrum.  The manual rejection of bad pixels was performed using the IDL routine IRSCLEAN\footnote{
http://ssc.spitzer.caltech.edu/postbcd/irsclean.html} provided by the SSC. The values of bad pixels were 
interpolated from those of  their neighbors. We extracted the spectrum of each nod position 
using the SSC package SPICE\footnote{http://ssc.spitzer.caltech.edu/postbcd/spice.html}.
For the extraction of the spectrum, we used the 'optimal' technique, which is recommended for faint source 
spectroscopy. This technique maximizes the signal-to-noise ratio of the spectrum by weighting the spatial pixels at any 
wavelength prior adding them. The code uses the spatial profile of a point source to compute the appropriate 
weights. The extracted one-dimensional spectrum and its uncertainty image are then flux calibrated. Aperture-loss 
and slit-loss corrections are applied based on a point-source profile. After extracting the spectrum for both nod
positions of each spectral order, we computed the average spectrum for both nod positions and trimmed all (noisy) 
edges. We merged the final spectra and interpolated the flux values in the wavelength range where different orders 
overlap to the wavelength values of the order with the lower spectral resolution. 

\subsection{Analysis techniques} 
\begin{figure*}
\centering
\includegraphics[width=17cm]{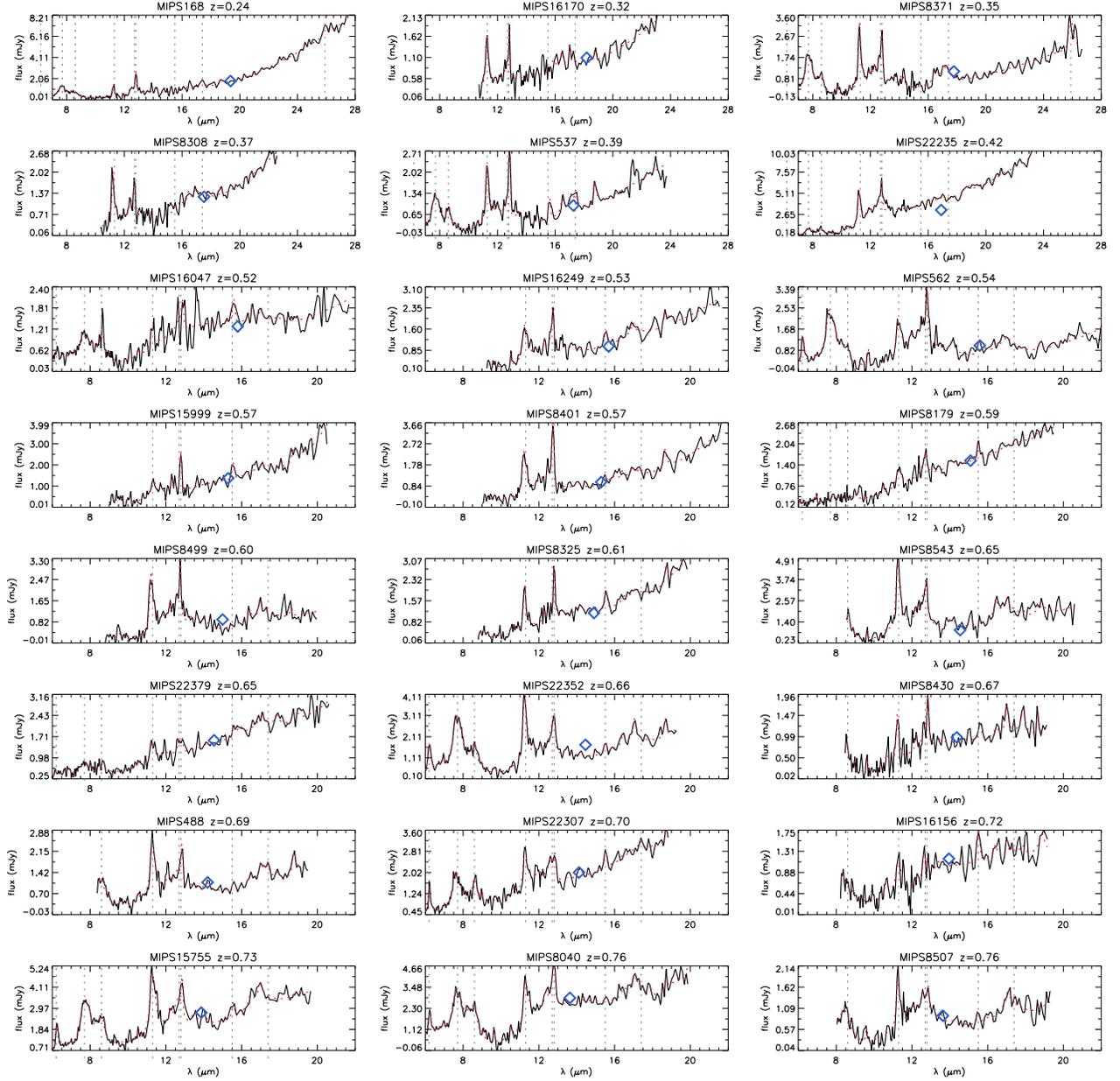}
\caption{\label{fig:spec} Rest-frame IRS spectra of 24-\micron\ selected, IR-luminous galaxies
with reliable redshift measurements. These are the sources with confidence classification {\it a, b, } or 
{\it c} in Table~\ref{tab:specz}. Each fully reduced spectrum is plotted using a solid line and its 
best-fit model, computed with PAHFIT, is overplotted using a dashed line. The MIPS 24 \micron\ flux of each
source, taken from the \cite{fadda06} catalog, is overplotted as a diamond. The dashed vertical 
lines correspond to the wavelengths where bright features or lines are expected to be, namely, at 6.2, 7.7, 
8.6, 11.3, 12.7, 12.8, 15.5, 17.4, and 25.9 \micron.
}
\end{figure*}

\begin{figure*}
\centering
\includegraphics[width=17cm]{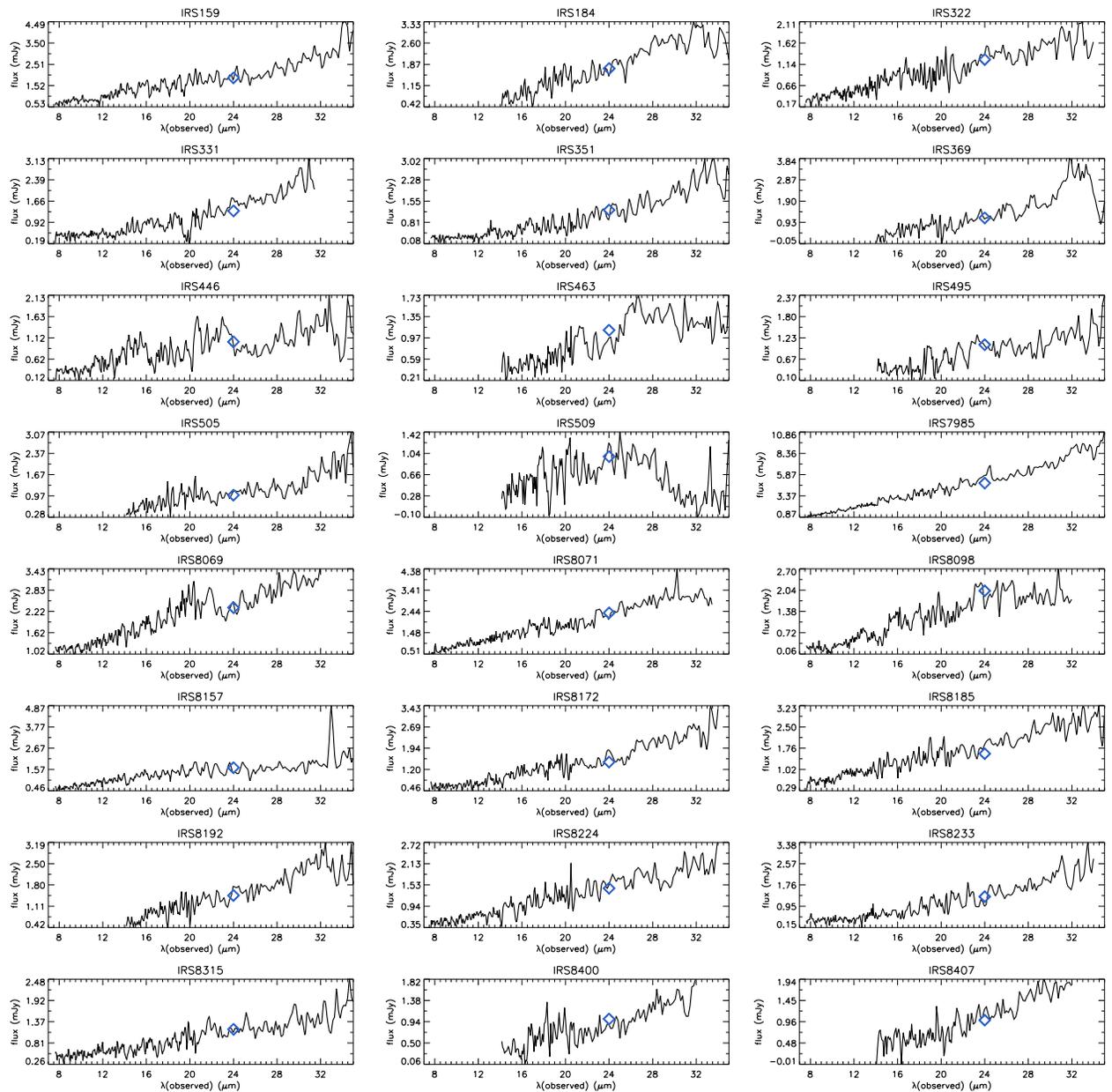} 
\caption{\label{fig:spec_noz} Observed-frame IRS spectra of 24-\micron\ selected, IR-luminous galaxies
of unknown or uncertain redshift (including sources with confidence classification {\it d, e, } or {\it f} in 
Table~\ref{tab:specz}). Each fully reduced spectrum is plotted using a solid line, and its
corresponding MIPS 24 \micron\ flux (\citealt{fadda06}) is overplotted as a diamond.  }
\end{figure*}

To compute the redshift of each source we first ran a code that simultaneously fits MIR fine-structure lines and 
PAHs using Gaussian and Lorentzian functions, respectively, using an initial redshift guess. After fitting the 
spectrum, the code returns the best $z$ solution by averaging the redshift of all 3-$\sigma$ detected lines and 
features that are within 0.2 \micron\ from their expected wavelength.  In Table~\ref{tab:specz}, we summarize 
the confidence and errors of all measured $z$ values. We define the $z$ measurement of a source reliable when 
its spectrum has more than two 3-$\sigma$ detected emission lines or features,  or silicate absorption at 9.7 \micron , 
or a combination of the two (see source confidence classifications {\it a}, {\it b}, and {\it c} in Table~\ref{tab:specz}). 
All sources with uncertain redshift measurement, e.g., because more than one redshift solutions are plausible, are 
excluded from the analysis in this paper. The fully reduced spectra are presented in the rest frame for the 95 sources 
with reliable $z$ measurement  (Fig.~\ref{fig:spec}),  and in the observed frame for the 55 sources with unknown or 
uncertain redshift (Fig.~\ref{fig:spec_noz}).

For sources with reliably measured redshift, we ran the fitting algorithm PAHFIT (\citealt{smith07}) to derive the 
fluxes of lines and features and the optical depth $\tau$ of the obscuring medium in the MIR. PAHFIT decomposes 
MIR spectra into emission originating from ionic and molecular lines, PAH features, dust continuum and stellar 
continuum.  All components are multiplied with an extinction curve prior to 
being added. The best-fit solution of the observed spectrum is computed using a $\chi ^2$ minimization method. 
The code returns the best-fit parameters for each of the components, including the line or feature fluxes 
and the optical depth of the silicate absorption feature at 9.7 \micron , $\tau$$_{9.7}$.  

The rest-frame spectrum of each source was used as input to PAHFIT. The spectral flux per frequency was 
divided by a factor of (1+z) so that all line fluxes are correctly measured at rest frame, where PAHFIT performs all line
flux computations. We added all available IRAC datapoints to the spectra, so that PAHFIT can best constrain the stellar 
continuum emission. Since the default PAHFIT parameters were optimized to fit local low-luminosity sources, we 
modified several parameters to make the code appropriate for high-z IR-bright galaxies. To include a rising
AGN continuum for $\lambda$$\lesssim$10 \micron , we allowed the dust grains to have temperatures of 400,600,
800,1000,1200,  or 1400K, which are higher than the default temperatures that PAHFIT uses ($\le$300K). By default, 
PAHFIT returns extinction-corrected line fluxes. Because the geometric distribution of dust grains in high-$z$ 
obscured galaxies is not known, we measured and used throughout this paper fluxes that are not extinction corrected. 
For this reason, we run PAHFIT assuming a screen Galactic center extinction (\citealt{chiar_tielens}), and we 
then multiplied the measured fluxes with the extinction curve value at the center of each line or PAH 
complex. The PAHFIT best-fit solution for each spectrum is presented in Fig.~\ref{fig:spec}.

To measure line or feature equivalent widths (EWs), we divided the best-fit profile of each individual line or feature,
which has no continuum,  by the combined continuum of dust and stars. We then integrated the result as a function 
of wavelength. Fluxes (or  3-$\sigma$ flux limits) and EWs of the most common bright lines and features in the IRS 
spectra are summarized in Tables ~\ref{tab:lines1} and~\ref{tab:lines2}. Fluxes of lines or features whose continuum 
cannot be well determined are not tabulated. For example,  we do not present an 8.6 \micron\ PAH flux measurement 
when silicate absorption is present and the continuum shortward of the 7.7 \micron\ PAH is outside the spectral range.

We note that variations can exist between the fluxes and the signal-to-noise ratios of features that are detected with the 
simple fitting algorithm that we used to derive redshifts (Table~\ref{tab:specz}) and those of features that are detected with 
PAHFIT (Tables ~\ref{tab:lines1} and~\ref{tab:lines2}) because of different continuum assumptions. To get an estimate
of the deviations in the fluxes provided by various codes, we also ran the fitting algorithm of \cite{sajina07} for all spectra 
with well-determined redshifts. While we found no significant differences for fluxes of PAHs at $\lambda$$>$10 \micron , 
systematic differences were observed for the short-wavelength features. 
The fluxes of both the 6.2 and the 7.7 \micron\ PAHs, when derived with PAHFIT, are on average 1.5 times higher  than 
those measured with the code of \cite{sajina07}. The discrepancies are high for the 7.7 \micron\ PAH because it is 
located at the edges of the 9.7 \micron\ absorption feature and because it is partially blended with the 8.6 \micron\ PAH. 
The statistics are poor for the 6.2 \micron\ PAH because it is either undetected or outside the spectral range in many 
sources.  Because we would like to use 
diagnostics that depend little on the choice of model (continuum) parameters and that are applicable to the majority of 
the sources in this sample, we used the EW of the 11.3 \micron\ feature, \hbox{EW$_{11.3}$}, to define sources with MIR 
spectra resembling those of AGN or starbursts (see also \citealt{desai07}). Despite that the 11.3 \micron\ feature can be 
heavily obscured, the \hbox{EW$_{11.3}$} measurements presented in this paper are independent of the selected 
extinction law because the continuum and the PAHs are equally obscured for a screen geometry. Different geometric 
distributions of the obscuring medium could affect  \hbox{EW$_{11.3}$}. Still, we opt to use it to distinguish the MIR 
spectral type of the sources because otherwise we would be restricted in using the EW of the 12.7 \micron\ PAH. This 
would render our classification highly uncertain because the 12.7 \micron\ PAH is blended with the \neii\ line at 12.81 \micron .

To determine the  EW$_{11.3}$  thresholds that we used to define AGN and starbursts, we ran PAHFIT for the local 
templates presented in \cite{armus07}.  We used the same PAHFIT parameters for the local templates as for the sources 
in our sample. Based on this calibration, we chose to classify as star-forming galaxies those with EW$_{11.3}$$>$0.8 \micron.  
This threshold selects sources with star-formation contribution to their MIR luminosity that is likely to exceed 
$\sim$90\%, as in NGC 6240, IRAS 12112+0305, and IRAS 14348-1447. Similarly, we define as AGN the sources with 
EW$_{11.3}$$\le$0.1 \micron , since this cutoff comprises sources with AGN contribution to their MIR luminosity that is likely 
to exceed $\sim$90\%, as in IRAS 05189-2524, IRAS08572+3915, Mrk 231, Mrk 463, and Mrk 1014 (\citealt{armus07}).

\section{Results}
\label{sec:results}

\subsection{Redshift distribution and estimated IR luminosities of the sources} 
\label{results:z_ir}
\begin{figure*}
\centering
\includegraphics[width=8cm]{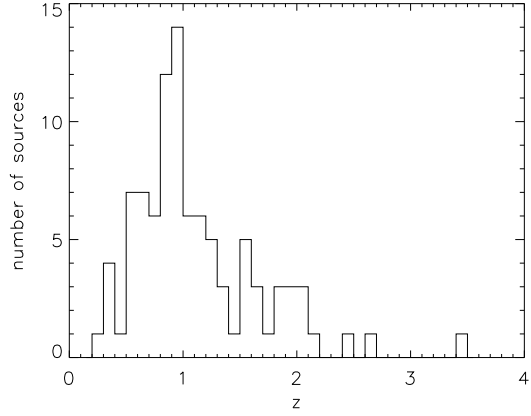}
\caption{\label{fig:z_dist} Redshift distribution of the 95 sources with reliable z measurements.}
\end{figure*}

Using redshifts derived from the IRS spectra, we find that the redshift distribution of the sources in this
sample peaks at $z$$=$1, with average and median $z$ values of 1.12 and 0.96,
respectively. The $z$ range that this sample spans is fairly high, with the lowest $z$ and the 
highest $z$ source being at $z$=0.3 and $z$$=$3.5, respectively (Fig.~\ref{fig:z_dist}). 

\begin{figure*}
\centering
\includegraphics[width=8cm]{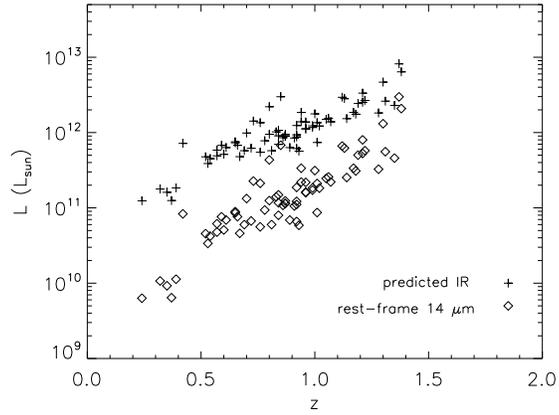} 

\caption{\label{fig:lum} Infrared luminosities of the sources  with reliable z measurements. The 
monochromatic, rest-frame 14 \micron\ luminosities of these sources are computed from the IRS 
spectra and are plotted as diamonds. The predicted bolometric IR luminosities (crosses) are 
estimated based on the relation between L$_{14}$ and L$_{IR}$ presented in \cite{sajina08}. Most 
sources are LIRGs at $z<1$ and ULIRGs at $z\gtrsim1$. }
\end{figure*}

To obtain an estimate of each source's bolometric IR luminosity, \lir,  we used the 14 $\micron$ continuum 
luminosity, which is available for most sources and which is little affected by PAH emission. We used the relation 
between L$_{14}$ and  \lir\ presented in \cite{sajina08} because of the similarities in the IRS spectra of the two 
IR-bright galaxy populations. We find that most sources at $z$$\gtrsim$1 are ULIRGs, with  \lir$\ge$10$^{12}$ \lsun . 
Most sources  at $z$$<$1 are luminous infrared galaxies (LIRGs), with 10$^{11}$$\le$\lir$<$10$^{12}$ \lsun\
(Fig.~\ref{fig:lum}). Because of the scatter in the relation that we used,  the computation of \lir\ from
 L$_{14}$ can be uncertain by a factor of at least 2.

\subsection{Detection of weak lines through spectral stacking}
\label{results:stack}
To investigate for weak lines that are hard to detect in individual spectra of high-$z$ galaxies, we performed 
a stacking analysis using sources 
with well determined redshifts. We began the stacking procedure by dividing the flux (per frequency) of 
each source by a factor of (1+$z$) in order to correctly measure line fluxes at the rest frame. We determined 
the continuum of each source by fitting  a spline function at feature-free wavelengths and subtracted it,
so that the weak features that we are trying to detect are not diluted into the continuum of the brightest 
galaxies. Moreover,  adding spectra with different MIR continuum slopes could create artificial bumps in 
the resulting spectrum at the wavelengths where the individual spectra start or end. We then stacked the 
continuum-subtracted spectra by computing their weighted average using their uncertainty images as weights.
To enhance the probability of detecting weak lines, we stacked sources that already have line detections. 
For this reason, we used the spectra of all sources with \neii\ emission, which  originates from an ion that is 
found both in star-forming regions and in AGN. 

\begin{figure*}
\centering
\includegraphics[width=8cm]{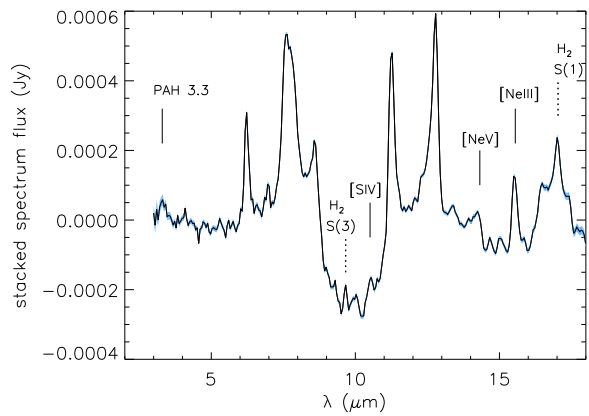}
\caption{\label{fig:neii_stack}
Stacked spectrum of sources with \neii\ detections.  Spectral stacking enables the significant detection of the PAH feature at 3.3 \micron , 
of H$_2$ lines, and of a line at 14.3 \micron\, which is likely to be \nev. The 1-$\sigma$ uncertainty around the spectrum is indicated with a 
filled area.}
\end{figure*}

The resulting stacked spectrum is presented in Fig.~\ref{fig:neii_stack}.
It contains molecular lines, i.e., the H$_2$ 0-0 S(1) and S(3) lines, detected with fluxes of 
2.65 ($\pm$0.30)$\times$10$^{-23}$ W cm$^{-2}$ and 2.42 ($\pm$0.41)$\times$10$^{-23}$ W cm$^{-2}$, 
respectively. It is unclear whether the line at 6.9 \micron\ originates from the H$_2$ 0-0 S(5) transition, since the 
H$_2$ line could be blended with [\ion{Ar}{2}] at 6.99 \micron. Assuming that the stacked spectrum is at $z$$=$0.85, 
which is the average redshift of the individual spectra used for its construction, we find that the S(1) and S(3) line 
luminosities are 2.42 ($\pm$0.27)$\times$$10^8$ \lsun\ and 2.21 ($\pm$0.37)$\times$$10^8$ \lsun ,  respectively. 
Using the median \lir\ value of the sources that were used for the computation of the stacked spectrum, 
2.2$\times$10$^{12}$ \lsun , we find that the ratio of the S(1) and S(3) line luminosities to \lir\ is of order 
10$^{-4}$, as in several local ULIRGs (\citealt{higdon06}). We computed the excitation temperature and the 
mass of the warm molecular hydrogen using the S(1) and S(3) line luminosities and the technique and assumptions 
of \cite{higdon06}. We find that the H$_2$ gas has an excitation temperature of 360 ($\pm$5) K and a mass of 
2.06 ($\pm$ 0.26)$\times$10$^8$ \msun , similar to those in local IR-bright galaxies (\citealt{rigopoulou02}; 
\citealt{higdon06}).

\begin{figure*}
\centering
\includegraphics[width=8cm]{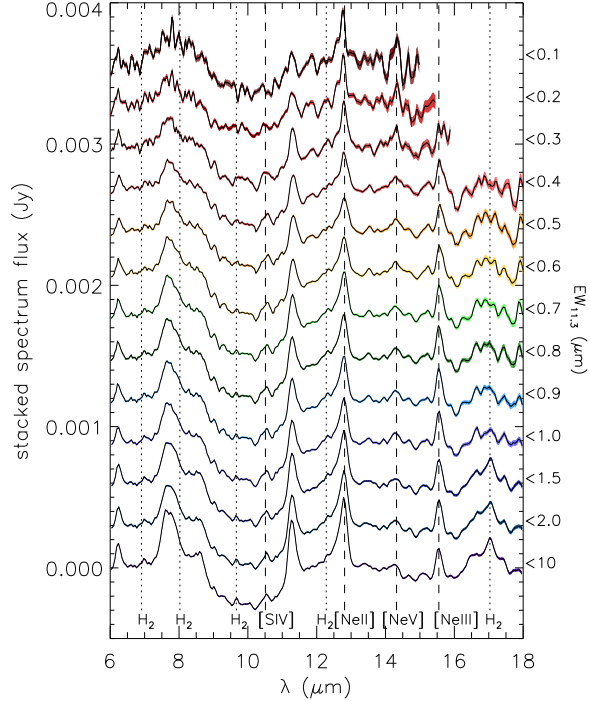}
\caption{\label{fig:ew_stack}
Stacked spectra of sources with 11.3 \micron\ PAH detections. Each stacked spectrum was created using sources
that have EW$_{11.3}$ below a certain threshold. We used an EW$_{11.3}$ threshold that increased from 0.1 to 10
\micron ,  leading to spectra that are shown in this figure from top to bottom, respectively. By construction, each stacked 
spectrum contains all individual galaxy spectra that were included in the computation of the stacked spectra displayed 
above it.  For viewing clarity, an arbitrary constant was set as the continuum of each spectrum.  The 1$\sigma$ uncertainty 
around each spectrum is indicated  with a filled area. The strength of the line at 14.3 \micron\ increases with decreasing 
EW$_{11.3}$  and it is highest for AGN-dominated sources, indicating that the emission is mostly originating from the 
high-ionization \nev\ line instead of its neighboring  \clii\ line. The H$_2$ emission is primarily detected in spectra that 
comprise sources with strong PAH emission. }
\end{figure*}

Spectral stacking also leads to the detection of a line at 14.32 \micron\ (Fig.~\ref{fig:neii_stack}),  which corresponds to \nev\ 
and which is frequently found in AGN. \nev\ can be blended with \clii\ at 14.37 \micron , which is primarily found in spectra of 
star-forming galaxies because  \ion{Cl}{2} only needs 23.81 eV to be ionized to \ion{Cl}{3}. To test whether this line 
primarily originates from \ion{Ne}{5} or \ion{Cl}{2} ions, we ran our stacking algorithm in a manner than can allow us to investigate 
how the stacked spectrum changes for different AGN and star-formation luminosities. For this reason, we stacked sources with 
different 11.3 \micron\ PAH strength (Fig.~\ref{fig:ew_stack}). We find that the flux of the line increases with decreasing 
EW$_{11.3}$ (or with increasing AGN contribution to the MIR luminosity). Specifically, the  line flux increases from 
4.02$\times$10$^{-23}$ W cm$^{-2}$  for sources with EW$_{11.3}$$<$0.5 \micron\ to 9.29$\times$10$^{-23}$ W cm$^{-2}$  
for sources with EW$_{11.3}$$<$0.1 \micron. This indicates that a significant fraction of the flux originates from an ion of high 
ionization potential, i.e., \nev . Since the ionization potential of \nev\ is 97.12 eV, its detection provides direct 
evidence for the presence of an AGN in some of the sources in this sample (\citealt{dasyra08b}). 
 
We also investigated whether we can detect the PAH feature at 3.3 \micron\ that has been seen in several local galaxies 
(\citealt{sturm00}; \citealt{imanishi02}; \citealt{imanishi06}; \citealt{risaliti06}). The PAH feature was detected at  $\sim$5$\sigma$ 
levels in the stacked spectrum of sources with \neii\ detections (Fig.~\ref{fig:neii_stack}) with a flux of 3.37$\times$10$^{-22}$ W cm$^{-2}$.  
Since the flux of the 6.2 \micron\ PAH is 7.50$\times$10$^{-22}$ W cm$^{-2}$, the flux ratio of the 6.2 \micron\ PAH to 
the 3.3 \micron\ PAH is 2.22, which is plausible for PAHs that are mainly neutral (\citealt{draine_li07}).  Using $z$$=$0.85 and 
\lir $=$2.2$\times$10$^{12}$\lsun\ for the stacked spectrum,  we find that the 3.3 \micron\ feature luminosity, 
L$_{3.3}$, is 3.07$\times$$10^9$\lsun\ or  $\sim$0.001$\times$\lir . The 6.2 \micron\  feature luminosity, L$_{6.2}$, is 
6.84$\times$10$^9$\lsun . The ratio L$_{6.2}$/\lir\  is $\sim$0.003, which is typical for local starbursts  (\citealt{peeters04}).  The 
3.3 \micron\ feature has also been detected with comparable L$_{3.3}$/\lir\ and  L$_{6.2}$/L$_{3.3}$ ratios in an XFLS source 
with deep IRS spectra (Sajina et al. 2009, in preparation).  Still, the L$_{3.3}$/\lir\ ratio is a factor of 3 higher than the highest 
L$_{3.3}$/\lir\ ratio found in a local ULIRG (Imanishi et al. 2006; 2008). This difference could be due to uncertainties in the 
determination of \lir\  or due to changes in the relative PAH strengths with $z$ (see \S~\ref{sec:discussion}).

\subsection{AGN vs star-formation properties}
\label{results:agn_vs_sf}

Using the EW$_{11.3}$ threshold of 0.8 \micron, we determined that the number of sources that have MIR spectra 
resembling those of starbursts is 39 (or 26\%). Similarly, the number of sources that are continuum-dominated is 49 (or 
33\%). Of these 49 sources, 31 have featureless MIR spectra and no $z$ measurement.

\begin{figure*} 
\centering
\includegraphics[width=8cm]{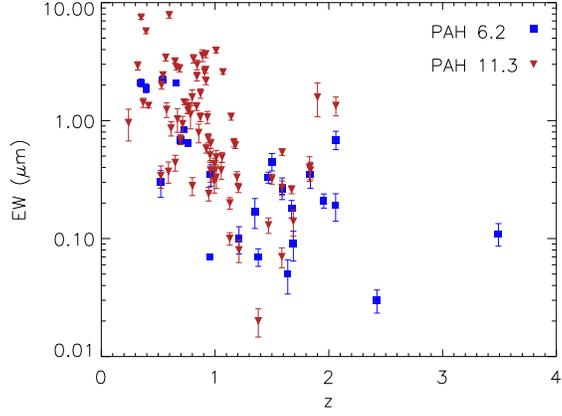}
\caption{\label{fig:ew_z} Equivalent width of PAH features vs $z$. The EW 
decreases with increasing $z$, indicating a brighter continuum underlying the features.
}
\end{figure*}

\begin{figure*}
\centering
\includegraphics[width=16cm]{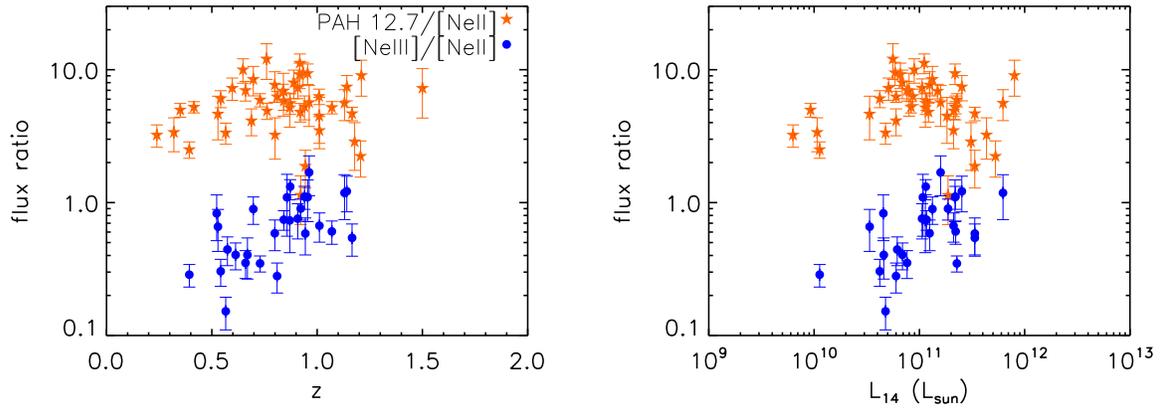}
\caption{\label{fig:line_ratios_z} Line and feature flux ratios as a function of redshift and 14 \micron\ continuum 
luminosity. The ratio of the 12.7 \micron\  PAH flux to the \neii\ flux is independent of redshift or luminosity. 
However, the \neiii $/$\neii\ flux ratio increases, implying that the radiation field becomes harder with increasing $z$ and L$_{14}$. 
}
\end{figure*}

The AGN contribution to the MIR luminosity of the sources increases with $z$. This can be demonstrated 
by the decreasing EW of PAHs with $z$ (Fig~\ref{fig:ew_z}).  Moreover, the \neiii /\neii\ flux ratio increases with $z$, indicating 
that the radiation field in the most distant sources is harder than that in the most nearby sources (Fig.~\ref{fig:line_ratios_z}). 
This effect is also seen as a function of luminosity possibly due to the flux-limited selection of the sample. 

\begin{figure*}
\centering
\includegraphics[width=8cm]{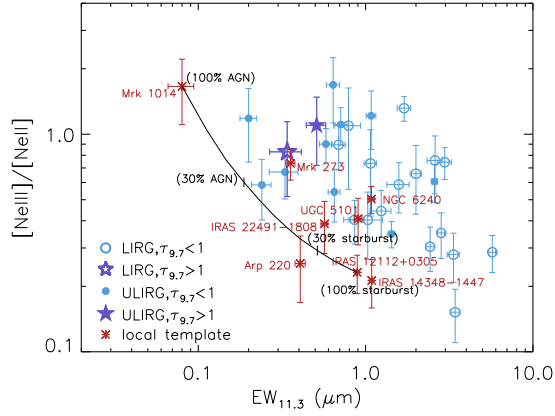}
\caption{\label{fig:ew_ratio} 
Diagnostic diagram for IR-luminous galaxy samples of wide $z$ range. LIRGs are plotted with open symbols and ULIRGs 
are plotted with filled symbols. Circles are used for sources with $\tau_{9.7}$$<$1, while five-pointed stars are used for 
sources with $\tau_{9.7}$$\ge$1. Local galaxy templates as plottes as plain stars. The solid line indicates how the position 
of a source changes  when the fraction of AGN or starburst contribution to L$_{14}$ increases from 0\% to 100\%. For 
this reason, we used  Mrk 1014 as an AGN template and IRAS 12112+0305 as a starburst template. The radiation field 
in many distant star-forming sources is harder than that in their local analogues. Moreover, their EW$_{11.3}$ values 
are often higher than those of local ULIRGs. }
\end{figure*}

In Fig.~\ref{fig:ew_ratio}, we constructed a diagnostic diagram between the \neiii /\neii\ flux ratio and EW$_{11.3}$, which 
determines the powering mechanism of a source's MIR emission. While this diagram does not use lines with large difference 
in ionic excitation potential (e.g., \citealt{genzel98}, \citealt{peeters04}, \citealt{dale06}, \citealt{armus07}) or two different states 
of the same ion (e.g., \citealt{sturm02}; \citealt{verma03}; \citealt{farrah07}), it is useful because it uses lines in the range 
11$<$$\lambda$$<$16 \micron . Hence, it can be applied to samples that span a wide $z$ range, i.e., 0$\le$$z$$\le$1.3 for 
samples observed with IRS. 
We used the local templates Mrk 1014 and IRAS 12112+0305 to demonstrate how the position of a source on this diagram 
changes when the AGN (or star-formation) fraction of L$_{14}$ increases from 0 to 100\%. We also plotted the positions of several
local templates to investigate for differences between distant IR-bright galaxies and their local analogues.  Because ionic 
lines are hard to detect in distant AGN, most of the sources in our sample that are used to populate this diagram are either 
composite or starburst dominated, i.e., they have EW$_{11.3}$ $>$0.1 and $>$0.8 \micron , respectively.  We find that, at 
intermediate and high $z$,  IR-bright galaxies often have a radiation field that is harder than that in their local analogues, with 
ULIRGs being more AGN dominated than LIRGs. Based on their locus on this diagram, several distant star-forming galaxies 
could resemble NGC 6240, which has a buried active nucleus (\citealt{komossa03}; \citealt{armus06}; \citealt{farrah07}).  Buried 
AGN have been discovered in X-ray data of $z$$\sim$2 starbursts (\citealt{huang09}). Moreover, the EW$_{11.3}$ values of 
many distant starbursts are higher than those of local ULIRGs, indicating a possible increase in star formation activity with 
$z$ (\citealt{weedman_houck}).  The few heavily obscured sources in the MIR that are seen on this diagram may be hosting 
an AGN. However, the dominant mechanism of their MIR emission is unclear. 

\subsection{IR color-color diagrams vs MIR spectral type}
\label{results:color_color}
Color-color diagrams in the IR (\citealt{lacy04}; \citealt{stern05}) have provided a way to select AGN, allowing IR samples 
to largely complement  $X$-ray samples in the identification of type-2 QSOs (\citealt{barmby06}; Poletta et al.~2006, 2007; 
\citealt{donley07}; \citealt{georgantopoulos07};  \citealt{lacy07};  \citealt{barger08}; \citealt{cardamone08}; \citealt{gorjian08}). 
These diagrams use 3.6$-$8.0 \micron\ broad-band photometry to determine whether the NIR$-$MIR SED of extragalactic sources 
resemble those of AGN. 

\begin{figure*}
\centering
\includegraphics[width=16cm]{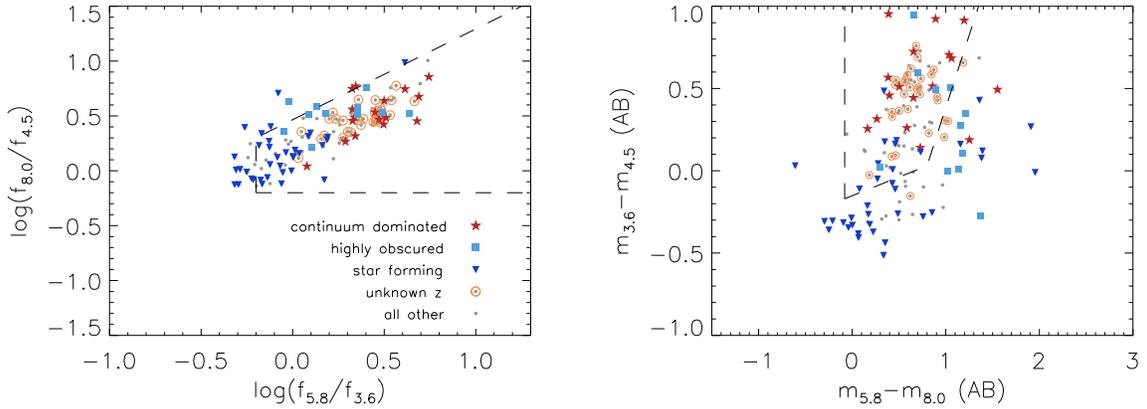}
\caption{\label{fig:mir_selection} Comparison between the IRS spectral classification and the position of sources on 
(observed-frame) color-color diagrams that are used for the selection of AGN based on broad-band IR data
(\citealt{lacy04}, left panel; \citealt{stern05}, right panel). In both panels, the dashed line sets the limits of the AGN wedge.
The criterion used for the classification of continuum-dominated sources (stars) and star-forming sources (triangles)
was that the EW$_{11.3}$ is $\le$0.1 and $>$0.8 \micron , respectively. Obscured systems with $\tau_{9.7}$$\ge$1 
(and EW$<$0.8 \micron ) are plotted as squares. While most sources with a continuum-dominated IRS spectrum lie 
inside the AGN wedge, so does a large fraction of star-forming and highly-obscured systems.}
\end{figure*}

Using the updated IRAC fluxes of the sources in this sample (Table~\ref{tab:obs}) and the AGN and star-forming 
galaxy classification based on EW$_{11.3}$, we investigated how well the MIR spectral type classification agrees
with the color-color diagram classification.  For this purpose, we constructed the \cite{lacy04} and \cite{stern05} diagrams for 
the 141 sources with detections in all four IRAC channels (Fig.~\ref{fig:mir_selection}). We displayed on these diagrams 
sources with AGN and star-formation dominated spectra, and sources that are highly obscured in the MIR (with 
$\tau _{9.7}$$\ge$1). Sources without $z$ have featureless MIR continua and are therefore also classified as AGN. 
We find that the number of sources that lie inside the AGN wedge, which is the area enclosed by the dashed lines, was 
122 (or 87\%) for the \cite{lacy04} diagram and 88 (or 62\%) for the \cite{stern05} diagram. In most cases, spectroscopically 
determined AGN with or without $z$ lie inside the AGN wedge. However, some star-forming galaxies and many heavily 
obscured systems are also located inside the same wedge. A plausible explanation why  this diagram could lead to the 
misidentification of starbursts as AGN is $z$ effects. The use of rest-frame IRAC fluxes can bring 
sources outside the AGN wedge boundaries  or close to them  (\citealt{barmby06}; \citealt{donley07}; 
\citealt{higdon08}; \citealt{yun08}).

\subsection{MIR spectral properties vs NIR radial extent}
\label{results:reff}

Previous NIR imaging of distant 24 \micron\ selected sources indicated that $z$$\sim$2 ULIRGs with large radial 
extents typically have MIR spectra resembling those of star-forming galaxies. Sources with continuum-dominated 
MIR spectra are often compact, while the extent of systems with high MIR obscuration can vary (\citealt{dasyra08a};
\citealt{melbourne09}). Using this large sample, we populated the redshift range 0$\lesssim$$z$$\lesssim$2  to 
investigate whether this trend is typical of  IR-luminous galaxies at several epochs. 

\begin{figure*}
\centering
\includegraphics[width=8cm]{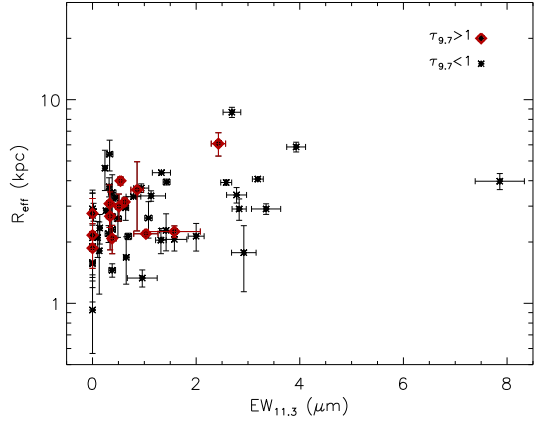}
\caption{\label{fig:extent}  Effective radii of IR-luminous galaxies in the rest-frame optical or NIR wavelengths vs EW of their 
11.3 \micron\ PAH emission.  The strongest starbursts, i.e., the sources with EW$_{11.3}$$>$2 \micron , are typically more extended 
than $\sim$3 kpc. The radial extent of the sources that are heavily obscured in the MIR can also be large.
}
\end{figure*}

We confirm that composite sources and sources that are heavily obscured in the MIR (i.e., sources with $\tau _{9.7}$$\ge$1)  
can either have a compact or an extended stellar distribution (Fig.~\ref{fig:extent}).  One of the heavily obscured sources in the 
MIR, MIPS 562, is one of the most extended sources in the $\it HST$ images with two (or possibly more) closely interacting 
gas-rich components. We also find that the sources with the strongest EW$_{11.3}$ values, i.e., EW$_{11.3}$$\ge$2 \micron , 
are typically more extended than $\sim$3 kpc (Fig.~\ref{fig:extent}). However,  there is no strong correlation between 
EW$_{11.3}$ and the observed-frame NIR radial extent of the sources. 

Scenarios of local gas-rich galaxy mergers suggest that the IR SED becomes warmer (\citealt{sanders88}), the IR 
luminosity increases (\citealt{veilleux06}), and the EW$_{\rm11.3}$ value decreases (\citealt{farrah09}) as a 
merger advances, approaching dynamical equilibrium.  In such scenarios,  the radial extent of interacting galaxies 
would correlate with EW$_{\rm 11.3}$. The lack of a strong correlation could be due to the large scatter in the merger 
timescales during which the peak of star-formation (or AGN) activity is observed (\citealt{rigopoulou99}). Moreover, 
several of the distant IR-bright galaxies may not be associated with interactions (\citealt{zheng04}; \citealt{dasyra08a}).

\section{Discussion: Changes in the MIR spectral properties with $z$ or $L$}
\label{sec:discussion}

\begin{figure*}
\centering
\includegraphics[width=8cm]{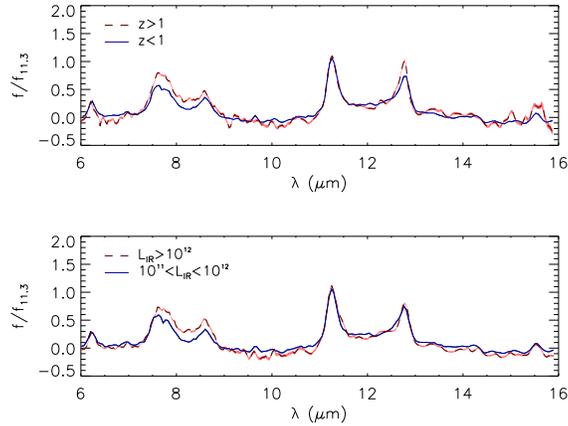}
\caption{\label{fig:lum_z_stack}
Stacked spectra of sources in different redshift and luminosity bins are shown in the top and bottom panels, respectively. The 
stacked spectrum of sources at $z$$<$1 or \lir $<$$10^{12}$\lsun\ is plotted with a solid line and that of sources at $z$$\ge$1 
or \lir $\ge$$10^{12}$\lsun\ is plotted with a dashed line. The filled area around each spectrum corresponds to its 1$\sigma$ 
uncertainty. All individual spectra were extinction corrected, continuum subtracted,  and normalized at the peak of the 
11.3 \micron\ feature so that changes in the relative shape of features can be investigated. For this reason, the stacked spectra were 
created using all sources with rest-frame 11.3 \micron\ data. We observe significant but small differences in the PAH flux ratios 
with $z$ or $L$.}
\end{figure*}

Having a flux-limited sample of IR-luminous galaxies spanning a wide $z$ range, we investigated for differences in 
the properties of spectral features with look-back time by comparing the stacked spectra of sources in different $z$ bins.  
To create a representative spectrum for each $z$ bin, we used a stacking algorithm that was similar  to that presented 
in \S~\ref{results:stack}. We first divided each spectrum by its extinction curve so that the line ratios in the stacked spectrum
are not affected by differences in the silicate depth among individual galaxies. For each extinction-corrected spectrum, we then
subtracted the continuum and divided the flux at all wavelengths by its 11.3 \micron\ value, so that all sources have similar 
11.3 \micron\ PAH strength. The resulting stacked spectrum, which was computed using all galaxy spectra with 
rest-frame 11.3 \micron\ data, was found by averaging the individual spectra using their uncertainties as weights. It is shown 
in Fig.~\ref{fig:lum_z_stack} for sources at $z$$\ge$1 and $z$$<$1.  We observe a small
but significant ($>$3$\sigma$) change in the fluxes of several PAH features with $z$. For example, the normalized flux of the 7.7 
\micron\ PAH is 22\% higher at $z$$\ge$1 than at $z$$<$1. Such flux variations are frequently seen within local galaxies 
(\citealt{galliano08}). The ratio of the 11.3 \micron\ PAH flux, f$_{11.3}$, to the flux of the PAHs at 7.7 \micron\ and 8.6 \micron\ 
decreases with increasing redshift. Changes in the PAH radius or excitation state  can reproduce this trend. For example, an 
increasing fraction of ionized PAHs leads to stronger PAH emission at 6-9 \micron\ than at 11.3 \micron\ (\citealt{draine_lee84}; 
\citealt{allamandola99}; \citealt{draine_li07};  \citealt{farrah08}). 

\begin{figure*}
\centering
\includegraphics[width=8cm]{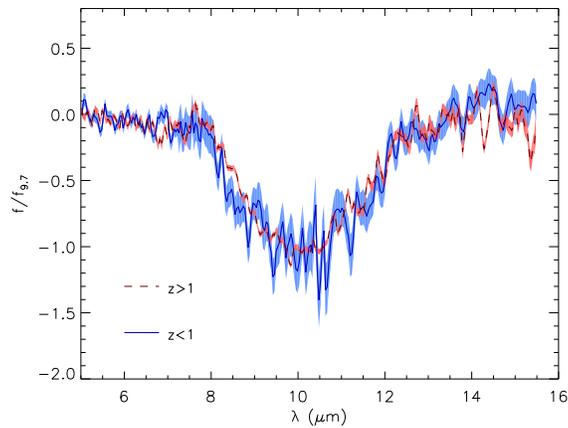}
\caption{\label{fig:ext_stack}
Stacked extinction curve spectra. The spectra for $z$$<$1 and $z$$\ge$1 are plotted with a solid line and a dashed line, respectively.
The filled area around each spectrum corresponds to its 1$\sigma$ uncertainty. We removed the continuum and emission features of 
each spectrum and divided it by its 9.7 \micron\ flux prior to computing the stacked spectrum in order to investigate for differences 
in the extinction properties with $z$. We used sources with rest-frame 9.7 \micron\ data that have moderate or high obscuration, i.e., 
$\tau_{9.7}$$\ge$0.5, and EW$_{11.3}$$<$0.8 \micron\  to avoid starbursts that could have emission-line residuals affecting  the 
shape of the silicate absorption. We find no significant ($>$3$\sigma$) change in the extinction curve properties with $z$.
}
\end{figure*}

To investigate whether some of the observed trends could be attributed to a change in the extinction curve shape 
with $z$, we computed the stacked spectrum of the 9.7 \micron\ absorption profile at $z$$<$1 and $z$$\ge$1. For this reason, 
we removed both the continuum and the best-fit profile of all lines and features in each individual spectrum. We normalized all
spectra at 9.7 \micron , and computed the average spectrum using only sources with intermediate and high extinction, i.e. 
$\tau_{9.7}$$\ge$0.5. We also used only sources with EW$_{11.3}$$<$0.8 \micron , so that possible residuals from the removal 
of lines and features do not affect the absorption feature profile. We find no significant change in the shape of the extinction at 
$z$$\ge$1 and $z$$<$1 (Fig.~\ref{fig:ext_stack}). Moreover, there is no correlation between $z$ and the depth of the silicate 
absorption feature, as computed by PAHFIT (see Table~\ref{tab:specz}), indicating that any changes in the spectral properties 
with $z$ are primarily due to differences in the PAH properties.

What remains to be tested is whether the observed changes in the spectra of IR-bright galaxies are purely 
dependent on redshift. Because of the flux-limited nature of the sample, the sources become brighter with increasing 
$z$. When computing the stacked spectra of sources as a function of luminosity, we also find that the flux ratio of 
the 11.3 \micron\ PAH over the 7.7 or the 8.6 \micron\ complex decreases with the transition from LIRGs to ULIRGs 
(Fig.~\ref{fig:lum_z_stack}; see also \citealt{farrah08}). The use of further datasets that extend the sample's \lir\ range 
in each $z$ bin will help us break this degeneracy by enabling us to compare the spectra of LIRGs and ULIRGs 
at $z$$\ge$1 and $z$$<$1. For this purpose, we will need to use 
sources from flux-limited samples of different f$_{24}$ thresholds. Including such samples will also help us to avoid 
possible biases associated with a single flux threshold, and possibly,  cosmic variance. This will be the focus of a forthcoming paper. 

\section{Conclusions}
\label{sec:conclusions}
We presented a MIR spectroscopic catalog of 150 IR-luminous galaxies in the \spi\ extragalactic first look survey. Our program 
aimed to collect data with the IRS spectrograph on board \spi\ for a purely flux-limited sample (with 
f$_{24}$$\gtrsim$0.9 mJy and m$_{\rm R}$$>$19 Vega mags) of IR-luminous galaxies in order to investigate 
for possible evolution in their MIR spectral properties with $z$.  This catalog is complemented by a large number of ancillary datasets, 
including ground-based $R$-band images, {\it HST} $H$-band imaging,  \spi\ IRAC (3.6, 4.5, 5.0, and 8.0 \micron ) and MIPS 
(24 and 70 \micron ) photometry. Our basic findings are summarized as follows.
\begin{enumerate}
\item
Of the 150 observed sources, 31 have featureless MIR continua and 24 have potential but uncertain $z$ measurements. 
Reliable spectroscopic redshifts were derived for the remaining 95 sources from MIR features. These 95 sources span a 
wide $z$ range, 0.3$\le$z$\le$3.5, with a peak at $z$$=$1.  Most of these sources are estimated to be LIRGs at 
$z$$<$1 and ULIRGs at $z$$\gtrsim$1, based on their monochromatic 14 \micron\ luminosities.
\item
We used the EW of the 11.3 \micron\ PAH to classify sources as AGN dominated (EW$_{11.3}$$\le$ 0.1 \micron  ) or star forming 
(EW$_{11.3}$$>$0.8 \micron ). We found that the sample comprises 39 star-forming galaxies, 9 of which are simultaneously 
highly obscured systems (with 9.7 \micron\ optical depth that exceeds unity). The sample also comprises 13 more highly obscured 
systems with EW$_{11.3}$$\le$0.8  \micron . The galaxies with an AGN-related power-law continuum are 49, including the 31 
sources that have no $z$ measurement.
\item
The AGN continuum emission becomes stronger with increasing $z$, as indicated by 
the decreasing EW of PAH features. Moreover, the interstellar radiation field becomes harder with increasing $z$, as determined
by the increasing \neiii /\neii\ flux ratio.  These results could be attributed to the flux-limited nature of the sample, since it comprises 
more luminous, and possibly more AGN-dominated systems, at earlier epochs.  Distant ULIRGs are more AGN-dominated than 
distant LIRGs based on a \neiii /\neii\  versus EW$_{\rm 11.3}$ diagram that is used as a diagnostic between AGN and star formation
activity. Based on their high EW$_{\rm 11.3}$ values, several distant starbursts can have an elevated star formation activity with respect 
to their local analogues.
\item
We tested how well the power source of the MIR emission, i.e., an AGN or a starburst, as inferred from the IRS spectra, 
agrees with the position of sources on 3.6$-$8 \micron\ color-color diagrams. While continuum-dominated sources fall inside 
the AGN wedge of color-color diagrams, so do some starbursts possibly due to redshift effects. Since most of the heavily 
obscured sources (with $\tau_{9.7}$$\ge$1) also fall inside the AGN wedge, they would be selected as type-2 AGN, even if 
they were undergoing nuclear starbursts.
\item
Stacking analysis enabled the detection of weak features that are often undetected in high-$z$ sources. Such an example is
the 14.32 \micron\ \nev\ line, which constitutes direct evidence for the existence of an AGN because it is emitted by an ion 
with high ionization potential.  The \nev\ line was detected in stacked spectra of sources with low EW$_{11.3}$, and its flux 
increased as the EW$_{11.3}$ decreased. The H$_2$ 0-0 S(1) and S(3) lines were observed in stacked spectra of sources 
with \neii\ emission, which comprised starburst galaxies. The excitation temperature of the warm molecular hydrogen is 360 
K and its mass is 2$\times$10$^{8}$ \msun. The luminosities of the S(1) and S(3) lines were $\sim$10$^{-4}$$\times$ \lir .
The 3.3 \micron\ PAH feature was also detected in the stacked spectra of sources with \neii\ emission. Its luminosity was of 
order $10^{-3}$$\times$\lir\ and $\sim$2 times lower than the luminosity of the 6.2 \micron\ PAH feature. 
\item
A small (but significant) change in the ratios of PAH features is observed in the stacked spectra of IR-luminous galaxies with $z$. 
The 11.3 \micron\ feature flux decreases with respect to the fluxes of the 7.7 and 8.6 \micron\ complexes from $z$$<$1 to 
$z$$\ge$1, while there is no evidence for change in the profile or the depth of the extinction curve with redshift. The same 
trend is seen as a function of luminosity. It is possible that this effect is due to changes in the PAH excitation properties.
\item
We investigated whether the MIR spectra of IR-luminous galaxies are related to their radial extents in the NIR. While we found
no strong correlation between EW$_{11.3}$  and \reff , as measured from {\it HST} NICMOS 
$H$-band images, the strongest starbursts, i.e., the sources with EW$_{11.3}$$>$2 \micron\ typically have 
\reff $\gtrsim$3 kpc. The extent of sources with composite MIR spectra or with high MIR obscuration can vary. 

\end{enumerate}

\acknowledgments
This work was based on observations made with the \spi\ Space Telescope, which is operated by JPL/Caltech 
under a contract with NASA, and on observations made with the NASA/ESA {\it Hubble} Space Telescope, obtained
at the Space Telescope Science Institute, which is operated by the Association of Universities for Research in 
Astronomy, Incorporated, under NASA contract NAS5-26555. The observations are associated with the \spi\ programs 
20629 and 40025, and the {\it HST} program 11142. The authors acknowledge support by NASA through awards 
issued by JPL/Caltech and through the grant HST-GO-11142.06-A awarded by the Space Telescope Science Institute.

\begin{figure*}
\centering
\includegraphics[width=17cm]{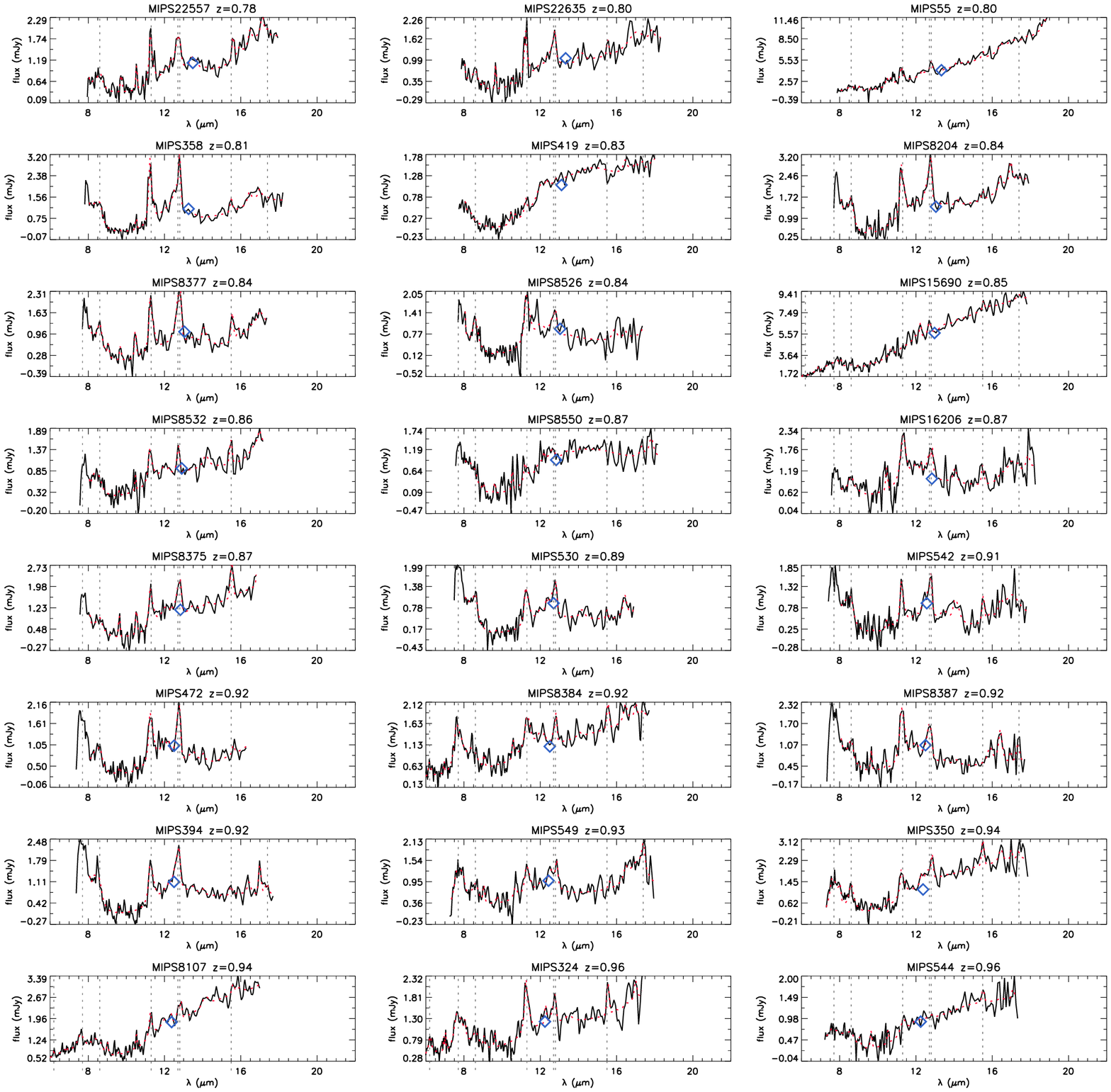} 
Figure 3$-$ continued.
\end{figure*}
\begin{figure*}
\centering
\includegraphics[width=17cm]{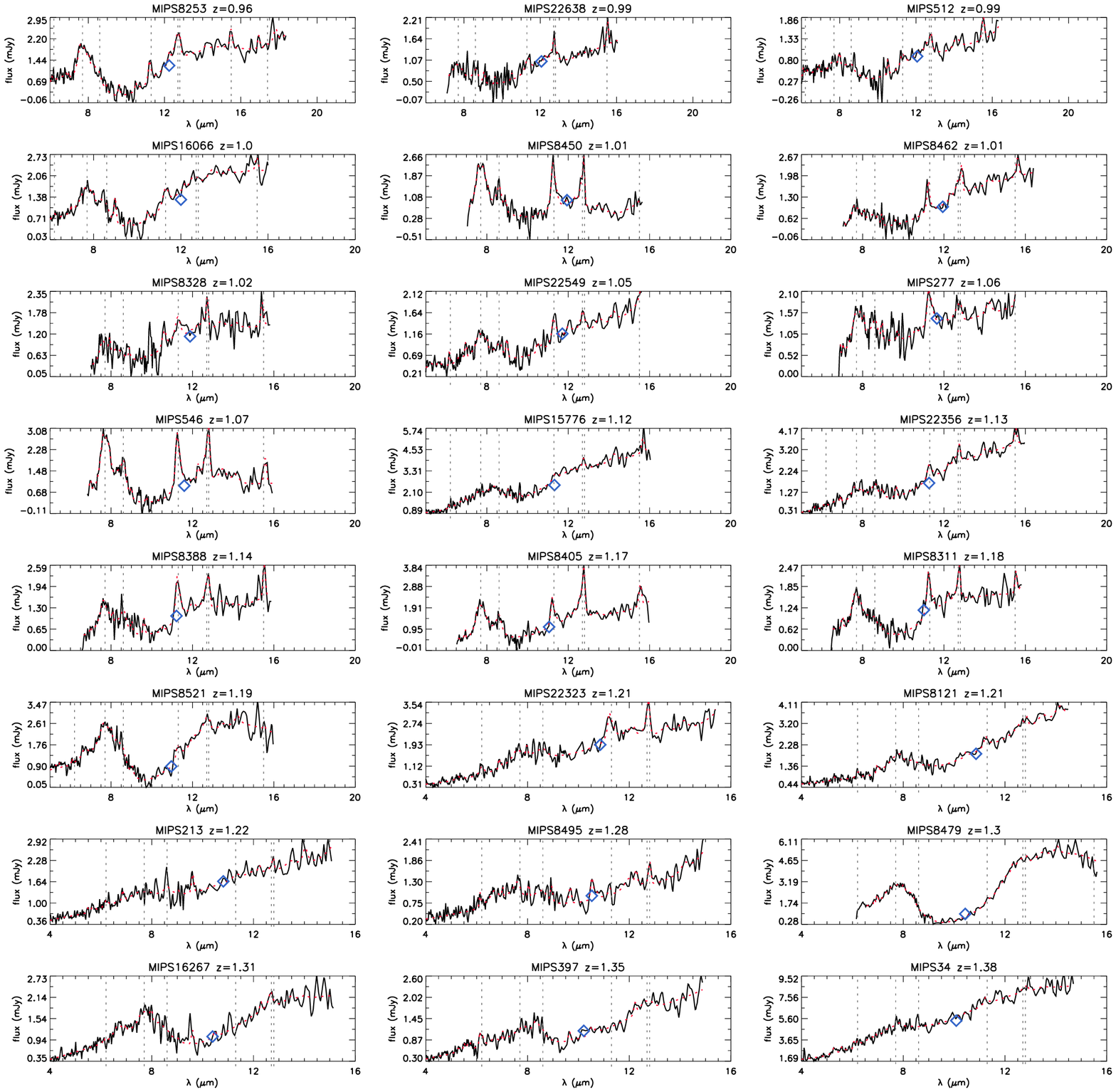}
Figure 3$-$ continued.
\end{figure*}
\begin{figure*}
\centering
\includegraphics[width=17cm]{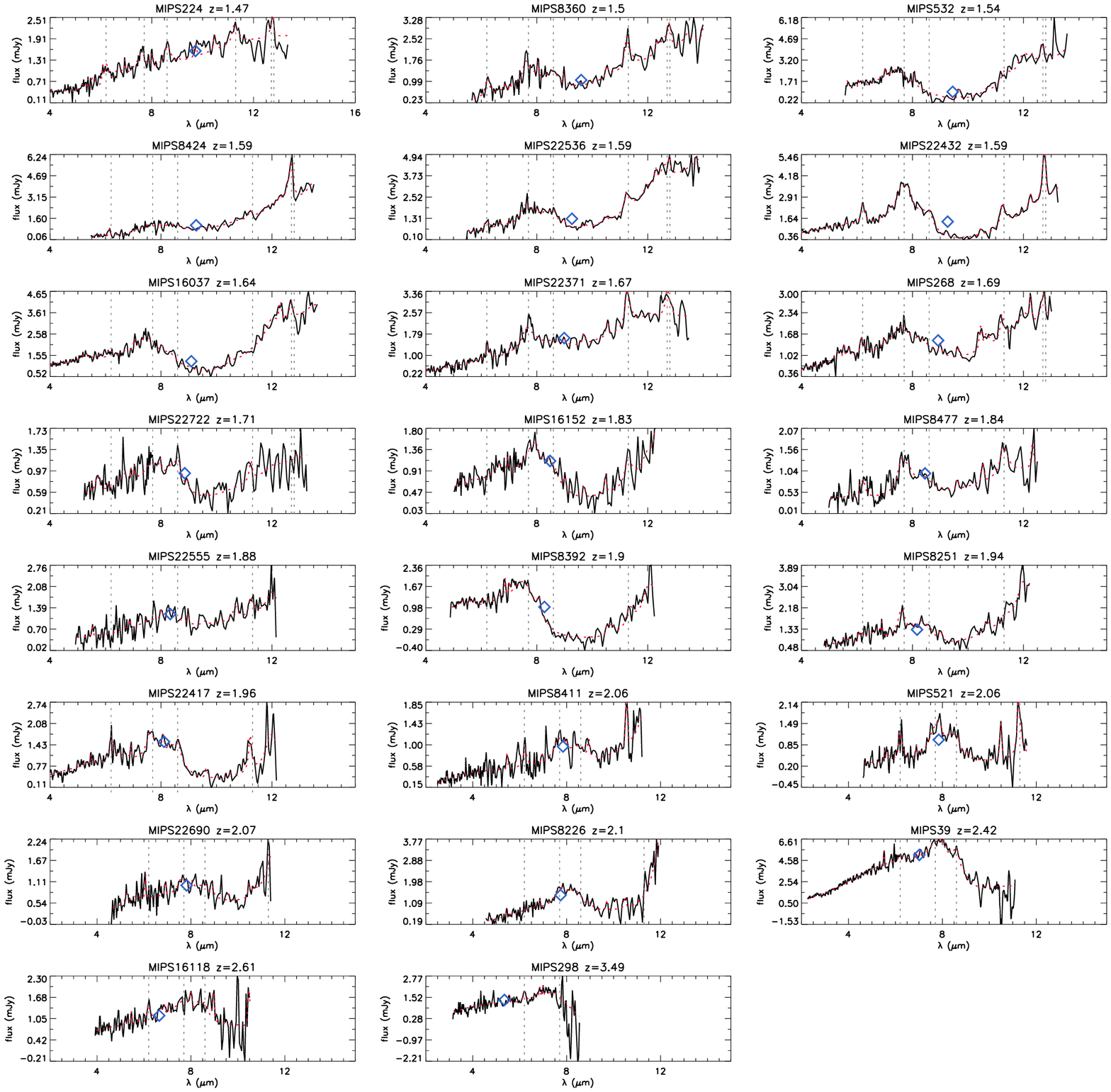}
Figure 3$-$ continued.
\end{figure*}

\begin{figure*}
\centering
\includegraphics[width=17cm]{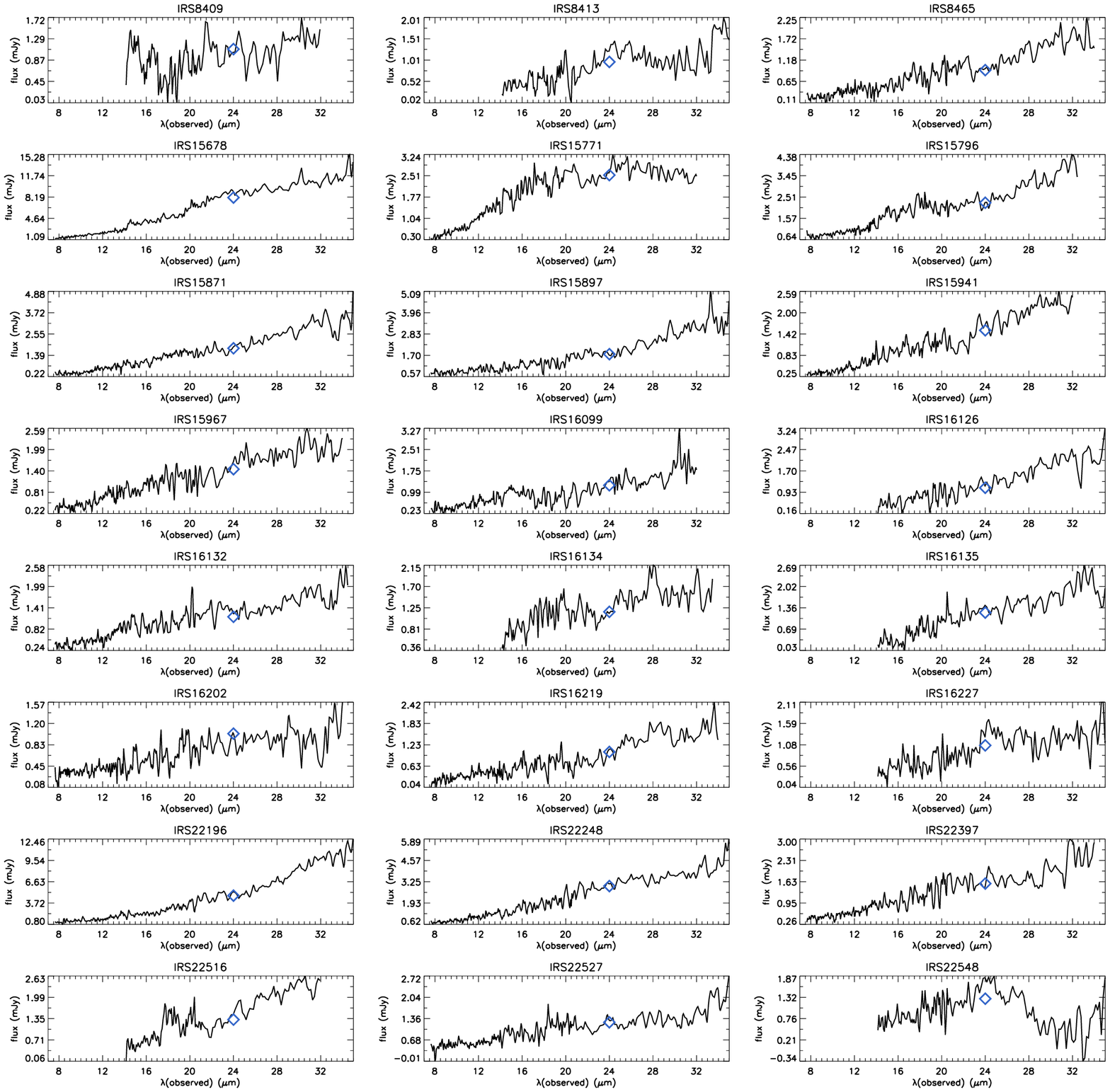}
Figure 4$-$ continued.
\end{figure*}
\begin{figure*}
\centering
\includegraphics[width=17cm]{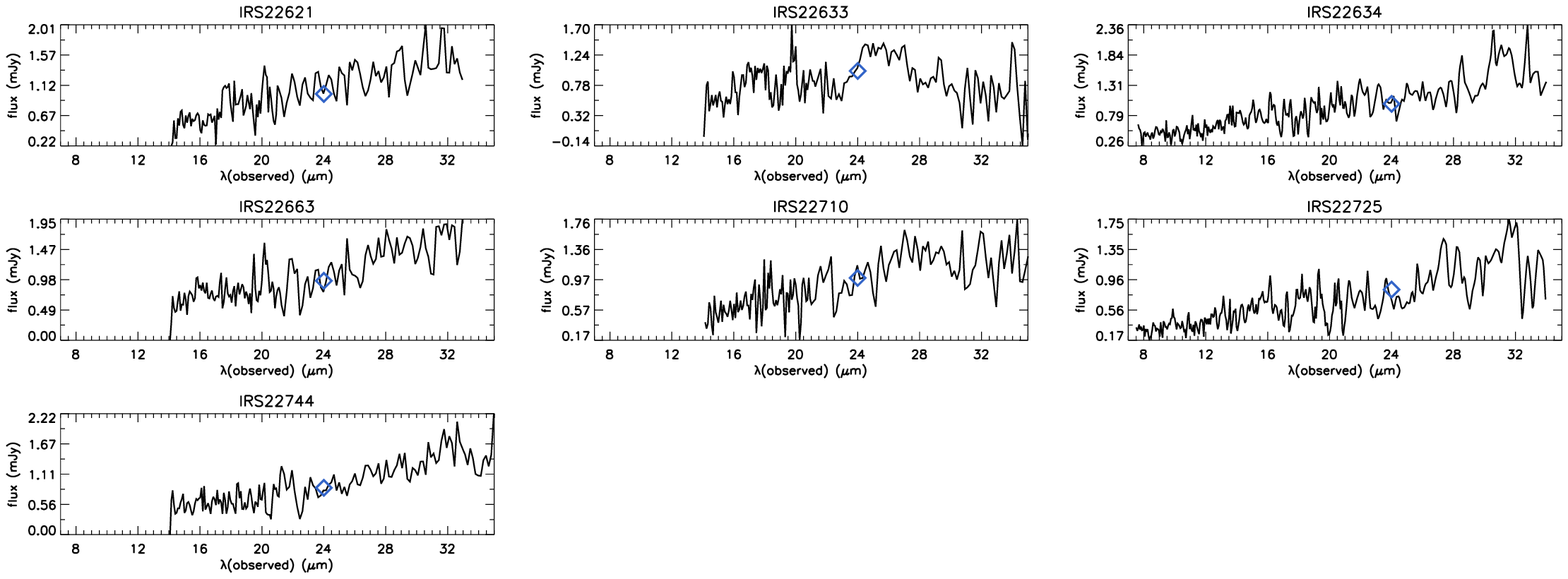}
Figure 4$-$ continued.
\end{figure*}

\clearpage

\clearpage
\oddsidemargin=-0.07in 
\topmargin=-0.35in 
\include{tab1}
\clearpage
 \oddsidemargin=-0.1in 
\include{tab2}

\clearpage
\oddsidemargin=-0.1in 
\include{tab3}
\clearpage
\oddsidemargin=-0.1in 
\include{tab4}

\clearpage
\oddsidemargin=0.0in
\topmargin=0.0in


\end{document}

%% file: tab1.tex
\begin{centering}
\begin{deluxetable}{cccccccccc}
\tablecolumns{10}
\tabletypesize{\tiny}
\tablewidth{0pt}
\tablecaption{\label{tab:obs} Coordinates, integration times, and photometric properties of the sources.}
\tablehead{ \colhead{Galaxy} & \colhead{coordinates} & \colhead{t$_{\rm int}$(SL1, LL2, LL1) }  
& \colhead{m$_{\rm R}$} & \colhead{f$_{3.6}$} & \colhead{f$_{4.5}$} & \colhead{f$_{5.8}$} & \colhead{f$_{8.0}$} & \colhead{f$_{24}$} & \colhead{\reff} \\
\colhead{} & \colhead{(J2000)} & \colhead{(s)} & \colhead{(Vega)} 
& \colhead{($\mu$Jy)} & \colhead{($\mu$Jy)} & \colhead{($\mu$Jy)} & \colhead{($\mu$Jy)} & \colhead{(mJy)} & \colhead{(kpc)} \\
\colhead{(1)} & \colhead{(2)} & \colhead{(3)} & \colhead{(4)} & \colhead{(5)} & \colhead{(6)} & \colhead{(7)} & \colhead{(8)} & \colhead{(9)} & \colhead{(10)} 
}
\startdata
   MIPS34 & 17:17:08.7+59:13:41.1 & 483, 487, 487 &  21.05 &  445.8 $\pm$45.1 &  680.1 $\pm$68.5 &  980.4 $\pm$100.6 & 1413.9 $\pm$142.9 &  5.42 $\pm$0.06 & 2.82 $\pm$0.28 \\
   MIPS39 & 17:17:50.7+58:47:45.3 & 483, 487, 731 &  23.27 &   58.3 $\pm$6.7 &  135.2 $\pm$15.0 &  322.9 $\pm$36.6 &  971.9 $\pm$100.7 &  5.10 $\pm$0.07 & \nodata \\
   MIPS55 & 17:15:06.6+58:39:39.7 & \nodata, 487, 487 &  20.75 &  189.8 $\pm$19.4 &  144.3 $\pm$15.3 &  174.1 $\pm$20.6 &  275.8 $\pm$30.6 &  4.13 $\pm$0.06 & \nodata \\
  MIPS159 & 17:13:33.4+59:27:47.2 & 483, 1462, 1706 &  22.08 &  168.7 $\pm$17.5 &  277.6 $\pm$28.7 &  499.8 $\pm$53.6 &  841.1 $\pm$87.4 &  1.88 $\pm$0.06 & \nodata \\
  MIPS168 & 17:17:55.1+59:28:45.3 & 483, 975, 1218 &  20.18 &   96.4 $\pm$10.8 &   95.5 $\pm$11.2 &   80.4 $\pm$16.1 &  486.2 $\pm$52.0 &  1.82 $\pm$0.06 & 1.33 $\pm$0.13\\
  MIPS184 & 17:14:04.5+59:03:03.7 & \nodata, 1462, 1706 &  21.56 &  130.6 $\pm$14.1 &  113.4 $\pm$12.9 &  145.6 $\pm$22.3 &  257.7 $\pm$30.9 &  1.73 $\pm$0.06 & \nodata \\
  MIPS213 & 17:13:35.7+59:27:21.8 & 483, 1462, 1706 &  23.42 &   34.3 $\pm$3.9 &   65.6 $\pm$7.4 &  140.7 $\pm$16.9 &  365.0 $\pm$38.8 &  1.65 $\pm$0.06 & 1.90 $\pm$0.52\\
  MIPS224 & 17:16:55.8+59:25:40.4 & 483, 1462, 1706 &  20.54 &  149.6 $\pm$16.0 &  248.7 $\pm$26.0 &  276.5 $\pm$35.1 &  493.1 $\pm$52.7 &  1.57 $\pm$0.06 & 1.81 $\pm$0.70 \\
  MIPS268 & 17:12:08.9+59:30:25.0 & 483, 1462, 1706 &  22.93 &   59.7 $\pm$7.0 &  116.9 $\pm$12.5 &  187.7 $\pm$25.2 &  366.2 $\pm$39.3 &  1.48 $\pm$0.06 & 2.34 $\pm$0.39 \\
  MIPS277 & 17:14:20.8+58:55:02.5 & \nodata, 1462, 1706 &  21.31 &  134.4 $\pm$14.2 &  105.3 $\pm$11.1 &   76.7 $\pm$14.7 &  139.0 $\pm$16.3 &  1.43 $\pm$0.06 & \nodata \\
  MIPS298 & 17:14:48.9+59:28:47.4 & \nodata, 1462, 1706 &  23.95 &   $<$6.5 \tablenotemark{g} &    $<$10.5 \tablenotemark{g} &   41.2 $\pm$12.8 &  163.5 $\pm$21.6 &  1.37 $\pm$0.06 & \nodata \\
  MIPS322 & 17:14:24.9+59:29:48.0 & 483, 1462, 1706 &  23.40 &   81.6 $\pm$9.2 &  131.8 $\pm$14.7 &  277.6 $\pm$34.8 &  424.2 $\pm$46.1 &  1.25 $\pm$0.06 & \nodata \\
  MIPS324 & 17:17:15.7+59:06:03.3 & 483, 1462, 1706 &  20.79 &  171.9 $\pm$18.2 &  211.1 $\pm$22.4 &  286.9 $\pm$35.6 &  271.8 $\pm$32.0 &  1.23 $\pm$0.06 & 2.94 $\pm$0.39  \\
  MIPS331 & 17:15:05.7+59:18:20.2 & 483, 1462, 1706 &  22.04 &  161.8 $\pm$17.2 &  200.1 $\pm$21.3 &  270.9 $\pm$32.9 &  455.4 $\pm$49.0 &  1.32 $\pm$0.07 & \nodata \\
  MIPS350 & 17:15:50.7+59:23:15.9 & \nodata, 1462, 1706 &  22.85 &  102.5 $\pm$10.8 &   86.2 $\pm$10.0 &  100.3 $\pm$14.5 &  159.8 $\pm$21.9 &  1.15 $\pm$0.06  \tablenotemark{d} & \nodata \\
  MIPS351 & 17:16:59.2+59:08:11.7 & 483, 1462, 1706 &  22.04 &  104.5 $\pm$11.6 &  146.2 $\pm$15.8 &  209.2 $\pm$27.8 &  331.9 $\pm$36.8 &  1.24 $\pm$0.06 & \nodata \\
  MIPS358 & 17:18:07.1+59:29:37.4 & \nodata, 1462, 1706 &  22.24 &  108.5 $\pm$11.5 &   76.3 $\pm$8.3 &   81.7 $\pm$13.8 &   86.8 $\pm$11.7 &  1.11 $\pm$0.05 & 2.91 $\pm$0.17  \\
  MIPS369 & 17:14:50.9+59:26:33.3 & \nodata, 1462, 1706 &  22.32 &   94.5 $\pm$10.2 &   93.8 $\pm$10.7 &   58.7 $\pm$11.8 &  106.9 $\pm$17.3 &  1.13 $\pm$0.06 & \nodata \\
  MIPS394 & 17:13:15.7+59:12:08.5 & \nodata, 1462, 1706 &  22.08 &  134.0 $\pm$14.0 &  105.3 $\pm$12.1 &   69.2 $\pm$12.6 &  108.5 $\pm$17.4 &  1.10 $\pm$0.06 & \nodata \\
  MIPS397 & 17:15:05.9+59:09:16.9 & 966, 2924, 3412 &  23.80 &   65.2 $\pm$7.7 &   74.2 $\pm$8.6 &  138.1 $\pm$20.9 &  270.0 $\pm$30.1 &  1.12 $\pm$0.06 & 2.79 $\pm$0.69 \\
  MIPS419 & 17:15:45.2+59:05:38.7 & \nodata, 2924, 3412 &  22.85 &   33.1 $\pm$4.6 &   42.6 $\pm$6.0 &   83.8 $\pm$16.6 &  243.7 $\pm$29.6 &  1.06 $\pm$0.06 & 1.87 $\pm$0.38 \\
  MIPS446 & 17:14:09.0+59:17:48.4 & 483, 1462, 1706 &  21.82 &  142.7 $\pm$14.8 &  140.7 $\pm$14.7 &  175.6 $\pm$21.2 &  283.3 $\pm$30.8 &  1.03 $\pm$0.06 & \nodata \\
  MIPS463 & 17:14:45.2+58:54:55.1 & \nodata, 2924, 3412 &  \nodata &   44.3 $\pm$6.0  \tablenotemark{h} &   58.2 $\pm$7.4  \tablenotemark{h} &   78.9 $\pm$16.1 &  169.1 $\pm$21.4  \tablenotemark{h} &  1.11 $\pm$0.06 & \nodata \\
  MIPS472 & 17:18:16.7+59:19:36.1 & \nodata, 1462, 1706 &  22.21 &   70.7 $\pm$16.0  \tablenotemark{d,i} &   41.7  $\pm$13.1  \tablenotemark{d,i}  &    $<$32.9 \tablenotemark{g} &   62.0 $\pm$12.6 &  1.03 $\pm$0.06 & \nodata \\
  MIPS488 & 17:15:11.9+58:49:35.0 & \nodata, 2924, 3412 &  21.16 &  119.2 $\pm$12.7 &   88.3 $\pm$10.1 &   98.4 $\pm$16.4 &  116.9 $\pm$16.4 &  1.08 $\pm$0.06 & 3.40 $\pm$0.30  \\
  MIPS495 & 17:14:33.0+58:58:19.8 & \nodata, 1462, 1706 &  21.84  \tablenotemark{b} &  159.3 $\pm$16.9 &  155.6 $\pm$16.9 &  171.4 $\pm$23.9 &  202.8 $\pm$25.1 &  1.05 $\pm$0.06 & \nodata \\
  MIPS505 & 17:15:59.4+59:27:33.4 & \nodata, 1462, 1706 &  24.91 &   17.2 $\pm$2.8 &   31.5 $\pm$5.4 &   63.1 $\pm$11.5 &  187.8 $\pm$23.8 &  1.00 $\pm$0.06 & \nodata \\
  MIPS509 & 17:14:22.1+59:28:14.3 & \nodata, 1462, 1706 & \nodata &   $<$6.9  \tablenotemark{g} &   17.3 $\pm$3.9 &  $<$36.3 \tablenotemark{g} &  136.4 $\pm$20.0 &  0.98 $\pm$0.06 & \nodata \\
  MIPS512 & 17:16:55.7+59:10:46.1 & 483, 1462, 1706 &  23.21 &  117.7 $\pm$12.7 &  162.7 $\pm$17.4 &  173.4 $\pm$24.0 &  272.0 $\pm$31.1 &  0.89 $\pm$0.06 & 2.19 $\pm$0.21 \\
  MIPS521 & 17:13:41.4+58:57:03.7 & \nodata, 1462, 1706 &  \nodata \tablenotemark{b} &   64.5 $\pm$7.4 \tablenotemark{d,i} &  75.8 $\pm$8.4 \tablenotemark{d,i} & 93.7 $\pm$19.6 \tablenotemark{d,i} & 129.1 $\pm$18.7 \tablenotemark{d,i} &  1.00 $\pm$0.06 & \nodata \\
  MIPS530 & 17:13:37.1+58:46:37.5 & \nodata, 2924, 3412 &  22.18 &  120.4 $\pm$12.7 &  123.8 $\pm$8.6 \tablenotemark{g} &  179.2 $\pm$21.2 \tablenotemark{g} &   102.291 $\pm$16.1 \tablenotemark{g} &  0.91 $\pm$0.06 & \nodata \\
  MIPS532 & 17:15:26.1+58:56:32.4 & \nodata, 2924, 3412 &  22.70 &   51.5 $\pm$6.7 &   56.9 $\pm$7.5 &   62.9 $\pm$15.1 &  185.8 $\pm$23.1 &  0.98 $\pm$0.06 & \nodata \\
  MIPS537 & 17:17:59.3+59:21:56.3 & 483, 1462, 1706 &  19.90 \tablenotemark{a,b} &   88.7 $\pm$10.1 &  113.6 $\pm$12.6  &   48.6 $\pm$14.1 &  282.2 $\pm$32.5 &  0.94 $\pm$0.06 & \nodata \\
  MIPS542 & 17:12:45.7+59:32:14.4 & \nodata, 1462, 1706 &  22.06 &  119.7 $\pm$13.0 &  108.1 $\pm$12.1 &   87.2 $\pm$16.8 &   93.6 $\pm$15.4 &  0.90 $\pm$0.05 & \nodata \\
  MIPS544 & 17:13:07.7+58:44:13.3 & \nodata, 1706, 1950 &  23.03 &   49.0 $\pm$5.8 &   67.7 $\pm$8.2 &   74.2 $\pm$13.4 &  227.2 $\pm$28.2 &  0.91 $\pm$0.06 & 2.08 $\pm$0.33 \\
  MIPS546 & 17:12:16.1+59:11:22.0 & \nodata, 1462, 1706 &  22.08 &  172.5 $\pm$18.1 &  141.8 $\pm$15.0 &  103.0 $\pm$17.7 &  119.7 $\pm$15.9 &  0.93 $\pm$0.06 & 3.92 $\pm$0.11\\
  MIPS549 & 17:16:28.9+59:29:21.0 & \nodata, 1462, 1706 &  21.54 &   80.2 $\pm$9.1 &   62.8 $\pm$7.8 &   82.6 $\pm$15.3 &   96.8 $\pm$16.5 &  0.98 $\pm$0.05 & \nodata \\
  MIPS562 & 17:12:39.6+58:59:55.1 & 483, 1462, 1706 &  20.90 &  214.0 $\pm$43.8 \tablenotemark{d,i} &  165.5 $\pm$24.8 \tablenotemark{d,i} &  140.2 $\pm$21.8 &  281.6 $\pm$31.2 &  1.01 $\pm$0.06 & 6.08 $\pm$0.80 \\
 MIPS7985 & 17:13:25.5+60:07:20.2 & 483, 487, 487 &  22.57 &  168.6 $\pm$17.5 &  269.1 $\pm$28.1 &  493.1 $\pm$53.0 &  947.0 $\pm$98.5 &  4.87 $\pm$0.06 & \nodata \\
  MIPS8040 & 17:13:12.0+60:08:40.2 & 483, 731, 975 &  \nodata \tablenotemark{j} &  \nodata \tablenotemark{j} &  357.7 $\pm$36.5 &  352.1 $\pm$41.7 &  529.6 $\pm$54.9 &  2.88 $\pm$0.07 & \nodata \\
 MIPS8069 & 17:15:00.4+59:56:11.7 & 483, 975, 975 & 19.12 \tablenotemark{c} &  386.9 $\pm$39.2 &  545.3 $\pm$54.8  \tablenotemark{e} &  695.0 $\pm$72.6 &  972.0 $\pm$97.8  \tablenotemark{e} &  2.33 $\pm$0.03 & \nodata \\
 MIPS8071 & 17:12:43.5+60:06:50.0 & 483, 1462, 1706 &  22.10 &  120.8 $\pm$13.0 &  216.9 $\pm$22.9 &  368.8 $\pm$41.9 &  782.4 $\pm$82.3 &  2.38 $\pm$0.06 & \nodata \\
 MIPS8098 & 17:16:08.0+60:06:38.8 & 483, 1462, 1706 &  21.83 &  167.8 $\pm$17.8 &  194.5 $\pm$20.2 &  189.1 $\pm$25.3 &  292.6 $\pm$31.9 &  2.02 $\pm$0.06 & \nodata \\
 MIPS8107 & 17:16:38.7+59:49:44.5 & 483, 14	62, 1706 &  22.17 &   104.1 $\pm$10.6 \tablenotemark{e} &  132.8 $\pm$13.5 \tablenotemark{e} &  217.8 $\pm$23.1 \tablenotemark{e} &  449.2 $\pm$45.0 \tablenotemark{e} &  1.84 $\pm$0.03 & 4.61 $\pm$1.03  \\
 MIPS8121 & 17:13:22.8+60:10:44.6 & 483, 1462, 1706 &  23.43 &   153.4 $\pm$31.5  \tablenotemark{d,i} & 258.7 $\pm$41.2 \tablenotemark{d,i} &  481.4 $\pm$53.8  &  685.9 $\pm$73.1  &  1.89 $\pm$0.06 & \nodata \\
 MIPS8157 & 17:18:40.6+60:09:35.3 & 483, 1462, 1706 &  23.31 &  142.9 $\pm$15.3 &  189.6 $\pm$19.9 &  226.4 $\pm$30.1 &  563.4 $\pm$58.5 &  1.66 $\pm$0.06 & \nodata \\
 MIPS8172 & 17:16:49.1+59:49:18.6 & 483, 1462, 1706 &  23.06 &  \nodata \tablenotemark{j} &  \nodata \tablenotemark{j} &  473.1 $\pm$53.2 &  798.7 $\pm$83.0 \tablenotemark{f}  &  1.46 $\pm$0.03 & \nodata \\
 MIPS8179 & 17:16:15.0+60:13:32.4 & 483, 1462, 1706 &  20.87 &  123.5 $\pm$13.4 &  138.8 $\pm$14.9 &  243.6 $\pm$31.5 &  274.2 $\pm$30.8 &  1.53 $\pm$0.06 & 2.68 $\pm$0.52 \\
 MIPS8185 & 17:14:22.8+60:28:34.6 & 483, 1218, 1462 &  23.80 &   87.5 $\pm$9.7 &  149.9 $\pm$15.8 &  262.4 $\pm$33.5 &  572.0 $\pm$59.1 &  1.57 $\pm$0.06  \tablenotemark{d} & \nodata \\
 MIPS8192 & 17:12:53.9+60:05:00.8 & \nodata, 1462, 1706 &  23.09 &   52.3 $\pm$5.9 &   63.1 $\pm$7.2 &   86.6 $\pm$13.7 &  214.2 $\pm$24.9 &  1.46 $\pm$0.06 & \nodata \\
 MIPS8204 & 17:14:40.9+59:58:19.2 & \nodata, 1462, 1706 &  22.83 &  163.1 $\pm$16.8 &  152.0 $\pm$15.8 &  149.9 $\pm$19.5 &  222.6 $\pm$24.8 &  1.39 $\pm$0.03 & 2.04 $\pm$0.29 \\
 MIPS8224 & 17:16:49.0+59:53:35.5 & 483, 1462, 1706 &  23.53 &   94.5 $\pm$9.6 \tablenotemark{e} &  167.6 $\pm$17.1 \tablenotemark{e} &  260.9 $\pm$27.4 \tablenotemark{e} &  465.5 $\pm$47.1 \tablenotemark{e} &  1.44 $\pm$0.03 & \nodata \\
 MIPS8226 & 17:12:04.7+60:16:32.0 & \nodata, 1462, 1706 &  20.95 &   30.3 $\pm$4.3 &   72.8 $\pm$8.3 &  144.6 $\pm$22.2 &  206.9 $\pm$23.9 &  1.42 $\pm$0.06 & 0.93$\pm$0.36 \\
 MIPS8233 & 17:17:31.5+59:40:33.8 & 483, 1462, 1706 &  21.92 &  143.0 $\pm$14.4 \tablenotemark{e} &  198.8 $\pm$20.2 \tablenotemark{e} &  270.0 $\pm$28.2 \tablenotemark{e} &  398.4 $\pm$40.5 \tablenotemark{e} &  1.32 $\pm$0.03 & \nodata \\
 MIPS8251 & 17:12:21.7+60:03:33.6 & \nodata, 1462, 1706 &  23.22 &   28.2 $\pm$4.2 &   44.5 $\pm$5.8 &   59.6 $\pm$14.8 &  249.0 $\pm$28.8 &  1.31 $\pm$0.06 & 2.10 $\pm$0.76 \\
 MIPS8253 & 17:15:07.8+60:13:52.9 & 483, 1462, 1706 &  21.78 &  113.4 $\pm$12.4 &  114.4 $\pm$12.7 &  154.1 $\pm$22.3 &  440.4 $\pm$47.0 &  1.25 $\pm$0.06 & 2.99 $\pm$0.45 \\
 MIPS8308 & 17:16:34.4+60:15:44.0 & \nodata, 1462, 1706 &  20.29 &   89.4 $\pm$10.1 &   96.0 $\pm$10.8 &   67.8 $\pm$15.5 &  242.2 $\pm$28.0 &  1.26 $\pm$0.06 & 2.28 $\pm$0.47 \\
 MIPS8311 & 17:15:27.7+60:07:26.3 & \nodata, 1462, 1706 &  22.08 &  100.4 $\pm$11.2 &  100.2 $\pm$11.3 &   89.6 $\pm$17.7 &  229.1 $\pm$28.5 &  1.17 $\pm$0.06 & 3.15 $\pm$0.15 \\
 MIPS8315 & 17:13:57.0+59:44:37.6 & 483, 1462, 1706 &  23.88 &   80.6 $\pm$9.3 &  135.2 $\pm$15.0 &  245.7 $\pm$29.7 &  482.7 $\pm$51.8 &  1.17 $\pm$0.04 & \nodata \\
 MIPS8325 & 17:16:43.7+60:20:13.6 & \nodata, 1462, 1706 &  21.05 &   69.8 $\pm$8.0 &   62.9 $\pm$7.6 &   84.9 $\pm$16.8 &  129.6 $\pm$18.1 &  1.13 $\pm$0.06 & 3.61 $\pm$1.34 \\
 MIPS8328 & 17:16:06.7+59:44:54.6 & \nodata, 1462, 1706 &  22.33 &   95.5 $\pm$9.7 \tablenotemark{e} &   89.9 $\pm$9.4 \tablenotemark{e} &   76.9 $\pm$9.3 \tablenotemark{e} &  129.5 $\pm$14.1 \tablenotemark{e} &  1.14 $\pm$0.03 & 2.60 $\pm$0.49 \\
 MIPS8360 & 17:16:52.5+59:35:31.9 & \nodata, 1462, 1706 &  22.07 &   96.7 $\pm$9.8 \tablenotemark{e} &  107.0 $\pm$11.2 \tablenotemark{e} &   88.1 $\pm$10.5 \tablenotemark{e} &  120.9 $\pm$13.1 \tablenotemark{e} &  1.04 $\pm$0.03 & 3.73 $\pm$0.40 \\
 MIPS8371 & 17:13:25.0+59:45:57.8 & 483, 1462, 1706 & 20.03 \tablenotemark{b,c} &  176.8 $\pm$39.2  \tablenotemark{d,i}  &  205.2 $\pm$19.6  \tablenotemark{d,i} &  131.5 $\pm$18.4 &  381.6 $\pm$24.5  \tablenotemark{d,i} &  1.12 $\pm$0.06 & \nodata \\
 MIPS8375 & 17:14:33.8+59:52:21.8 & \nodata, 1462, 1706 &  21.06 \tablenotemark{b,c} &  143.5 $\pm$4.7 \tablenotemark{h, g} &  144.5 $\pm$5.9 \tablenotemark{h, g,e} &   72.6 $\pm$16.0  \tablenotemark{d} &   108.2 $\pm$12.2  \tablenotemark{e,d}  &  1.16 $\pm$0.03 \tablenotemark{d} & \nodata \\
  MIPS8377 & 17:17:33.5+59:46:40.7 & \nodata, 1462, 1706 &  22.17  \tablenotemark{b} &  101.7 $\pm$33.1 \tablenotemark{d,i,e} &   76.8 $\pm$22.4 \tablenotemark{d,i,e} &   89.9 $\pm$9.8 \tablenotemark{e} &   74.1 $\pm$8.4 \tablenotemark{e,d} &  1.03 $\pm$0.03 & \nodata \\
 MIPS8384 & 17:16:47.5+60:17:57.2 & 483, 1462, 1706 &  22.39 &  232.3 $\pm$23.7 &  238.8 $\pm$24.6 &  242.8 $\pm$27.7 &  340.1 $\pm$36.4 &  1.09 $\pm$0.06 & \nodata \\
 MIPS8387 & 17:18:02.8+60:15:22.0 & \nodata, 1462, 1706 &  21.33 \tablenotemark{a, c} &  280.1 $\pm$28.8 &  201.1 $\pm$21.2 &  190.8 $\pm$26.2 &  151.7 $\pm$20.5 &  1.06 $\pm$0.06 & 8.67 $\pm$0.50  \\
 MIPS8388 & 17:14:00.4+60:14:09.5 & \nodata, 1462, 1706 &  21.42 &  112.4 $\pm$12.3 &  130.6 $\pm$14.2 &  113.4 $\pm$18.9 &  173.1 $\pm$21.6 &  1.05 $\pm$0.06 & 2.62 $\pm$0.53  \\
 MIPS8392 & 17:13:43.9+59:57:14.5 & \nodata, 2924, 3412 &  \nodata &   15.5 $\pm$2.7 &   23.1 $\pm$4.3 &   63.7 $\pm$12.6 &  223.0 $\pm$26.8 &  1.01 $\pm$0.06 &  2.25 $\pm$0.15 \\
 MIPS8400 & 17:13:13.9+59:55:06.2 & \nodata, 1462, 1706 &  20.11 &  115.5 $\pm$12.6 &  125.1 $\pm$14.0 &  165.3 $\pm$23.3 &  243.9 $\pm$29.7 &  1.00 $\pm$0.06 & \nodata \\
 MIPS8401 & 17:12:38.5+59:42:33.4 & \nodata, 1462, 1706 &  20.42 &  122.2 $\pm$13.0 &   88.9 $\pm$9.7 &  132.7 $\pm$18.7 &  128.0 $\pm$16.7 &  1.02 $\pm$0.06 & \nodata \\
 MIPS8405 & 17:15:27.7+60:17:31.2 & \nodata, 1462, 1706 &  21.28 &  144.3 $\pm$15.5 &  160.9 $\pm$17.4 &  108.6 $\pm$18.5 &  123.1 $\pm$18.5 &  1.05 $\pm$0.06 & 1.68 $\pm$0.44 \\
 MIPS8407 & 17:16:12.8+60:19:49.0 & \nodata, 1462, 1706 &  24.26 &   54.3 $\pm$6.4 &   83.3 $\pm$9.6 &  124.3 $\pm$18.6 &  242.3 $\pm$28.5 &  1.00 $\pm$0.06 & \nodata \\
 MIPS8409 & 17:17:36.0+60:15:40.1 & \nodata, 1462, 1706 &  20.73  \tablenotemark{b} &  105.5 $\pm$11.7 \tablenotemark{h} &   80.5 $\pm$9.7 \tablenotemark{h} &   75.5 $\pm$16.1  &  100.9 $\pm$16.0 \tablenotemark{h} &  1.09 $\pm$0.06 \tablenotemark{h} & \nodata \\
 MIPS8411 & 17:14:03.3+60:16:56.6 & 483, 1462, 1706 &  23.69 &   48.3 $\pm$6.3 &   90.8 $\pm$9.8 &  105.0 $\pm$18.0 &  279.1 $\pm$30.2 &  0.97 $\pm$0.06 & \nodata \\
 MIPS8413 & 17:15:45.7+59:51:56.4 & \nodata, 1462, 1706 &   \nodata &   22.0 $\pm$2.3 \tablenotemark{e} &   31.5 $\pm$3.9 \tablenotemark{e}  &   73.7 $\pm$8.5 \tablenotemark{e} &  139.4 $\pm$15.0 \tablenotemark{e} &  0.97 $\pm$0.03 & \nodata \\
 MIPS8424 & 17:16:21.9+59:56:52.2 & \nodata, 1706, 1950 &  22.95 &   44.7 $\pm$5.7 &   56.6 $\pm$6.2  \tablenotemark{e}  &   53.5 $\pm$12.7 &   62.3 $\pm$8.1  \tablenotemark{e}  &  1.10 $\pm$0.04 & \nodata \\
 MIPS8430 & 17:13:33.2+60:10:19.5 & \nodata, 1462, 1706 &  22.03 & 95.9 $\pm$6.91  \tablenotemark{d,i} &  106.4 $\pm$12.2 &  119.9 $\pm$18.5 &  235.2 $\pm$28.8 &  0.97 $\pm$0.06 & 2.20 $\pm$0.05\\
 MIPS8450 & 17:18:26.0+59:53:53.5 & \nodata, 1462, 1706 &  21.38 &  124.7 $\pm$12.6 \tablenotemark{e} &   88.5 $\pm$9.2 \tablenotemark{e} &   70.0 $\pm$8.9 \tablenotemark{e} &   86.0 $\pm$10.1 \tablenotemark{e} &  0.96 $\pm$0.03 & 5.86 $\pm$0.32 \\
 MIPS8462 & 17:13:09.5+60:08:16.0 & \nodata, 1462, 1706 &  21.64 &   90.7 $\pm$10.2 &   73.9 $\pm$8.3 &   53.3 $\pm$14.8 &  130.6 $\pm$17.0 &  0.99 $\pm$0.06 & 5.40$\pm$0.93 \\
 MIPS8465 & 17:13:01.8+59:37:22.4 & 483, 1462, 1706 &  22.61 &   65.0 $\pm$7.2 &  108.0 $\pm$11.4 &  194.3 $\pm$23.9 &  298.5 $\pm$32.1 &  0.93 $\pm$0.06 & \nodata \\
 MIPS8477 & 17:13:44.4+60:15:31.5 & \nodata, 1706, 1950 &  22.16 &   91.9 $\pm$10.2 \tablenotemark{h} &   84.0 $\pm$9.7 \tablenotemark{h} &  124.6 $\pm$19.5 \tablenotemark{h} &  251.1 $\pm$30.3 \tablenotemark{h} &  0.98 $\pm$0.06 \tablenotemark{h} & 1.45 $\pm$0.11 \\
 MIPS8479 & 17:13:02.1+59:55:00.2 & \nodata, 2924, 3412 &  24.23 &   19.6 $\pm$3.2 &   23.4 $\pm$4.3 &   $<$38.9 \tablenotemark{g} &   70.5 $\pm$15.1 &  0.99 $\pm$0.06 & 2.76 $\pm$0.51 \\
 MIPS8495 & 17:14:49.0+59:53:38.8 & 483, 1462, 1706 &  22.79 &  100.0 $\pm$10.6 &  127.1 $\pm$13.2  \tablenotemark{e} &  212.3 $\pm$24.4 &  363.7 $\pm$37.0  \tablenotemark{e} &  0.94 $\pm$0.03 & \nodata \\
 MIPS8499 & 17:12:16.5+59:57:10.9 & \nodata, 1462, 1706 &  20.51 &  261.2 $\pm$26.9 &  192.6 $\pm$19.8 &  159.3 $\pm$20.8 &  159.7 $\pm$18.7 &  0.91 $\pm$0.06 & 3.98 $\pm$0.36 \\
 MIPS8507 & 17:13:29.9+59:44:34.2 & \nodata, 1706, 1950 &  20.89 &  150.5 $\pm$15.8 &  113.4 $\pm$12.8 &  148.3 $\pm$20.8 &  112.9 $\pm$17.7 &  0.91 $\pm$0.06 & 4.38 $\pm$0.02 \\
 MIPS8521 & 17:12:18.9+60:10:03.8 & 483, 1462, 1706 &   \nodata &  107.6 $\pm$11.7 &  169.3 $\pm$18.0 &  243.4 $\pm$29.6 &  555.4 $\pm$58.2 &  0.89 $\pm$0.06 & 3.09 $\pm$0.73 \\
 MIPS8526 & 17:16:28.4+59:44:22.9 & \nodata, 1462, 1706 &  22.44 &  131.5 $\pm$13.3 \tablenotemark{e} &  98.4 $\pm$10.4 \tablenotemark{e} &   88.0 $\pm$10.4 \tablenotemark{e} &   80.7 $\pm$9.5 \tablenotemark{e} &  0.92 $\pm$0.03 & \nodata \\
 MIPS8532 & 17:16:23.3+59:36:46.9 & \nodata, 1462, 1706 &  21.73 &   93.1 $\pm$9.5 \tablenotemark{e} &   69.9 $\pm$7.3 \tablenotemark{e} &   62.8 $\pm$8.2 \tablenotemark{e} &   73.2 $\pm$8.5 \tablenotemark{e} &  0.91 $\pm$0.03 & 3.34 $\pm$0.04 \\
 MIPS8543 & 17:18:12.7+59:39:22.6 & \nodata, 1462, 1706 &  20.65 &   96.3 $\pm$9.8 \tablenotemark{e} &   73.9 $\pm$7.8 \tablenotemark{e} &   82.2 $\pm$10.0 \tablenotemark{e} &   81.7 $\pm$9.6 \tablenotemark{e} &  0.94 $\pm$0.03 & 4.07 $\pm$0.02 \\
 MIPS8550 & 17:18:14.6+59:56:05.7 & \nodata, 1462, 1706 &  22.66 &   76.1 $\pm$8.9 &   59.2 $\pm$6.4 \tablenotemark{e} &   72.5 $\pm$15.9 &  255.5 $\pm$26.4 \tablenotemark{e} &  0.92 $\pm$0.04 & \nodata \\
 MIPS15678 & 17:23:28.4+59:29:47.3 & 483, 487, 487 &  22.15 &  339.9 $\pm$14.1  \tablenotemark{d,i} &  612.2 $\pm$62.3 &  972.4  $\pm$102.6 & 1699.6  $\pm$172.6 &  8.10 $\pm$0.07 & \nodata \\
MIPS15690 & 17:19:22.4+60:14:56.2 & 483, 487, 487 &  22.20 &  125.4 $\pm$13.6 &  149.1 $\pm$16.3 &  277.3 $\pm$34.8 &  879.1 $\pm$91.6 &  5.66 $\pm$0.07 & 2.91 $\pm$0.69 \\
MIPS15755 & 17:18:34.9+59:45:34.1 & 483, 731, 975 &  20.72 &  262.0 $\pm$26.3 \tablenotemark{e} &  272.7 $\pm$27.6 \tablenotemark{e} &  403.9 $\pm$41.0 \tablenotemark{e} &  521.5 $\pm$53.0 \tablenotemark{e} &  2.74 $\pm$0.03 & 3.94 $\pm$0.13 \\
MIPS15771 & 17:20:11.4+59:36:00.9 & 483, 731, 975 &  24.00 &   20.3 $\pm$3.2 &   38.2 $\pm$4.8 &  111.1 $\pm$18.0 &  387.5 $\pm$41.3 &  2.53 $\pm$0.03 & \nodata \\
MIPS15776 & 17:20:50.7+59:32:54.6 & 483, 1462, 1706 &  22.19 &  204.8 $\pm$21.5 &  328.0 $\pm$33.8 &  607.2 $\pm$66.0 &  965.3 $\pm$99.8 &  2.51 $\pm$0.06 & 2.10 $\pm$0.42 \\
MIPS15796 & 17:18:44.9+60:00:26.2 & 483, 731, 975 &  24.16 &  191.3 $\pm$19.5 &  313.1 $\pm$32.0 &  520.4 $\pm$54.0 &  822.9 $\pm$84.7 &  2.26 $\pm$0.06 & \nodata \\
MIPS15871 & 17:19:17.3+59:29:29.3 & 483, 1462, 1706 &  23.26 &   48.7 $\pm$6.0 &   98.0 $\pm$11.0 &  225.0 $\pm$30.3 &  422.1 $\pm$45.7 &  1.77 $\pm$0.05 & \nodata \\
MIPS15897 & 17:20:17.0+59:27:33.1 & 483, 1462, 1706 &  22.20 &  129.8 $\pm$13.6 &  223.9 $\pm$23.6 &  380.5 $\pm$42.6 &  647.7 $\pm$67.6 &  1.76 $\pm$0.05 & \nodata \\
MIPS15941 & 17:22:47.8+60:09:46.7 & 483, 1462, 1706 &  23.44 &   52.9 $\pm$6.8 \tablenotemark{d,i} & 97.9 $\pm$14.9 \tablenotemark{d,i} &  162.6 $\pm$20.8  &  316.7 $\pm$30.3 \tablenotemark{f,i} &  1.52 $\pm$0.06 & \nodata \\
MIPS15967 & 17:19:24.7+59:28:58.7 & 483, 1462, 1706 &  20.93 &  147.3 $\pm$15.7 &  160.6 $\pm$16.9 &  219.4 $\pm$28.1 &  336.5 $\pm$36.2 &  1.45 $\pm$0.05 & \nodata \\
MIPS15999 & 17:19:35.0+59:42:41.3 & \nodata, 1462, 1706 &  22.27 &  34.5 $\pm$5.1 \tablenotemark{e,d,i} &   40.9 $\pm$5.9 &   53.6 $\pm$7.6 \tablenotemark{e} &   83.0 $\pm$16.1 &  1.37 $\pm$0.03 & \nodata \\
MIPS16037 & 17:21:33.9+59:50:46.9 & 483, 1462, 1706 &  24.31 &   89.9 $\pm$9.5 &  214.6 $\pm$22.7 &  388.5 $\pm$42.3 &  711.6 $\pm$73.8 &  1.26 $\pm$0.06 & 2.16 $\pm$0.30 \\
MIPS16047 & 17:23:01.4+59:40:54.0 & 483, 1462, 1706 &  20.95 &  164.1 $\pm$17.1 &  167.9 $\pm$18.1 &  209.6 $\pm$26.2 &  275.3 $\pm$32.3 &  1.29 $\pm$0.06 & 2.68 $\pm$0.85 \\
MIPS16066 & 17:21:23.7+59:56:17.4 & 483, 1462, 1706 &  21.99  \tablenotemark{a,c} &  185.1 $\pm$19.6 &  171.0 $\pm$17.9 &  211.7 $\pm$28.0 &  515.1 $\pm$54.8 &  1.30 $\pm$0.06 & \nodata \\
MIPS16099 & 17:19:37.5+60:01:24.9 & 483, 1462, 1706 &  23.11 &   74.2 $\pm$8.4 &  129.7 $\pm$14.5 &  225.7 $\pm$28.8 &  421.1 $\pm$45.8 &  1.24 $\pm$0.06 & \nodata \\
MIPS16118 & 17:22:16.3+59:47:00.8 & \nodata, 1462, 1706 &  \nodata &   52.3 $\pm$6.8 &   35.6 $\pm$5.1 &   $<$35.9 \tablenotemark{g} &  146.4 $\pm$19.8 &  1.14 $\pm$0.06 & 2.70 $\pm$0.77 \\
MIPS16126 & 17:22:27.5+60:04:30.8 & \nodata, 1462, 1706 &  22.20 \tablenotemark{a} &   28.9 $\pm$4.3 &   38.2 $\pm$5.5 &   66.2 $\pm$15.4 &  170.1 $\pm$22.2 &  1.09 $\pm$0.06 & \nodata \\
MIPS16132 & 17:21:08.3+59:54:16.0 & 483, 1462, 1706 &  22.00 &  164.3 $\pm$10.4  \tablenotemark{d,i} &  222.3 $\pm$23.5 &  338.8 $\pm$38.0 &  502.3 $\pm$53.6 &  1.16 $\pm$0.06 & \nodata \\
MIPS16134 & 17:20:46.7+59:37:42.7 & \nodata, 1462, 1706 &  23.44 &   28.5 $\pm$4.2 &   52.1 $\pm$6.1 &  101.3 $\pm$18.7 &  239.6 $\pm$26.8 &  1.17 $\pm$0.05 & \nodata \\
MIPS16135 & 17:22:10.0+59:40:05.5 & \nodata, 1462, 1706 &  22.51 &   53.9 $\pm$17.3  \tablenotemark{d, i} &   60.9 $\pm$11.9  \tablenotemark{d, i} &   96.4 $\pm$17.7 &  174.4 $\pm$22.2 &  1.21 $\pm$0.06 & \nodata \\
MIPS16152 & 17:20:55.8+59:35:41.8 & \nodata, 1462, 1706 &  23.01 &   44.2 $\pm$5.6  \tablenotemark{d, i} &  70.4 $\pm$10.8  \tablenotemark{d, i} &   99.9 $\pm$17.8 &  264.5 $\pm$30.8 &  1.12 $\pm$0.06 & \nodata \\
MIPS16156 & 17:22:56.8+59:30:58.7 & \nodata, 1462, 1706 &  21.47 \tablenotemark{b} &  128.3 $\pm$36.0 \tablenotemark{d, i} &  101.5 $\pm$22.4 \tablenotemark{d, i} &   61.8 $\pm$15.1 &  135.6 $\pm$18.6 &  1.16 $\pm$0.06 & 3.68 $\pm$0.18 \\
MIPS16170 & 17:22:33.3+60:15:14.4 & \nodata, 1462, 1706 &  20.52 &   99.3 $\pm$10.9 &   91.6 $\pm$10.8 &   $<$41.7 \tablenotemark{g}  &  225.1 $\pm$27.4 &  1.09 $\pm$0.06 & 1.77 $\pm$0.63 \\
MIPS16202 & 17:22:02.1+59:32:55.6 & 483, 1462, 1706 &  22.64 &   66.9 $\pm$8.1 &   94.5 $\pm$10.7 &  162.8 $\pm$22.0 &  281.6 $\pm$32.9 &  1.02 $\pm$0.06 & \nodata \\
MIPS16206 & 17:21:10.4+60:06:08.2 & \nodata, 1462, 1706 &  \nodata &   94.5 $\pm$10.4  \tablenotemark{d} &  147.7 $\pm$15.7  \tablenotemark{d} &   82.1 $\pm$15.4   &  112.3 $\pm$15.9 &  0.98 $\pm$0.06  \tablenotemark{d} & \nodata \\
MIPS16219 & 17:20:13.2+60:10:28.3 & 483, 1462, 1706 &  22.88 &   58.4 $\pm$6.9 &  110.5 $\pm$12.3 &  165.7 $\pm$23.5 &  315.1 $\pm$36.0 &  1.03 $\pm$0.06 & \nodata \\
MIPS16227 & 17:19:20.2+60:14:48.7 & \nodata, 1462, 1706 &  23.92 &   40.3 $\pm$5.1 &   63.3 $\pm$7.8 &  109.4 $\pm$18.9 &  233.0 $\pm$28.0 &  1.06 $\pm$0.06 & \nodata \\
MIPS16249 & 17:22:08.4+59:45:54.4 & \nodata, 1462, 1706 &  20.47 &   54.3 $\pm$5.9 &   60.7 $\pm$7.3 &   36.7 $\pm$9.6 &  132.4 $\pm$17.9 &  0.99 $\pm$0.06 & 2.14 $\pm$0.33 \\
MIPS16267 & 17:19:10.3+60:13:16.0 & 483, 1462, 1706 &  24.36 &   47.9 $\pm$5.6 &   93.4 $\pm$11.0 &  155.0 $\pm$20.1 &  283.6 $\pm$33.3 &  1.04 $\pm$0.06 & 1.56 $\pm$0.37 \\
MIPS22196 & 17:22:45.2+59:03:28.3 & 483, 487, 487 &  21.00 & 215.8 $\pm$19.6  \tablenotemark{d, i}  & 358.9 $\pm$29.9  \tablenotemark{d, i} &  618.3 $\pm$66.1 & 1041.1  $\pm$106.5 &  4.73 $\pm$0.06 & \nodata \\
MIPS22235 & 17:20:17.2+59:16:37.4 & 483, 731, 975 &  20.57 \tablenotemark{b} &  583.6 $\pm$58.8  \tablenotemark{d} &  401.2 $\pm$41.0  \tablenotemark{d} &  281.5 $\pm$32.6 &  299.6 $\pm$33.1  \tablenotemark{d} &  3.16 $\pm$0.07  \tablenotemark{d}& \nodata \\
MIPS22248 & 17:20:03.6+59:19:08.0 & 483, 731, 975 &  21.98 &  114.4 $\pm$12.4 &  180.5 $\pm$19.1 &  321.2 $\pm$38.1 &  609.7 $\pm$64.9 &  2.98 $\pm$0.07 & \nodata \\
MIPS22307 & 17:19:51.4+58:42:22.8 & 483, 975, 1218 &  20.50 &  171.9 $\pm$17.7 &  197.7 $\pm$21.0 &  254.9 $\pm$29.5 &  403.6 $\pm$44.1 &  2.02 $\pm$0.06 & 2.14 $\pm$0.08 \\
MIPS22323 & 17:22:34.1+59:10:01.7 & 483, 731, 975 &  21.60 &  251.7 $\pm$25.9 &  230.8 $\pm$24.3 &  238.6 $\pm$28.8 &  403.7 $\pm$44.2 &  1.93 $\pm$0.07 & \nodata \\
MIPS22352 & 17:21:47.7+58:53:55.9 & 483, 1462, 1706 &  20.52 &  228.8 $\pm$23.8 &  218.5 $\pm$22.6 &  229.5 $\pm$30.8 &  294.4 $\pm$32.7 &  1.72 $\pm$0.06 & 2.91 $\pm$0.35 \\
MIPS22356 & 17:20:14.6+59:14:43.6 & 483, 1462, 1706 &  23.37 &  110.8 $\pm$12.2 &  127.3 $\pm$13.9 &  200.7 $\pm$26.5 &  334.0 $\pm$37.1 &  1.69 $\pm$0.06 & \nodata \\
MIPS22371 & 17:23:00.5+59:11:07.9 & 483, 1462, 1706 &   \nodata &   72.4 $\pm$8.4 &  115.9 $\pm$12.8 &  173.9 $\pm$23.8 &  340.7 $\pm$37.1 &  1.64 $\pm$0.06 & 2.84 $\pm$0.05 \\
MIPS22379 & 17:19:18.6+59:00:42.6 & 483, 1462, 1706 &  20.87 &  227.3 $\pm$23.4 &  176.6 $\pm$18.4 &  256.7 $\pm$29.9 &  358.0 $\pm$38.3 &  1.58 $\pm$0.06 & 3.29 $\pm$0.11 \\
MIPS22397 & 17:20:06.0+59:17:45.3 & 483, 1462, 1706 &  23.12 &   99.5 $\pm$11.1 &  169.9 $\pm$17.7 &  284.4 $\pm$35.0 &  489.7 $\pm$50.7 &  1.57 $\pm$0.06 & \nodata \\
MIPS22417 & 17:23:01.6+59:15:01.5 & 483, 1462, 1706 &  22.36 &   57.8 $\pm$6.4 &   86.9 $\pm$9.5 &  162.5 $\pm$20.5 &  296.8 $\pm$31.9 &  1.52 $\pm$0.06 & \nodata \\
MIPS22432 & 17:21:36.9+59:16:07.8 & 483, 1462, 1706 &  21.80 \tablenotemark{a, c}  &   84.8 $\pm$9.5 &  146.6 $\pm$16.0 &  261.1 $\pm$32.3 &  500.2 $\pm$53.3 &  1.44 $\pm$0.06 & 4.00 $\pm$0.19 \\
MIPS22516 & 17:21:33.7+59:07:28.1 & \nodata, 1462, 1706 &  22.26 &  67.7 $\pm$8.1 \tablenotemark{h} &   69.0 $\pm$8.1  \tablenotemark{h} &   67.0 $\pm$15.5 &   97.2 $\pm$14.4 &  1.32 $\pm$0.06 & \nodata \\
MIPS22527 & 17:18:15.5+59:12:00.7 & 483, 1462, 1706 &  22.87 \tablenotemark{b,c}  &   90.9 $\pm$9.7 &  139.7 $\pm$14.5 &  259.9 $\pm$29.9 &  433.3 $\pm$44.5 &  1.24 $\pm$0.06 & \nodata \\
MIPS22536 & 17:21:08.3+59:01:16.0 & \nodata, 1462, 1706 & \nodata \tablenotemark{b} &  109.6 $\pm$11.6 &  157.7 $\pm$16.9 &   69.5 $\pm$12.5 &  115.1 $\pm$15.6 &  1.28 $\pm$0.06 & \nodata \\
MIPS22548 & 17:23:30.5+58:45:44.8 & \nodata, 1462, 1706 &   \nodata &    9.2 $\pm$2.8 &   36.6 $\pm$5.4 &   45.8 $\pm$13.0 &  228.7 $\pm$27.4 &  1.28 $\pm$0.06 & \nodata \\
MIPS22549 & 17:22:13.9+59:17:40.5 & 483, 1462, 1706 &  23.64 &   81.6 $\pm$8.8 &  134.2 $\pm$14.8 &  232.0 $\pm$27.9 &  427.8 $\pm$46.4 &  1.17 $\pm$0.06 & 2.30 $\pm$0.01 \\
MIPS22555 & 17:21:31.0+58:42:49.0 & \nodata, 1462, 1706 &  23.71 \tablenotemark{a, c} &   78.4 $\pm$9.1 \tablenotemark{h} &  104.8 $\pm$12.1 \tablenotemark{h} &  152.3 $\pm$22.0  \tablenotemark{h}  &  194.1 $\pm$25.7  \tablenotemark{h}  &  1.19 $\pm$0.06  \tablenotemark{h}  & \nodata \\
MIPS22557 & 17:20:25.5+59:13:59.1 & \nodata, 1462, 1706 &  20.88 &  128.8 $\pm$13.5 &   86.1 $\pm$9.9 &   63.2 $\pm$12.6 &   87.4 $\pm$16.0 &  1.12 $\pm$0.06 & 3.37 $\pm$0.20 \\
MIPS22621 & 17:23:04.2+58:57:05.0 & \nodata, 1462, 1706 &  \nodata &   18.3 $\pm$3.3 &   23.9 $\pm$3.6 &   62.6 $\pm$15.7 &  136.9 $\pm$17.1 &  0.99 $\pm$0.06 & \nodata \\
MIPS22633 & 17:22:57.5+59:16:07.3 & \nodata, 1462, 1706 & 22.81 &   59.7 $\pm$7.4  \tablenotemark{h} &   77.0 $\pm$9.5   \tablenotemark{h} &   62.8 $\pm$15.1  \tablenotemark{h} &  116.0 $\pm$17.8   \tablenotemark{h} &  1.00 $\pm$0.06   \tablenotemark{h} & \nodata \\
MIPS22634 & 17:21:13.6+59:22:15.4 & 483, 1462, 1706 &  22.05 &  119.5 $\pm$12.5 &  147.6 $\pm$23.9  \tablenotemark{d, i}  &  206.3  $\pm$ 24.1 &  438.4  $\pm$ 47.5 &  0.99  $\pm$ 0.06 & \nodata \\
MIPS22635 & 17:20:55.0+59:10:39.8 & \nodata, 1462, 1706 &  22.64 &   49.8 $\pm$5.4 &   31.0 $\pm$ 5.0 &   37.2 $\pm$8.7 &   50.7 $\pm$13.7 &  1.04 $\pm$0.06 & 2.06 $\pm$0.25 \\
MIPS22638 & 17:21:54.7+58:54:36.6 & \nodata, 1462, 1706 &  23.04 &   67.9 $\pm$8.1  \tablenotemark{f} &   57.7 $\pm$7.3  \tablenotemark{f} &    $<$34.4 \tablenotemark{g} &  104.9 $\pm$18.1 &  1.04 $\pm$0.06 & 3.49 $\pm$0.78 \\
MIPS22663 & 17:23:20.7+59:03:43.2 & \nodata, 1462, 1706 &  23.15 &   39.3 $\pm$5.5 &   58.2 $\pm$6.8 &  111.0 $\pm$18.7 &  257.2 $\pm$29.1 &  0.96 $\pm$0.06 & \nodata \\
MIPS22690 & 17:19:49.0+58:54:10.2 & \nodata, 1462, 1706 &   \nodata &   23.3 $\pm$3.3 &   37.4 $\pm$5.0 &   73.6 $\pm$13.2 &  162.5  $\pm$20.4 &  1.01 $\pm$0.06 & 1.58 $\pm$0.56 \\
MIPS22710 & 17:23:28.5+58:52:25.4 & \nodata, 1462, 1706 &  24.25 &   30.3 $\pm$4.0 &   53.5 $\pm$ 6.5 &  105.2 $\pm$16.5 &  234.8 $\pm$27.4 &  0.99 $\pm$0.06 & \nodata \\
MIPS22722 & 17:18:45.4+58:51:22.8 & \nodata, 1706, 1950 &  23.33 &    7.7 $\pm$2.3 \tablenotemark{g} &  17.8 $\pm$ 3.1  \tablenotemark{g} &   37.3 $\pm$11.5   \tablenotemark{g}&   84.7 $\pm$15.8 &  0.93 $\pm$0.06 & \nodata \\
MIPS22725 & 17:19:03.7+59:26:57.4 & 483, 1462, 1706 &  22.80 &   87.7 $\pm$9.8 &  140.0 $\pm$15.4 &  207.1 $\pm$26.1 &  365.6 $\pm$40.6 &  0.83 $\pm$0.05 & \nodata \\
MIPS22744 & 17:23:33.2+59:06:30.5 & \nodata, 1462, 1706 &  22.31 &   37.8 $\pm$5.0 &   57.2 $\pm$7.2 &   71.2 $\pm$15.5 &  164.4 $\pm$22.2 &  0.86 $\pm$0.06 & \nodata \\
\enddata
\tablecomments{Columns 2 and 9 are the 24 \micron\ coordinates and fluxes of the targets, respectively, taken from the catalog of \cite{fadda04}. Column 4 lists the $R$-band magnitudes of the sources in units of Vega mags. The magnitudes are taken from \cite{fadda06}, unless otherwise noted. Undetected sources in the $R$ band are fainter than 25.5 Vega magnitudes. Columns 5,6,7, and 8 list the IRAC fluxes of the sources, taken from \cite{lacy04}, unless otherwise noted. Because the \cite{lacy04} catalog limits are computed based on completeness levels (instead of noise levels), we remeasured the flux limits for all sources that were undetected in any of the IRAC channels. Column 10 lists the $H$ band effective radii of the sources measured with Sextractor from {\it HST} NICMOS images. In this table, we present \reff\ measurements for sources with known redshift that can be used for the comparison of their radial extents with their MIR spectral feature properties. The full dataset will be presented in Zamojski et al. (2009, in preparation). For consistency with the photometry in all other bands, closely interacting systems have been treated as a single object. \\ 
Notes on individual objects: }
\tablenotetext{a}{Two or more possibly interacting sources constitute the $R$-band counterpart of the 24$\micron$ source.}
\tablenotetext{b}{The identification of the optical counterpart is uncertain.}
\tablenotetext{c}{We remeasured these $R$-band magnitudes from the catalog of \cite{fadda04} due to a problem in the detection or identification of the optical counterpart.}
\tablenotetext{d}{Two or more sources are blended in this IRAC or MIPS channel.}
\tablenotetext{e}{The lRAC flux is taken from the deep catalog of \cite{lacy05}.}
\tablenotetext{f}{The flux measurement of the IRAC counterpart of the 24 \micron\ source is uncertain due to the blending of two sources.}
\tablenotetext{g}{The flux or limit was remeasured (often due to a misidentification).}
\tablenotetext{h}{Two or more sources that could be at similar redshifts are comprised in the IRAC or MIPS flux.}
\tablenotetext{i}{To recover the flux of this source, we first deblended it from its neighbour(s) using point-spread-funtion fitting that was carried out with galfit. Once the best fitting parameters were determined, we removed the neighbour(s) and measured the flux of the source using Sextractor, as for all other sources.}
\tablenotetext{j}{This source is blended with a star. Its flux cannot be recovered.}
\end{deluxetable}
\end{centering}

%% file: tab2.tex
\evensidemargin=-0.0in
\begin{centering}
\footskip=65pt
\begin{deluxetable}{lllllll}
\tablecolumns{7}
\tabletypesize{\tiny}
\tablewidth{0pt}
\tablecaption{\label{tab:specz} Redshift, absorption strength, and rest-frame 14 \micron\ monochromatic luminosity measurements. }
\tablehead{ \colhead{Galaxy} & \colhead{z} & \colhead{confidence} & \colhead{emission lines or features} & \colhead{absorption feature} & \colhead{$\tau_{9.7 \micron}$} & \colhead{$\nu$L$_{\nu}$(14 \micron )}  \\
\colhead{(1)} & \colhead{(2)} & \colhead{(3)} & \colhead{(4)} & \colhead{(5)} & \colhead{(6)}  & \colhead{(7)} 
}
\startdata
MIPS34 & 1.38$\pm$0.03 & a & \psix,\psev,\peig,\neiitab & silicate & 0.14 & 2.08$\times$10$^{12}$ \\
MIPS39 & 2.42 & c & \peig & silicate (partial) & 0.87 & \nodata \\
MIPS55 & 0.80$\pm$0.03 & a & \pele,\ptwe & silicate & 0.48 & 4.34$\times$10$^{11}$ \\
MIPS159 & \nodata & \nodata & \nodata & \nodata & \nodata & \nodata \\
MIPS168 & 0.24$\pm$0.01 & a & \psev,\pele,\ptwe,\pts & silicate & 0.74 & 6.31$\times$10$^{9}$ \\
MIPS184 & \nodata & \nodata & \nodata & \nodata & \nodata & \nodata \\
MIPS213 & 1.22 & b & \psix & silicate & 0.20 & 5.75$\times$10$^{11}$ \\
MIPS224 & 1.47$\pm$0.02 & a & \psix,\psev,\peig,\ptwe & \nodata & 0.00 & \nodata \\
MIPS268 & 1.69 & b & \psev & silicate & 0.67 & \nodata \\
MIPS277 & 1.06$\pm$0.02 & a & \psev,\pele,\ptwe & \nodata & 0.00 & 2.58$\times$10$^{11}$ \\
MIPS298 & 3.49 & b & \psix & silicate & 3.54 & \nodata \\
MIPS322 & \nodata & \nodata & \nodata & \nodata & \nodata & \nodata \\
MIPS324 & 0.96$\pm$0.01 & a & \psix,\psev,\pele,\ptwe,\neiiitab & silicate & 0.19 & 1.59$\times$10$^{11}$ \\
MIPS331 & 1.03 & d,f & \nodata & silicate & \nodata & \nodata \\
MIPS350 & 0.94$\pm$0.02 & a & \peig,\pele,\ptwe,\neiiitab & silicate & 0.45 & 2.20$\times$10$^{11}$ \\
MIPS351 & \nodata & \nodata & \nodata & \nodata & \nodata & \nodata \\
MIPS358 & 0.81$\pm$0.02 & a & \pele,\ptwe & silicate & 0.02 & 6.00$\times$10$^{10}$ \\
MIPS369 & 3.21 & e,f & \psix & silicate (partial) & \nodata & \nodata \\
MIPS394 & 0.92$\pm$0.01 & a & \pele,\ptwe & silicate & 1.08 & 1.11$\times$10$^{11}$ \\
MIPS397 & 1.35 & b & \nodata & silicate & 0.12 & 4.58$\times$10$^{11}$ \\
MIPS419 & 0.83 & c & \nodata & silicate & 2.60 & 1.40$\times$10$^{11}$ \\
MIPS446 & 0.82$\pm$0.01 & d & \peig,\ptwe,\neiiitab & \nodata & \nodata & \nodata \\
MIPS463 & 2.44 & e & \psix & \nodata & \nodata & \nodata \\
MIPS472 & 0.92$\pm$0.02 & a & \pele,\ptwe & \nodata & \nodata & 1.22$\times$10$^{11}$ \\
MIPS488 & 0.69$\pm$0.02 & a & \pele,\ptwe,\pts & silicate & 0.81 & 6.00$\times$10$^{10}$ \\
MIPS495 & \nodata & \nodata & \nodata & \nodata & \nodata & \nodata \\
MIPS505 & \nodata & \nodata & \nodata & \nodata & \nodata & \nodata \\
MIPS509 & $\ge$2.28 & f & \nodata & silicate (partial) & \nodata & \nodata \\
MIPS512 & 0.99$\pm$0.01 & a & \ptwe,\neiiitab & silicate & 0.84 & 1.84$\times$10$^{11}$ \\
MIPS521 & 2.06$\pm$0.01 & a & \psix,\psev,\peig & \nodata & 0.24 & \nodata \\
MIPS530 & 0.89$\pm$0.02 & a & \pele,\ptwe & silicate & 1.32 & 6.89$\times$10$^{10}$ \\
MIPS532 & 1.54 & b & \nodata & silicate & 1.51 & \nodata \\
MIPS537 & 0.39$\pm$0.01 & a &  \psix,\psev,\pele,\ptwe,\neiiitab,\pts & silicate & 0.32 & 1.13$\times$10$^{10}$ \\
MIPS542 & 0.91$\pm$0.01 & a & \pele,\ptwe & silicate & 0.09 & 1.06$\times$10$^{11}$ \\
MIPS544 & 0.96 & b & \nodata & silicate & 1.12 & 1.62$\times$10$^{11}$ \\
MIPS546 & 1.07$\pm$0.01 & a & \psev,\pele,\ptwe & silicate & 0.47 & 2.21$\times$10$^{11}$ \\
MIPS549 & 0.93$\pm$0.01 & a & \pele,\ptwe,\pts & \nodata & 0.00 & 5.87$\times$10$^{10}$ \\
MIPS562 & 0.54$\pm$0.01 & a & \psix,\psev,\pele,\ptwe & silicate & 1.06 & 4.21$\times$10$^{10}$ \\
MIPS7985 & \nodata & \nodata & \nodata & \nodata & \nodata & \nodata \\
MIPS8040 & 0.76$\pm$0.01 & a & \psix,\psev,\pele,\ptwe,\pts & silicate & 1.73 & 2.12$\times$10$^{11}$ \\
MIPS8069 & 1.45 & e & \nodata & silicate & 0.28 & \nodata \\
MIPS8071 & 0.98 & d & \ptwe & \nodata & \nodata & \nodata \\
MIPS8098 & 1.07 & d & \psix,\ptwe & \nodata & \nodata & \nodata \\
MIPS8107 & 0.94$\pm$0.03 & a & \psev,\pele,\neiiitab & silicate & 0.33 & 3.36$\times$10$^{11}$ \\
MIPS8121 & 1.21$\pm$0.03 & a & \psix,\psev & silicate & 0.25 & 8.00$\times$10$^{11}$ \\
MIPS8157 & \nodata & \nodata & \nodata & \nodata & \nodata & \nodata \\
MIPS8172 & 1.11$\pm$0.03 & d & \psix,\pele & \nodata & \nodata & \nodata \\
MIPS8179 & 0.59$\pm$0.01 & a & \pele,\ptwe,\neiiitab & silicate & 0.55 & 7.62$\times$10$^{10}$ \\
MIPS8185 & \nodata & \nodata & \nodata & \nodata & \nodata & \nodata \\
MIPS8192 & \nodata & \nodata & \nodata & \nodata & \nodata & \nodata \\
MIPS8204 & 0.84$\pm$0.02 & a & \pele,\ptwe & \nodata & \nodata & 1.49$\times$10$^{11}$ \\
MIPS8224 & \nodata & \nodata & \nodata & \nodata & \nodata & \nodata \\
MIPS8226 & 2.10 & c & \psev & silicate & 0.41 & \nodata \\
MIPS8233 & \nodata & \nodata & \nodata & \nodata & \nodata & \nodata \\
MIPS8251 & 1.94 & b & \psev & silicate & 0.59 & \nodata \\
MIPS8253 & 0.96$\pm$0.01 & a &  \psev,\pele,\ptwe,\neiiitab & silicate & 2.03 & 2.18$\times$10$^{11}$ \\
MIPS8308 & 0.37$\pm$0.01 & a & \pele,\ptwe & \nodata & \nodata & 6.41$\times$10$^{9}$ \\
MIPS8311 & 1.18$\pm$0.01 & a & \psev,\pele,\ptwe & silicate & 1.13 & 3.09$\times$10$^{11}$ \\
MIPS8315 & \nodata & \nodata & \nodata & \nodata & \nodata & \nodata \\
MIPS8325 & 0.61$\pm$0.02 & a & \pele,\ptwe,\neiiitab & silicate & 1.14 & 6.91$\times$10$^{10}$ \\
MIPS8328 & 1.02 & b & \pele & silicate & 0.26 & 1.83$\times$10$^{11}$ \\
MIPS8360 & 1.50$\pm$0.02 & a & \psix,\psev,\pele,\ptwe & silicate & 0.57 & \nodata \\
MIPS8371 & 0.35$\pm$0.01 & a &  \psix,\psev,\pele,\ptwe,\pts & silicate & 0.43 & 9.24$\times$10$^{9}$ \\
MIPS8375 & 0.87$\pm$0.01 & a & \pele,\ptwe,\neiiitab & silicate & 0.92 & 1.15$\times$10$^{11}$ \\
MIPS8377 & 0.84$\pm$0.01 & a & \pele,\ptwe,\neiiitab & silicate & 1.67 & 1.17$\times$10$^{11}$ \\
MIPS8384 & 0.92$\pm$0.01 & a & \psix,\psev,\pele,\ptwe,\neiiitab & silicate & 0.01 & 1.87$\times$10$^{11}$ \\
MIPS8387 & 0.92$\pm$0.01 & a & \pele,\ptwe & \nodata & 0.00 & 6.60$\times$10$^{10}$ \\
MIPS8388 & 1.14$\pm$0.02 & a & \pele,\ptwe & silicate & 0.11 & 2.53$\times$10$^{11}$ \\
MIPS8392 & 1.90 & b & \nodata & silicate & 4.56 & \nodata \\
MIPS8400 & \nodata & \nodata & \nodata & \nodata & \nodata & \nodata \\
MIPS8401 & 0.57$\pm$0.01 & a & \pele,\ptwe & silicate & 0.52 & 4.77$\times$10$^{10}$ \\
MIPS8405 & 1.17$\pm$0.01 & a & \psev,\pele,\ptwe & \nodata & 0.00 & 3.36$\times$10$^{11}$ \\
MIPS8407 & \nodata & \nodata & \nodata & \nodata & \nodata & \nodata \\
MIPS8409 & 0.74$\pm$0.04 & d & \ptwe,\pts & \nodata & \nodata & \nodata \\
MIPS8411 & 2.06 & c & \psev & silicate & 0.26 & \nodata \\
MIPS8413 & 2.18 & d,f & \psix & silicate & \nodata & \nodata \\
MIPS8424 & 1.59 & b & \ptwe & silicate & 0.20 & \nodata \\
MIPS8430 & 0.67$\pm$0.01 & a & \pele,\ptwe,\neiiitab & silicate & 1.15 & 4.59$\times$10$^{10}$ \\
MIPS8450 & 1.01$\pm$0.01 & a & \psev,\pele,\ptwe & \nodata & 0.00 & 8.67$\times$10$^{10}$ \\
MIPS8462 & 1.01$\pm$0.02 & a & \pele,\ptwe & silicate & 0.19 & 2.11$\times$10$^{11}$ \\
MIPS8465 & \nodata & \nodata & \nodata & \nodata & \nodata & \nodata \\
MIPS8477 & 1.84$\pm$0.01 & a & \psev,\pele & silicate & 0.41 & \nodata \\
MIPS8479 & 1.30 & b & \nodata & silicate & 2.36 & 1.31$\times$10$^{12}$ \\
MIPS8495 & 1.28 & b & \peig & silicate & 0.08 & 3.26$\times$10$^{11}$ \\
MIPS8499 & 0.60$\pm$0.02 & a & \pele,\ptwe & \nodata & \nodata & 5.10$\times$10$^{10}$ \\
MIPS8507 & 0.76$\pm$0.02 & a & \pele,\ptwe,\pts & silicate & 0.75 & 5.69$\times$10$^{10}$ \\
MIPS8521 & 1.19$\pm$0.02 & a & \psev,\ptwe & silicate & 1.90 & 5.02$\times$10$^{11}$ \\
MIPS8526 & 0.84 & b,g & \ptwe & silicate & 1.36 & 7.98$\times$10$^{10}$ \\
MIPS8532 & 0.86$\pm$0.01 & a & \pele,\ptwe,\neiiitab & silicate & 0.59 & 1.08$\times$10$^{11}$ \\
MIPS8543 & 0.65$\pm$0.01 & a & \pele,\ptwe & silicate & 0.34 & 8.94$\times$10$^{10}$ \\
MIPS8550 & 0.87 & b & \nodata & silicate & 2.24 & 1.24$\times$10$^{11}$ \\
MIPS15678 & 1.37 & d & \psix,\pele,\ptwe & \nodata & \nodata & \nodata \\
MIPS15690 & 0.85 & b & \ptwe & silicate & 0.04 & 6.79$\times$10$^{11}$ \\
MIPS15755 & 0.73$\pm$0.02 & a & \psix,\psev,\pele,\ptwe,\neiiitab,\pts & \nodata & 0.00 & 2.27$\times$10$^{11}$ \\
MIPS15771 & 1.32 & d,e,f & \nodata & silicate & \nodata & \nodata \\
MIPS15776 & 1.12 & b & \ptwe & silicate & 0.48 & 6.63$\times$10$^{11}$ \\
MIPS15796 & 1.45 & f & \nodata & silicate & \nodata & \nodata \\
MIPS15871 & \nodata & \nodata & \nodata & \nodata & \nodata & \nodata \\
MIPS15897 & 1.62 & d & \pele,\ptwe & \nodata & \nodata & \nodata \\
MIPS15941 & 1.23 & e,f & \nodata & silicate & \nodata & \nodata \\
MIPS15967 & \nodata & \nodata & \nodata & \nodata & \nodata & \nodata \\
MIPS15999 & 0.57$\pm$0.01 & a & \ptwe,\neiiitab & silicate & 0.76 & 6.16$\times$10$^{10}$ \\
MIPS16037 & 1.64 & b & \ptwe & silicate & 1.04 & \nodata \\
MIPS16047 & 0.52$\pm$0.02 & a & \psev,\neiiitab & silicate & 1.13 & 4.55$\times$10$^{10}$ \\
MIPS16066 & 1.00 & b & \psev & silicate & 0.87 & 3.13$\times$10$^{11}$ \\
MIPS16099 & 0.95 & d & \neiiitab & silicate & \nodata & \nodata \\
MIPS16118 & 2.61 & c & \psix & silicate (partial) & \nodata & \nodata \\
MIPS16126 & \nodata & \nodata & \nodata & \nodata & \nodata & \nodata \\
MIPS16132 & \nodata & \nodata & \nodata & \nodata & \nodata & \nodata \\
MIPS16134 & 1.20 & d & \ptwe & silicate & \nodata & \nodata \\
MIPS16135 & 0.62 & d & \ptwe & \nodata & \nodata & \nodata \\
MIPS16152 & 1.83 & c & \psev & silicate & 1.19 & \nodata \\
MIPS16156 & 0.72 & b & \neiiitab & silicate & 0.48 & 6.70$\times$10$^{10}$ \\
MIPS16170 & 0.32$\pm$0.01 & a & \pele,\ptwe & \nodata & \nodata & 1.07$\times$10$^{10}$ \\
MIPS16202 & \nodata & \nodata & \nodata & \nodata & \nodata & \nodata \\
MIPS16206 & 0.87$\pm$0.02 & a & \pele,\ptwe,\neiiitab & silicate & 0.49 & 1.14$\times$10$^{11}$ \\
MIPS16219 & \nodata & \nodata & \nodata & \nodata & \nodata & \nodata \\
MIPS16227 & 2.15 & e & \psev & \nodata & \nodata & \nodata \\
MIPS16249 & 0.53$\pm$0.02 & a & \pele,\ptwe & \nodata & \nodata & 3.37$\times$10$^{10}$ \\
MIPS16267 & 1.31 & b & \ptwe & silicate & 0.61 & 5.57$\times$10$^{11}$ \\
MIPS22196 & \nodata & \nodata & \nodata & \nodata & \nodata & \nodata \\
MIPS22235 & 0.42$\pm$0.01 & a & \psev,\peig,\pele,\ptwe & silicate & 0.55 & 8.33$\times$10$^{10}$ \\
MIPS22248 & \nodata & \nodata & \nodata & \nodata & \nodata & \nodata \\
MIPS22307 & 0.70$\pm$0.01 & a & \psix,\psev,\peig,\pele & \nodata & 0.00 & 1.33$\times$10$^{11}$ \\
MIPS22323 & 1.21$\pm$0.01 & a & \psix,\psev,\ptwe & silicate & 0.10 & 5.28$\times$10$^{11}$ \\
MIPS22352 & 0.66$\pm$0.01 & a & \psix,\psev,\pele,\ptwe & silicate & 0.19 & 7.61$\times$10$^{10}$ \\
MIPS22356 & 1.13$\pm$0.01 & a & \peig,\pele,\ptwe & silicate & 0.39 & 6.22$\times$10$^{11}$ \\
MIPS22371 & 1.67$\pm$0.01 & a & \psix,\psev,\pele & silicate & 0.37 & \nodata \\
MIPS22379 & 0.65$\pm$0.02 & a & \peig,\pele & silicate & 0.13 & 8.55$\times$10$^{10}$ \\
MIPS22397 & \nodata & \nodata & \nodata & \nodata & \nodata & \nodata \\
MIPS22417 & 1.96$\pm$0.02 & a & \psix,\peig,\pele & silicate (partial) & 0.90 & \nodata \\
MIPS22432 & 1.59$\pm$0.01 & a & \psix,\pele,\ptwe & silicate & 1.18 & \nodata \\
MIPS22516 & 1.35 & e & \psev & \nodata & \nodata & \nodata \\
MIPS22527 & \nodata & \nodata & \nodata & \nodata & \nodata & \nodata \\
MIPS22536 & 1.59$\pm$0.02 & b & \psix,\pele & silicate & 0.51 & \nodata \\
MIPS22548 & $\ge$2.12 & f & \nodata & silicate (partial) & \nodata & \nodata \\
MIPS22549 & 1.05$\pm$0.01 & a & \pele,\ptwe & silicate & 0.01 & 2.46$\times$10$^{11}$ \\
MIPS22555 & 1.88 & b & \nodata & silicate & 0.08 & \nodata \\
MIPS22557 & 0.78$\pm$0.01 & a & \pele,\ptwe,\neiiitab & silicate & \nodata & 9.36$\times$10$^{10}$ \\
MIPS22621 & 0.59 & e & \ptwe & \nodata & \nodata & \nodata \\
MIPS22633 & 2.20 & e & \psix & \nodata & \nodata & \nodata \\
MIPS22634 & \nodata & \nodata & \nodata & \nodata & \nodata & \nodata \\
MIPS22635 & 0.80$\pm$0.01 & a & \pele,\ptwe & silicate & 0.78 & 1.25$\times$10$^{11}$ \\
MIPS22638 & 0.99 & b & \neiitab & silicate & 0.39 & 1.72$\times$10$^{11}$ \\
MIPS22663 & \nodata & \nodata & \nodata & \nodata & \nodata & \nodata \\
MIPS22690 & 2.07 & b & \nodata & silicate & 0.39 & \nodata \\
MIPS22710 & \nodata & \nodata & \nodata & \nodata & \nodata & \nodata \\
MIPS22722 & 1.71 & b & \peig & silicate & 0.55 & \nodata \\
MIPS22725 & \nodata & \nodata & \nodata & \nodata & \nodata & \nodata \\
MIPS22744 & \nodata & \nodata & \nodata & \nodata & \nodata & \nodata \\

\enddata
\tablecomments{
Column (2): Redshifts based on MIR spectra. Uncertainties are given for all sources that
have more than two 3$\sigma$ detected emission lines or features. For all other sources, the
plausible redshift uncertainty is in the range [0.01,0.3]. \\
Column (3): Classification of the confidence of the MIR redshift. The letters correspond to: \\
a: The reshift is confirmed by more than two emission lines or features that have $>3\sigma$ detections. \\
b: The redshift is either derived from a single emission line but confirmed from the 9.7 \micron\ silicate 
absorption feature or measured from the silicate absorption feature. The uncertainty in this redshift 
measurement is typically of order 0.1, but it can be as high as 0.3. \\
c: The redshift is measured from a single emission line and confirmed by the presence of part of the 9.7 \micron\
silicate feature in the spectrum. The accuracy of the measurement can be lower than that of category (b). \\
d: More than one solutions are possible. This redshift is uncertain. \\
e: Only one line or feature had a significant detection. This redshift is also uncertain. \\
f: Only part of the 9.7 \micron\ silicate absorption feature was seen in the MIPS spectral range, or the silicate feature
had a very small optical depth, $<$0.1. This redshift is uncertain. \\
g: This spectrum contains features of two spatially unresolved sources, one at $z$=0.84, and a somewhat
fainter at $z$=0.91. This could either be a binary merging system, or sources that neighbour each other
due to projection effects. \\
Column (4): Emission lines or dust features used for the computation of each source's redshift. 
These features have a signal-to-noise ratio greater than 3 when fitted with a Lorentzian profile. Unless 
otherwise noted, the 12.7 complex refers to the blended 12.6 and 12.7 \micron\ dust features 
and the [NeII] line. \\
Column (5): Indicates whether the 9.7 \micron\ silicate absorption feature was used in the redshift computation. 
Partially detected silicate absorption features were not used for that purpose. \\
Column (6): The 9.7 \micron\ silicate feature optical depth was measured with PAHFIT for sources with reliable redshifts, i.e.
with confidence classification a,b, or c in column 3. \\
Column (7):The rest-frame monochromatic 14 \micron\ luminosity is given in units of solar luminosity for sources with
reliable redshift measurements. \\
}

\end{deluxetable}
\end{centering}




%% file: tab3.tex
\begin{centering}

\textheight=7.7in 
\footskip=75pt 

\begin{deluxetable}{lllllllll}
\tablecolumns{9}
\tabletypesize{\tiny}
\tablewidth{0pt}

\tablecaption{\hbox{\label{tab:lines1}  Fluxes and EWs of emission lines or features in the range 6.0$<\lambda<$11.0 $\mu{\rm m}$. }}
\tablehead{ \colhead{Galaxy} &   \colhead{f$_{6.2}$} &   \colhead{$EW_{6.2 }$}
                             &   \colhead{f$_{7.7}$} &   \colhead{$EW_{7.7 }$}
                             &   \colhead{f$_{8.6}$} &   \colhead{$EW_{8.6 }$} 
                             &  \colhead{f$_{[SIV]}$} &   \colhead{$EW_{[SIV]}$} \\

 \colhead{(MIPS) }            &   \colhead{(Wcm$^{-2}$)}        &   \colhead{(\micron )}
                             &   \colhead{(Wcm$^{-2}$)}        &   \colhead{(\micron )}
                             &   \colhead{(Wcm$^{-2}$)}        &   \colhead{(\micron )} 
                             &   \colhead{(Wcm$^{-2}$)}        &   \colhead{(\micron )} 
}
\startdata  
    MIPS34 & 6.65($\pm$1.13)$\times$$10^{-22}$ &  0.07$\pm$0.01 & 3.71($\pm$0.41)$\times$$10^{-21}$ &  0.42$\pm$0.05 &  $<$3.31$\times$$10^{-22}$  &  \nodata  & $<$5.19$\times$$10^{-23}$  &  \nodata  \\ 
    MIPS39 & 3.02($\pm$0.67)$\times$$10^{-22}$ &  0.03$\pm$0.01 & 2.85($\pm$0.19)$\times$$10^{-21}$ &  0.42$\pm$0.03 & \nodata &  \nodata & \nodata &  \nodata  \\ 
    MIPS55 &  \nodata  &  \nodata  &  \nodata  &  \nodata  & 7.40($\pm$1.47)$\times$$10^{-22}$ &  0.22$\pm$0.04 & 1.99($\pm$0.46)$\times$$10^{-22}$ &  0.06$\pm$0.01 \\ 
   MIPS168 &  \nodata  &  \nodata  & 4.73($\pm$0.43)$\times$$10^{-21}$ &  9.70$\pm$0.88 & 8.10($\pm$1.61)$\times$$10^{-22}$ &  1.83$\pm$0.36 & $<$1.27$\times$$10^{-22}$  &  \nodata  \\ 
   MIPS213 & $<$3.98$\times$$10^{-22}$ &  \nodata  & $<$6.22$\times$$10^{-22}$ &  \nodata &  $<$2.59$\times$$10^{-22}$ &  \nodata  & $<$4.24$\times$$10^{-23}$ &  \nodata  \\ 
   MIPS224 & 5.30($\pm$0.57)$\times$$10^{-22}$ & 0.33$\pm$0.04 & 1.53($\pm$0.19)$\times$$10^{-21}$ & 0.92$\pm$0.11  &  5.01($\pm$0.79)$\times$$10^{-22}$ & 0.29$\pm$0.05  & 5.20($\pm$1.19)$\times$$10^{-23}$ &  0.03$\pm$0.01 \\ 
   MIPS268 & 2.82($\pm$0.80)$\times$$10^{-22}$ &  0.09$\pm$0.03 & 1.60($\pm$0.26)$\times$$10^{-21}$ &  0.64$\pm$0.10 &  $<$1.04$\times$$10^{-22}$  &  \nodata  & 7.73($\pm$1.17)$\times$$10^{-23}$ &  0.06$\pm$0.01 \\ 
   MIPS277 &  \nodata  &  \nodata  & 1.91($\pm$0.16)$\times$$10^{-21}$ &  1.20$\pm$0.10 & 5.67($\pm$1.08)$\times$$10^{-22}$ &  0.38$\pm$0.07 &  $<$6.15$\times$$10^{-23}$ &  \nodata  \\ 
   MIPS298 & 2.07($\pm$0.45)$\times$$10^{-22}$ & 0.11$\pm$0.02 & $<$2.15$\times$$10^{-21}$ &  \nodata &  \nodata  &  \nodata  & \nodata & \nodata \\ 
   MIPS324 & 5.41($\pm$1.11)$\times$$10^{-22}$ &  0.35$\pm$0.07 & 1.81($\pm$0.16)$\times$$10^{-21}$ &  1.40$\pm$0.12 & 4.01($\pm$1.00)$\times$$10^{-22}$ &  0.35$\pm$0.09 & $<$1.16$\times$$10^{-22}$  &  \nodata  \\ 
   MIPS350 &  \nodata  &  \nodata  & \nodata & \nodata & 1.18($\pm$0.11)$\times$$10^{-21}$ &  2.88$\pm$0.27 &  $<$1.34$\times$$10^{-22}$ &  \nodata  \\ 
   MIPS358 &  \nodata  &  \nodata  &  \nodata  &  \nodata  & \nodata & \nodata &  $<$8.97$\times$$10^{-23}$ &  \nodata  \\ 
   MIPS394 &  \nodata  &  \nodata  &  \nodata  &  \nodata  & \nodata &  \nodata &  $<$1.31$\times$$10^{-22}$  &  \nodata  \\ 
   MIPS397 & 3.17($\pm$0.90)$\times$$10^{-22}$ &  0.17$\pm$0.05 & 1.45($\pm$0.26)$\times$$10^{-21}$ &  0.97$\pm$0.17 &  $<$3.74$\times$$10^{-22}$ &  \nodata  &  $<$3.60$\times$$10^{-23}$ &  \nodata  \\ 
   MIPS419 &  \nodata  &  \nodata  &  \nodata  &  \nodata  &  \nodata  &  \nodata  &  $<$6.81$\times$$10^{-22}$ &  \nodata  \\ 
   MIPS472 &  \nodata  &  \nodata  &  \nodata  &  \nodata  & \nodata &  \nodata &  $<$1.09$\times$$10^{-22}$ &  \nodata  \\ 
   MIPS488 &  \nodata  &  \nodata  &  \nodata  &  \nodata  & \nodata & \nodata &  $<$1.87$\times$$10^{-22}$ &  \nodata  \\ 
   MIPS512 &  $<$2.96$\times$$10^{-22}$ &  \nodata  &  $<$9.35$\times$$10^{-22}$ &  \nodata  &  $<$3.95$\times$$10^{-22}$ &  \nodata  &  $<$1.41$\times$$10^{-22}$ &  \nodata  \\ 
   MIPS521 & 5.87($\pm$1.04)$\times$$10^{-22}$ &  0.69$\pm$0.12 & 1.87($\pm$0.28)$\times$$10^{-21}$ &  2.97$\pm$0.44 & 3.69($\pm$0.52)$\times$$10^{-22}$ &  0.75$\pm$0.11 & 1.12($\pm$0.17)$\times$$10^{-22}$ &  0.35 $\pm$0.05\\ 
   MIPS530 &  \nodata  &  \nodata  &  \nodata  &  \nodata  & \nodata &  \nodata &  $<$1.18$\times$$10^{-22}$  &  \nodata  \\ 
   MIPS532 & $<$2.91$\times$$10^{-22}$ & \nodata & 4.85($\pm$0.33)$\times$$10^{-21}$ & 2.34$\pm$0.16 & $<$2.17$\times$$10^{-22}$ & \nodata  &  $<$5.54$\times$$10^{-23}$  &  \nodata  \\ 
   MIPS537 & 1.14($\pm$0.10)$\times$$10^{-21}$ &  1.89$\pm$0.17 & 3.66($\pm$0.29)$\times$$10^{-21}$ &  7.31$\pm$0.58 & 1.01($\pm$0.12)$\times$$10^{-21}$ &  2.56$\pm$0.30 &  $<$1.06$\times$$10^{-22}$ &  \nodata  \\ 
   MIPS542 &  \nodata  &  \nodata  &  \nodata  &  \nodata  & \nodata &  \nodata &  $<$1.17$\times$$10^{-22}$ &  \nodata  \\ 
   MIPS544 &  \nodata  &  \nodata  & \nodata &  \nodata  &  $<$5.55$\times$$10^{-22}$ &  \nodata  &  $<$2.00$\times$$10^{-22}$ &  \nodata  \\ 
   MIPS546 &  \nodata  &  \nodata  & 5.69($\pm$0.32)$\times$$10^{-21}$ & 19.26$\pm$1.08 & 1.36($\pm$0.12)$\times$$10^{-21}$ &  4.44$\pm$0.39 &  $<$7.52$\times$$10^{-23}$  &  \nodata  \\ 
   MIPS549 &  \nodata  &  \nodata  & \nodata &  \nodata & 5.94($\pm$1.11)$\times$$10^{-22}$ &  1.19$\pm$0.22 &  $<$1.33$\times$$10^{-22}$  &  \nodata  \\ 
   MIPS562 & 1.72($\pm$0.12)$\times$$10^{-21}$ &  2.23$\pm$0.16 & 5.96($\pm$0.41)$\times$$10^{-21}$ &  7.25$\pm$0.50 & 5.09($\pm$1.67)$\times$$10^{-22}$ &  0.78$\pm$0.26 &  $<$2.00$\times$$10^{-22}$ &  \nodata  \\ 
  MIPS8040 & 1.74($\pm$0.09)$\times$$10^{-21}$ &  0.64$\pm$0.03 & 5.70($\pm$0.47)$\times$$10^{-21}$ &  2.30$\pm$0.19 & 1.80($\pm$0.13)$\times$$10^{-21}$ &  1.06$\pm$0.08 & 9.95($\pm$3.01)$\times$$10^{-23}$ &  0.09$\pm$0.03 \\ 
  MIPS8107 &  $<$4.51$\times$$10^{-22}$ &  \nodata  & 1.99($\pm$0.45)$\times$$10^{-21}$ &  1.06$\pm$0.24 & 3.15($\pm$0.91)$\times$$10^{-22}$ &  0.21$\pm$0.06 &  $<$1.21$\times$$10^{-22}$ &  \nodata  \\ 
  MIPS8121 & 2.88($\pm$0.75)$\times$$10^{-22}$ &  0.10$\pm$0.03 & 2.00($\pm$0.24)$\times$$10^{-21}$ &  0.77$\pm$0.09 &  $<$3.44$\times$$10^{-22}$ &  \nodata  &  $<$4.77$\times$$10^{-23}$ &  \nodata  \\ 
  MIPS8179 &  $<$2.53$\times$$10^{-22}$ &  \nodata  &  $<$3.70$\times$$10^{-22}$ &  \nodata  &  $<$4.37$\times$$10^{-22}$ &  \nodata  &  $<$1.11$\times$$10^{-22}$ &  \nodata  \\ 
  MIPS8204 &  \nodata  &  \nodata  &  \nodata  &  \nodata  & \nodata &  \nodata &  $<$1.32$\times$$10^{-22}$   &  \nodata  \\ 
  MIPS8226 &  $<$2.96$\times$$10^{-22}$ &  \nodata  & 1.07($\pm$0.26)$\times$$10^{-21}$ &  0.69$\pm$0.17 &  $<$2.04$\times$$10^{-22}$ &  \nodata  &  $<$7.47$\times$$10^{-23}$ &  \nodata  \\ 
  MIPS8251 &  $<$3.16$\times$$10^{-22}$ &  \nodata  & 1.14($\pm$0.08)$\times$$10^{-21}$ &  0.72$\pm$0.05 & 2.22($\pm$0.52)$\times$$10^{-22}$ &  0.17$\pm$0.04 &  $<$6.60$\times$$10^{-23}$  &  \nodata  \\ 
  MIPS8253 & 1.92($\pm$0.10)$\times$$10^{-22}$ &  0.07$\pm$0.00$^*$ & 2.48($\pm$0.08)$\times$$10^{-21}$ &  1.39$\pm$0.04 & 4.51($\pm$0.87)$\times$$10^{-22}$ &  0.46$\pm$0.09 & 3.82($\pm$1.07)$\times$$10^{-23}$ &  0.08$\pm$0.02 \\ 
  MIPS8308 &  \nodata  &  \nodata  &  \nodata  &  \nodata  &  \nodata  &  \nodata  &  \nodata  &  \nodata  \\ 
  MIPS8311 &  \nodata  &  \nodata  & 1.77($\pm$0.12)$\times$$10^{-21}$ &  1.27$\pm$0.09 &  $<$5.81$\times$$10^{-22}$ &  \nodata  & $<$5.89$\times$$10^{-23}$   &  \nodata  \\ 
  MIPS8325 &  \nodata  &  \nodata  &  \nodata  &  \nodata  &  \nodata  &  \nodata  &  $<$1.80$\times$$10^{-22}$ &  \nodata  \\ 
  MIPS8328 &  \nodata  &  \nodata  & \nodata &  \nodata &  $<$4.08$\times$$10^{-22}$ &  \nodata  &  $<$9.84$\times$$10^{-23}$ &  \nodata  \\ 
  MIPS8360 & 6.03($\pm$1.01)$\times$$10^{-22}$ &  0.45$\pm$0.08 & 1.42($\pm$0.34)$\times$$10^{-21}$ &  0.96$\pm$0.23 & 3.68($\pm$0.86)$\times$$10^{-22}$ &  0.26$\pm$0.06 & 4.42($\pm$1.18)$\times$$10^{-23}$ &  0.03$\pm$0.01 \\ 
  MIPS8371 & 1.59($\pm$0.12)$\times$$10^{-21}$ &  2.09$\pm$0.16 & 5.32($\pm$0.37)$\times$$10^{-21}$ &  9.81$\pm$0.68 & 1.10($\pm$0.12)$\times$$10^{-21}$ &  2.69$\pm$0.29 &  $<$1.25$\times$$10^{-22}$ &  \nodata  \\ 
  MIPS8375 &  \nodata  &  \nodata  &  \nodata  &  \nodata  & \nodata & \nodata & 1.44($\pm$0.32)$\times$$10^{-22}$ &  0.38$\pm$0.09 \\ 
  MIPS8377 &  \nodata  &  \nodata  &  \nodata  &  \nodata  & \nodata & \nodata & 1.60($\pm$0.37)$\times$$10^{-22}$ &  0.67$\pm$0.16 \\ 
  MIPS8384 &  $<$3.40$\times$$10^{-22}$ &  \nodata  & 2.20($\pm$0.12)$\times$$10^{-21}$ &  1.98$\pm$0.11 &  $<$2.91$\times$$10^{-22}$ &  \nodata  &  $<$9.87$\times$$10^{-23}$ &  \nodata  \\ 
  MIPS8387 &  \nodata  &  \nodata  &  \nodata  &  \nodata  & \nodata & \nodata &  \nodata &  \nodata  \\ 
  MIPS8388 &  \nodata  &  \nodata  & 2.29($\pm$0.39)$\times$$10^{-21}$ &  2.18$\pm$0.37 & 6.72($\pm$1.25)$\times$$10^{-22}$ &  0.84$\pm$0.16 &  $<$6.45$\times$$10^{-23}$ &  \nodata  \\ 
  MIPS8392 &  $<$1.21$\times$$10^{-21}$ &  \nodata  & 2.73($\pm$0.42)$\times$$10^{-21}$ &  1.91$\pm$0.29 &  $<$1.22$\times$$10^{-22}$  &  \nodata  &  $<$1.44$\times$$10^{-21}$ &  \nodata  \\ 
  MIPS8401 &  \nodata  &  \nodata  &  \nodata  &  \nodata  &  \nodata  &  \nodata  &  $<$1.21$\times$$10^{-22}$ &  \nodata  \\ 
  MIPS8405 &  \nodata  &  \nodata  & 3.92($\pm$0.29)$\times$$10^{-21}$ &  8.01$\pm$0.59 & 9.81($\pm$1.17)$\times$$10^{-22}$ &  1.63$\pm$0.19 &  $<$5.19$\times$$10^{-23}$ &  \nodata  \\ 
  MIPS8411 & 2.78($\pm$0.73)$\times$$10^{-22}$ &  0.19$\pm$0.05 & 7.23($\pm$1.82)$\times$$10^{-22}$ &  0.62$\pm$0.16 & 1.27($\pm$0.42)$\times$$10^{-22}$ &  0.12$\pm$0.04 & \nodata & \nodata \\ 
  MIPS8424 &  $<$2.58$\times$$10^{-22}$ &  \nodata  & 1.62($\pm$0.26)$\times$$10^{-21}$ &  1.61$\pm$0.26 & 2.01($\pm$0.66)$\times$$10^{-22}$ &  0.20$\pm$0.06 &  $<$3.96$\times$$10^{-23}$  &  \nodata  \\ 
  MIPS8430 &  \nodata  &  \nodata  &  \nodata  &  \nodata  &  \nodata  &  \nodata  &  $<$9.78$\times$$10^{-23}$  &  \nodata  \\ 
  MIPS8450 &  \nodata  &  \nodata  & 4.97($\pm$0.28)$\times$$10^{-21}$ &  7.11$\pm$0.40 & 1.08($\pm$0.11)$\times$$10^{-21}$ &  2.07$\pm$0.21 &  $<$7.08$\times$$10^{-23}$ &  \nodata  \\ 
  MIPS8462 &  \nodata  &  \nodata  & 1.98($\pm$0.60)$\times$$10^{-21}$ &  4.49$\pm$1.36 & 3.64($\pm$1.16)$\times$$10^{-22}$ &  0.75$\pm$0.24 &  $<$7.62$\times$$10^{-23}$ &  \nodata  \\ 
  MIPS8477 & 3.63($\pm$0.86)$\times$$10^{-22}$ &  0.35$\pm$0.08 & 1.09($\pm$0.33)$\times$$10^{-21}$ &  1.19$\pm$0.36 &  $<$1.86$\times$$10^{-22}$ &  \nodata  & 4.64($\pm$1.12)$\times$$10^{-23}$ &  0.06$\pm$0.02 \\ 
  MIPS8479 &  \nodata  &  \nodata  & \nodata &  \nodata &  $<$1.25$\times$$10^{-21}$ &  \nodata  &  $<$5.57$\times$$10^{-23}$  &  \nodata  \\ 
  MIPS8495 & $<$2.21$\times$$10^{-22}$ & \nodata & 1.49($\pm$0.23)$\times$$10^{-21}$ &  1.00$\pm$0.15 &  $<$3.04$\times$$10^{-22}$ &  \nodata  & 7.15($\pm$1.36)$\times$$10^{-23}$ &  0.06$\pm$0.01 \\ 
  MIPS8499 &  \nodata  &  \nodata  &  \nodata  &  \nodata  &  \nodata  &  \nodata  &  $<$1.01$\times$$10^{-22}$  &  \nodata  \\ 
  MIPS8507 &  \nodata  &  \nodata  &  \nodata  &  \nodata  & \nodata & \nodata & $<$1.04$\times$$10^{-22}$  &  \nodata  \\ 
  MIPS8521 &  $<$6.27$\times$$10^{-22}$ &  \nodata  & 4.30($\pm$0.29)$\times$$10^{-21}$ &  1.86$\pm$0.13 & 5.72($\pm$1.04)$\times$$10^{-22}$ &  0.43$\pm$0.08 & $<$5.57$\times$$10^{-23}$  &  \nodata  \\ 
  MIPS8526 &  \nodata  &  \nodata  &  \nodata  &  \nodata  & \nodata & \nodata &  $<$1.74$\times$$10^{-22}$   &  \nodata  \\ 
  MIPS8532 &  \nodata  &  \nodata  &  \nodata  &  \nodata  & \nodata & \nodata &  $<$1.33$\times$$10^{-22}$ &  \nodata  \\ 
  MIPS8543 &  \nodata  &  \nodata  &  \nodata  &  \nodata  & \nodata & \nodata &  $<$1.39$\times$$10^{-22}$ &  \nodata  \\ 
  MIPS8550 &  \nodata  &  \nodata  &  \nodata  &  \nodata  & \nodata & \nodata &  $<$4.50$\times$$10^{-22}$ &  \nodata  \\ 
 MIPS15690 &  $<$2.59$\times$$10^{-22}$ &  \nodata  & 4.21($\pm$0.30)$\times$$10^{-21}$ &  0.71$\pm$0.05 &  $<$4.62$\times$$10^{-22}$ &  \nodata  & $<$1.39$\times$$10^{-22}$ & \nodata \\ 
 MIPS15755 & 1.89($\pm$0.11)$\times$$10^{-21}$ &  0.84$\pm$0.05 & 6.65($\pm$0.34)$\times$$10^{-21}$ &  3.36$\pm$0.17 & 1.68($\pm$0.15)$\times$$10^{-21}$ &  0.87$\pm$0.08 &  $<$1.06$\times$$10^{-22}$ &  \nodata  \\ 
 MIPS15776 &  $<$3.59$\times$$10^{-22}$ &  \nodata  & 1.44($\pm$0.29)$\times$$10^{-21}$ &  0.35$\pm$0.07 &  $<$3.14$\times$$10^{-22}$ &  \nodata  &   $<$6.02$\times$$10^{-23}$ &  \nodata  \\ 
 MIPS15999 &  \nodata  &  \nodata  &  \nodata  &  \nodata  &  \nodata  &  \nodata  &   $<$1.19$\times$$10^{-22}$  &  \nodata  \\ 
 MIPS16037 & 2.17($\pm$0.70)$\times$$10^{-22}$ &  0.05$\pm$0.02 & 3.14($\pm$0.25)$\times$$10^{-21}$ &  1.09$\pm$0.09 &  $<$1.27$\times$$10^{-22}$  &  \nodata  &  $<$6.42$\times$$10^{-23}$ &  \nodata  \\ 
 MIPS16047 & 4.93($\pm$1.26)$\times$$10^{-22}$ &  0.30$\pm$0.08 & 2.00($\pm$0.35)$\times$$10^{-21}$ &  1.36$\pm$0.24 & 9.89($\pm$1.18)$\times$$10^{-22}$ &  0.90$\pm$0.11 &  $<$1.81$\times$$10^{-22}$ &  \nodata  \\ 
 MIPS16066 &  $<$3.79$\times$$10^{-22}$ &  \nodata  & 3.04($\pm$0.30)$\times$$10^{-21}$ &  1.76$\pm$0.17 & 4.11($\pm$0.98)$\times$$10^{-22}$ &  0.35$\pm$0.08 &  $<$1.65$\times$$10^{-22}$ &  \nodata  \\ 
 MIPS16118 &  $<$2.38$\times$$10^{-22}$ &  \nodata  & \nodata & \nodata & \nodata & \nodata & \nodata & \nodata \\ 
 MIPS16152 &  \nodata  &  \nodata  &  $<$1.27$\times$$10^{-21}$ &  \nodata  & 2.10($\pm$0.52)$\times$$10^{-22}$ &  0.21$\pm$0.05 &  $<$9.93$\times$$10^{-23}$ &  \nodata  \\ 
 MIPS16156 &  \nodata  &  \nodata  &  \nodata  &  \nodata  & \nodata &  \nodata  &   $<$1.16$\times$$10^{-22}$  &  \nodata  \\ 
 MIPS16170 &  \nodata  &  \nodata  &  \nodata  &  \nodata  &  \nodata  &  \nodata  & \nodata & \nodata \\ 
 MIPS16206 &  \nodata  &  \nodata  &  \nodata  &  \nodata  & \nodata & \nodata & $<$1.02$\times$$10^{-22}$ &  \nodata \\ 
 MIPS16249 &  \nodata  &  \nodata  &  \nodata  &  \nodata  &  \nodata  &  \nodata  &  $<$8.61$\times$$10^{-23}$ &  \nodata  \\ 
 MIPS16267 &   $<$3.25$\times$$10^{-22}$  &  \nodata  & 1.72($\pm$0.35)$\times$$10^{-21}$ &  0.76$\pm$0.15 &  $<$4.36$\times$$10^{-22}$ &  \nodata  &  $<$5.19$\times$$10^{-23}$ &  \nodata  \\ 
 MIPS22235 & 3.35($\pm$1.11)$\times$$10^{-22}$ &  0.43$\pm$0.14 & 1.83($\pm$0.15)$\times$$10^{-21}$ &  2.07$\pm$0.17 & 8.33($\pm$1.79)$\times$$10^{-22}$ &  0.81$\pm$0.17 &  $<$1.87$\times$$10^{-22}$ &  \nodata  \\ 
 MIPS22307 & 1.31($\pm$0.12)$\times$$10^{-21}$ &  0.69$\pm$0.06 & 3.56($\pm$0.44)$\times$$10^{-21}$ &  2.07$\pm$0.26 & 1.28($\pm$0.11)$\times$$10^{-21}$ &  0.78$\pm$0.07 &  $<$9.21$\times$$10^{-23}$ &  \nodata  \\ 
 MIPS22323 &  $<$3.59$\times$$10^{-22}$ &  \nodata  & 1.43($\pm$0.33)$\times$$10^{-21}$ &  0.53$\pm$0.12 &  $<$3.98$\times$$10^{-22}$ &  \nodata  &  $<$6.41$\times$$10^{-23}$  &  \nodata  \\ 
 MIPS22352 & 1.92($\pm$0.11)$\times$$10^{-21}$ &  2.09$\pm$0.12 & 7.81($\pm$0.43)$\times$$10^{-21}$ & 11.27$\pm$0.62 & 1.49($\pm$0.12)$\times$$10^{-21}$ &  2.35$\pm$0.19 &  $<$9.48$\times$$10^{-23}$ &  \nodata  \\ 
 MIPS22356 &  $<$4.45$\times$$10^{-22}$ &  \nodata  & 1.50($\pm$0.46)$\times$$10^{-21}$ &  0.78$\pm$0.24 & 6.44($\pm$1.12)$\times$$10^{-22}$ &  0.35$\pm$0.06 &  $<$4.46$\times$$10^{-23}$  &  \nodata  \\ 
 MIPS22371 & 4.37($\pm$0.74)$\times$$10^{-22}$ &  0.18$\pm$0.03 & 1.20($\pm$0.32)$\times$$10^{-21}$ &  0.50$\pm$0.13 &  $<$1.86$\times$$10^{-22}$ &  \nodata  & 6.02($\pm$1.10)$\times$$10^{-23}$ &  0.03$\pm$0.01 \\ 
 MIPS22379 &  $<$3.05$\times$$10^{-22}$ &  \nodata  & 9.97($\pm$3.14)$\times$$10^{-22}$ &  0.80$\pm$0.25 & 4.59($\pm$1.27)$\times$$10^{-22}$ &  0.41$\pm$0.11 &  $<$8.73$\times$$10^{-23}$ &  \nodata  \\ 
 MIPS22417 & 5.02($\pm$0.69)$\times$$10^{-22}$ &  0.21$\pm$0.03 & 1.02($\pm$0.08)$\times$$10^{-21}$ &  0.67$\pm$0.05 & 5.20($\pm$0.48)$\times$$10^{-22}$ &  0.55$\pm$0.05 &  $<$5.67$\times$$10^{-23}$ &  \nodata  \\ 
 MIPS22432 & 9.25($\pm$0.80)$\times$$10^{-22}$ &  0.26$\pm$0.02 & 5.50($\pm$0.23)$\times$$10^{-21}$ &  2.37$\pm$0.10 & 8.01($\pm$0.62)$\times$$10^{-22}$ &  0.61$\pm$0.05 & 4.19($\pm$1.06)$\times$$10^{-23}$ &  0.06$\pm$0.02 \\ 
 MIPS22536 & 4.38($\pm$0.91)$\times$$10^{-22}$ &  0.27$\pm$0.06 & 2.55($\pm$0.33)$\times$$10^{-21}$ &  1.77$\pm$0.23 & 5.88($\pm$0.77)$\times$$10^{-22}$ &  0.48$\pm$0.06 &  $<$5.79$\times$$10^{-23}$ &  \nodata  \\ 
 MIPS22549 &  $<$2.20$\times$$10^{-22}$ &  \nodata  & 1.92($\pm$0.25)$\times$$10^{-21}$ &  1.59$\pm$0.21 & 3.69($\pm$1.04)$\times$$10^{-22}$ &  0.35$\pm$0.10 &  $<$5.73$\times$$10^{-23}$ &  \nodata  \\ 
 MIPS22555 &  \nodata  &  \nodata  &  $<$8.50$\times$$10^{-22}$ &  \nodata  &  $<$1.72$\times$$10^{-22}$ &  \nodata  &  $<$4.92$\times$$10^{-23}$ &  \nodata  \\ 
 MIPS22557 &  \nodata  &  \nodata  &  \nodata  &  \nodata  & \nodata & \nodata &  $<$1.21$\times$$10^{-22}$ &  \nodata  \\
 MIPS22635 &  \nodata  &  \nodata  &  \nodata  &  \nodata  & \nodata & \nodata &  $<$1.82$\times$$10^{-22}$ &  \nodata  \\ 
 MIPS22638 &  \nodata  &  \nodata  & \nodata & \nodata & $<$4.00$\times$$10^{-22}$ & \nodata &  $<$1.10$\times$$10^{-22}$ &  \nodata  \\ 
 MIPS22690 &  $<$2.62$\times$$10^{-22}$ &  \nodata  & 4.24($\pm$0.63)$\times$$10^{-22}$ &  0.31$\pm$0.05 & 1.80($\pm$0.45)$\times$$10^{-22}$ &  0.18$\pm$0.05 & $<$6.31$\times$$10^{-23}$ & \nodata \\ 
 MIPS22722 &  \nodata  &  \nodata  & 9.24($\pm$2.61)$\times$$10^{-22}$ & 0.71$\pm$0.20 & \nodata &  \nodata  &  \nodata  \\ 
\enddata
\tablecomments{ ~~ Fluxes may not be given for lines that are in the IRS wavelength range when their neighboring continuum cannot be well determined. \\
~~*~Error bars of 0 correspond to EW uncertainties that are lower than the selected measurement accuracy.}
\end{deluxetable}
\end{centering}

%% file: tab4.tex
\begin{centering}

\begin{deluxetable}{lllllllll}
\tablecolumns{9}
\tabletypesize{\tiny}
\tablewidth{0pt}

\textheight=7.7in 
\footskip=75pt 

\tablecaption{\hbox{ \label{tab:lines2} Fluxes and EWs of emission lines or features in the range 11.0$<\lambda<$16.0 $\mu{\rm m}$.} }

\tablehead{ \colhead{Galaxy} &   \colhead{f$_{11.3}$} &   \colhead{$EW_{11.3 }$}
                             &   \colhead{f$_{12.7}$} &   \colhead{$EW_{12.7 }$} 
                             &   \colhead{f$_{[NeII]}$} &   \colhead{$EW_{[NeII]}$}
                             &   \colhead{f$_{[NeIII]}$} &   \colhead{$EW_{[NeIII]}$} \\

 \colhead{(MIPS) }            &   \colhead{(Wcm$^{-2}$)}        &   \colhead{(\micron )}
                             &   \colhead{(Wcm$^{-2}$)}        &   \colhead{(\micron )}
                             &   \colhead{(Wcm$^{-2}$)}        &   \colhead{(\micron )} 
                             &   \colhead{(Wcm$^{-2}$)}        &   \colhead{(\micron )} 
}
\startdata
 
    MIPS34 & 2.01($\pm$0.54)$\times$$10^{-22}$ &  0.02$\pm$0.01 &  $<$1.12$\times$$10^{-22}$ &  \nodata  & 5.58($\pm$1.79)$\times$$10^{-23}$ &  0.01$\pm$0.00$^*$ &  \nodata  &  \nodata  \\ 
    MIPS39 &  \nodata  &  \nodata  &  \nodata  &  \nodata  &  \nodata  &  \nodata  &  \nodata  &  \nodata  \\ 
    MIPS55 & 8.97($\pm$1.56)$\times$$10^{-22}$ &  0.28$\pm$0.05 & 5.14($\pm$1.58)$\times$$10^{-22}$ &  0.13$\pm$0.04 & 1.59($\pm$0.24)$\times$$10^{-22}$ &  0.04$\pm$0.01 & $<$1.11$\times$$10^{-22}$ &  \nodata \\ 
   MIPS168 & 4.86($\pm$1.46)$\times$$10^{-22}$ &  0.96$\pm$0.28 & 9.68($\pm$1.57)$\times$$10^{-22}$ &  1.45$\pm$0.24 & 3.00($\pm$0.31)$\times$$10^{-22}$ &  0.45$\pm$0.05 &  $<$1.30$\times$$10^{-22}$ &  \nodata  \\ 
   MIPS213 &  $<$1.15$\times$$10^{-22}$ &  \nodata  &  $<$7.08$\times$$10^{-23}$ &  \nodata  &  $<$3.35$\times$$10^{-23}$ &  \nodata  &  \nodata  &  \nodata  \\ 
   MIPS224 & 2.34($\pm$0.35)$\times$$10^{-22}$ &  0.13$\pm$0.02 & $<$7.42$\times$$10^{-23}$ &  \nodata & $<$3.73$\times$$10^{-23}$ & \nodata &  \nodata  &  \nodata  \\ 
   MIPS268 & 1.86($\pm$0.47)$\times$$10^{-22}$ &  0.14$\pm$0.04 &  $<$2.01$\times$$10^{-22}$ &  \nodata  &  $<$8.71$\times$$10^{-23}$ &  \nodata  &  \nodata  &  \nodata  \\ 
   MIPS277 & 5.94($\pm$0.58)$\times$$10^{-22}$ &  0.49$\pm$0.05 & 2.84($\pm$0.78)$\times$$10^{-22}$ &  0.25$\pm$0.07 &  $<$3.44$\times$$10^{-23}$ &  \nodata  &  \nodata &  \nodata  \\ 
   MIPS298 &  \nodata  &  \nodata  &  \nodata  &  \nodata  &  \nodata  &  \nodata  &  \nodata  &  \nodata  \\ 
   MIPS324 & 6.57($\pm$0.59)$\times$$10^{-22}$ &  0.64$\pm$0.06 & 2.66($\pm$0.57)$\times$$10^{-22}$ &  0.26$\pm$0.06 & 4.69($\pm$1.32)$\times$$10^{-23}$ &  0.04$\pm$0.01 & 7.91($\pm$1.19)$\times$$10^{-23}$ &  0.09$\pm$0.01 \\ 
   MIPS350 & 6.18($\pm$0.55)$\times$$10^{-22}$ & 0.71$\pm$0.06 & 4.90($\pm$0.63)$\times$$10^{-22}$ & 0.42$\pm$0.05 & 9.30($\pm$1.41)$\times$$10^{-23}$ & 0.07$\pm$0.01 & 1.03($\pm$0.13)$\times$$10^{-22}$ & 0.07$\pm$0.01 \\ 
   MIPS358 & 1.19($\pm$0.10)$\times$$10^{-21}$ &  3.35$\pm$0.28 & 1.24($\pm$0.07)$\times$$10^{-21}$ &  2.72$\pm$0.15 & 1.98($\pm$0.15)$\times$$10^{-22}$ &  0.41$\pm$0.03 & 5.54($\pm$1.22)$\times$$10^{-23}$ &  0.08$\pm$0.02 \\ 
   MIPS394 & 8.87($\pm$0.60)$\times$$10^{-22}$ &  3.66$\pm$0.25 & 9.57($\pm$0.53)$\times$$10^{-22}$ &  2.48$\pm$0.14 & 8.52($\pm$1.25)$\times$$10^{-23}$ &  0.21$\pm$0.03 &  $<$5.67$\times$$10^{-23}$ &  \nodata  \\ 
   MIPS397 & $<$7.31$\times$$10^{-23}$  &  \nodata  & 2.14($\pm$0.40)$\times$$10^{-22}$ &  0.16$\pm$0.03 &  $<$3.07$\times$$10^{-23}$ &  \nodata  &  \nodata  &  \nodata  \\ 
   MIPS419 &  $<$2.91$\times$$10^{-22}$ &  \nodata  &  $<$2.32$\times$$10^{-22}$ &  \nodata  &  $<$4.61$\times$$10^{-23}$ &  \nodata  & $<$4.54$\times$$10^{-23}$ &  \nodata  \\ 
   MIPS472 & 8.60($\pm$0.62)$\times$$10^{-22}$ &  2.17$\pm$0.16 & 5.14($\pm$0.62)$\times$$10^{-22}$ &  1.35$\pm$0.16 & 1.07($\pm$0.12)$\times$$10^{-22}$ &  0.28$\pm$0.03 &  $<$4.17$\times$$10^{-23}$ &  \nodata  \\ 
   MIPS488 & 1.44($\pm$0.10)$\times$$10^{-21}$ &  2.78$\pm$0.19 & 6.69($\pm$1.21)$\times$$10^{-22}$ &  1.06$\pm$0.19 & 1.62($\pm$0.22)$\times$$10^{-22}$ &  0.26$\pm$0.04 &  $<$4.97$\times$$10^{-23}$ &  \nodata  \\ 
   MIPS512 & 2.13($\pm$0.46)$\times$$10^{-22}$ &  0.31$\pm$0.07 & 3.00($\pm$0.53)$\times$$10^{-22}$ &  0.37$\pm$0.07 &  $<$4.95$\times$$10^{-23}$ &  \nodata  & 5.74($\pm$1.26)$\times$$10^{-23}$ &  0.07$\pm$0.02 \\ 
   MIPS521 & 4.55($\pm$0.83)$\times$$10^{-22}$ &  1.34$\pm$0.24 &  \nodata  &  \nodata  &  \nodata  &  \nodata  &  \nodata  &  \nodata  \\ 
   MIPS530 & 6.61($\pm$0.76)$\times$$10^{-22}$ &  3.57$\pm$0.41 & 6.17($\pm$0.96)$\times$$10^{-22}$ &  2.12$\pm$0.33 & 7.72($\pm$1.44)$\times$$10^{-23}$ &  0.26$\pm$0.05 &  $<$6.67$\times$$10^{-23}$ &  \nodata  \\ 
   MIPS532 & $<$1.58$\times$$10^{-22}$ & \nodata & 2.22($\pm$0.17)$\times$$10^{-22}$ & 0.10$\pm$0.01  &  $<$9.37$\times$$10^{-23}$  &  \nodata  &  \nodata  &  \nodata  \\ 
   MIPS537 & 1.54($\pm$0.07)$\times$$10^{-21}$ &  5.72$\pm$0.26 & 8.35($\pm$0.97)$\times$$10^{-22}$ &  2.83$\pm$0.33 & 3.33($\pm$0.27)$\times$$10^{-22}$ &  1.16$\pm$0.09 & 9.55($\pm$1.64)$\times$$10^{-23}$ &  0.24$\pm$0.04 \\ 
   MIPS542 & 5.93($\pm$0.62)$\times$$10^{-22}$ &  2.60$\pm$0.27 & 6.55($\pm$0.65)$\times$$10^{-22}$ &  2.96$\pm$0.29 & 8.98($\pm$1.23)$\times$$10^{-23}$ &  0.42$\pm$0.06 & 6.80($\pm$1.35)$\times$$10^{-23}$ &  0.34$\pm$0.07 \\ 
   MIPS544 & 2.53($\pm$0.53)$\times$$10^{-22}$ &  0.38$\pm$0.08 &  $<$2.21$\times$$10^{-22}$ &  \nodata  &  $<$4.68$\times$$10^{-23}$ &  \nodata  &  $<$5.47$\times$$10^{-23}$ &  \nodata  \\ 
   MIPS546 & 1.22($\pm$0.05)$\times$$10^{-21}$ &  2.58$\pm$0.11 & 8.17($\pm$0.62)$\times$$10^{-22}$ &  1.37$\pm$0.10 & 1.57($\pm$0.12)$\times$$10^{-22}$ &  0.26$\pm$0.02 & 9.52($\pm$1.68)$\times$$10^{-23}$ &  0.16$\pm$0.03 \\ 
   MIPS549 & 5.70($\pm$0.67)$\times$$10^{-22}$ &  1.07$\pm$0.13 & 5.87($\pm$0.54)$\times$$10^{-22}$ &  1.14$\pm$0.10 & 6.18($\pm$1.21)$\times$$10^{-23}$ &  0.12$\pm$0.02 &  $<$3.52$\times$$10^{-23}$ &  \nodata  \\ 
   MIPS562 & 1.23($\pm$0.07)$\times$$10^{-21}$ &  2.43$\pm$0.14 & 1.48($\pm$0.11)$\times$$10^{-21}$ &  2.52$\pm$0.19 & 2.44($\pm$0.30)$\times$$10^{-22}$ &  0.43$\pm$0.05 & 7.42($\pm$1.25)$\times$$10^{-23}$ &  0.17$\pm$0.03 \\ 
  MIPS8040 & 1.85($\pm$0.12)$\times$$10^{-21}$ &  1.22$\pm$0.08 & 1.05($\pm$0.11)$\times$$10^{-21}$ &  0.47$\pm$0.05 & 2.14($\pm$0.20)$\times$$10^{-22}$ &  0.09$\pm$0.01 &  $<$5.96$\times$$10^{-23}$ &  \nodata  \\ 
  MIPS8107 & 3.63($\pm$0.48)$\times$$10^{-22}$ &  0.24$\pm$0.03 & 1.11($\pm$0.19)$\times$$10^{-22}$ &  0.06$\pm$0.01 & 5.90($\pm$0.99)$\times$$10^{-23}$ &  0.03$\pm$0.01 & 3.45($\pm$1.07)$\times$$10^{-23}$ &  0.02$\pm$0.01 \\ 
  MIPS8121 & 1.84($\pm$0.40)$\times$$10^{-22}$ &  0.08$\pm$0.02 & 2.97($\pm$0.61)$\times$$10^{-22}$ &  0.12$\pm$0.02 & 3.28($\pm$0.98)$\times$$10^{-23}$ &  0.01$\pm$0.00$^*$  &  \nodata  &  \nodata  \\ 
  MIPS8179 & 3.97($\pm$0.81)$\times$$10^{-22}$ &  0.37$\pm$0.08 & 3.75($\pm$1.20)$\times$$10^{-22}$ &  0.30$\pm$0.10 &  $<$1.21$\times$$10^{-22}$ &  \nodata  & 6.69($\pm$1.05)$\times$$10^{-23}$ &  0.05$\pm$0.01 \\ 
  MIPS8204 & 1.16($\pm$0.09)$\times$$10^{-21}$ &  1.32$\pm$0.10 & 9.52($\pm$0.68)$\times$$10^{-22}$ &  0.96$\pm$0.07 & 1.38($\pm$0.15)$\times$$10^{-22}$ &  0.13$\pm$0.02 &  $<$3.73$\times$$10^{-23}$ &  \nodata  \\ 
  MIPS8226 &  \nodata  &  \nodata  &  \nodata  &  \nodata  &  \nodata  &  \nodata  &  \nodata  &  \nodata  \\ 
  MIPS8251 &  \nodata  &  \nodata  &  \nodata  &  \nodata  &  \nodata  &  \nodata  &  \nodata  &  \nodata  \\ 
  MIPS8253 & 3.55($\pm$0.47)$\times$$10^{-22}$ &  0.51$\pm$0.07 & 5.67($\pm$0.59)$\times$$10^{-22}$ &  0.46$\pm$0.05 & 6.04($\pm$1.07)$\times$$10^{-23}$ &  0.04$\pm$0.01 & 6.64($\pm$1.11)$\times$$10^{-23}$ &  0.05$\pm$0.01 \\ 
  MIPS8308 & 9.38($\pm$0.84)$\times$$10^{-22}$ &  1.42$\pm$0.13 & 6.67($\pm$1.35)$\times$$10^{-22}$ &  1.01$\pm$0.20 &  $<$8.37$\times$$10^{-23}$ &  \nodata  &  $<$6.36$\times$$10^{-23}$ &  \nodata  \\ 
  MIPS8311 & 6.13($\pm$0.41)$\times$$10^{-22}$ &  0.62$\pm$0.04 & 1.84($\pm$0.58)$\times$$10^{-22}$ &  0.15$\pm$0.05 & 6.41($\pm$1.19)$\times$$10^{-23}$ &  0.05$\pm$0.01 &  $<$9.97$\times$$10^{-23}$ &  \nodata  \\ 
  MIPS8325 & 6.44($\pm$0.90)$\times$$10^{-22}$ &  0.86$\pm$0.12 &  $<$6.00$\times$$10^{-22}$ &  \nodata  & 2.05($\pm$0.33)$\times$$10^{-22}$ &  0.20$\pm$0.03 & 8.29($\pm$1.31)$\times$$10^{-23}$ &  0.08$\pm$0.01 \\ 
  MIPS8328 & 4.82($\pm$0.69)$\times$$10^{-22}$ &  0.49$\pm$0.07 & 3.24($\pm$0.99)$\times$$10^{-22}$ &  0.31$\pm$0.09 & 7.22($\pm$1.73)$\times$$10^{-23}$ &  0.07$\pm$0.02 &  $<$7.29$\times$$10^{-23}$ &  \nodata  \\ 
  MIPS8360 & 4.36($\pm$0.42)$\times$$10^{-22}$ &  0.32$\pm$0.03 & 4.11($\pm$0.87)$\times$$10^{-22}$ &  0.27$\pm$0.06 & 5.66($\pm$1.56)$\times$$10^{-23}$ &  0.04$\pm$0.01 &  \nodata  &  \nodata  \\ 
  MIPS8371 & 1.89($\pm$0.07)$\times$$10^{-21}$ &  7.50$\pm$0.28 & 1.38($\pm$0.11)$\times$$10^{-21}$ &  5.08$\pm$0.40 & 2.77($\pm$0.24)$\times$$10^{-22}$ &  1.06$\pm$0.09 &  $<$1.04$\times$$10^{-22}$  &  \nodata  \\ 
  MIPS8375 & 8.00($\pm$0.73)$\times$$10^{-22}$ &  1.70$\pm$0.16 & 5.82($\pm$0.80)$\times$$10^{-22}$ &  0.83$\pm$0.11 & 1.04($\pm$0.13)$\times$$10^{-22}$ &  0.14$\pm$0.02 & 1.37($\pm$0.12)$\times$$10^{-22}$ &  0.13$\pm$0.01 \\ 
  MIPS8377 & 9.57($\pm$0.90)$\times$$10^{-22}$ &  2.97$\pm$0.28 & 7.74($\pm$0.91)$\times$$10^{-22}$ &  1.70$\pm$0.20 & 1.33($\pm$0.13)$\times$$10^{-22}$ &  0.29$\pm$0.03 & 9.89($\pm$1.30)$\times$$10^{-23}$ &  0.29$\pm$0.04 \\ 
  MIPS8384 & 6.13($\pm$0.53)$\times$$10^{-22}$ &  0.58$\pm$0.05 & 7.90($\pm$2.13)$\times$$10^{-23}$ &  0.07$\pm$0.02 & 6.95($\pm$1.09)$\times$$10^{-23}$ &  0.06$\pm$0.01 & 6.26($\pm$1.13)$\times$$10^{-23}$ &  0.06$\pm$0.01 \\ 
  MIPS8387 & 1.10($\pm$0.07)$\times$$10^{-21}$ &  2.69$\pm$0.17 & 5.63($\pm$0.68)$\times$$10^{-22}$ &  1.51$\pm$0.18 & 6.11($\pm$1.20)$\times$$10^{-23}$ &  0.17$\pm$0.03 &  $<$4.68$\times$$10^{-23}$  &  \nodata  \\ 
  MIPS8388 & 7.52($\pm$0.50)$\times$$10^{-22}$ &  1.08$\pm$0.07 & 5.27($\pm$0.71)$\times$$10^{-22}$ &  0.67$\pm$0.09 & 7.09($\pm$1.24)$\times$$10^{-23}$ &  0.09$\pm$0.02 & 8.62($\pm$1.87)$\times$$10^{-23}$ &  0.09$\pm$0.02 \\ 
  MIPS8392 & 3.06($\pm$0.97)$\times$$10^{-22}$ &  1.58$\pm$0.50 &  \nodata  &  \nodata  &  \nodata  &  \nodata  &  \nodata  &  \nodata  \\ 
  MIPS8401 & 1.33($\pm$0.09)$\times$$10^{-21}$ &  3.43$\pm$0.23 & 1.09($\pm$0.14)$\times$$10^{-21}$ &  2.05$\pm$0.26 & 3.25($\pm$0.41)$\times$$10^{-22}$ &  0.60$\pm$0.08 & 4.94($\pm$1.29)$\times$$10^{-23}$ &  0.07$\pm$0.02 \\ 
  MIPS8405 & 5.90($\pm$0.48)$\times$$10^{-22}$ &  0.65$\pm$0.05 & 7.90($\pm$0.77)$\times$$10^{-22}$ &  0.80$\pm$0.08 & 1.69($\pm$0.11)$\times$$10^{-22}$ &  0.16$\pm$0.01 & 9.17($\pm$2.35)$\times$$10^{-23}$ &  0.08$\pm$0.02 \\ 
  MIPS8411 & \nodata & \nodata &  \nodata  &  \nodata  &  \nodata  &  \nodata  &  \nodata  &  \nodata  \\ 
  MIPS8424 & 1.04($\pm$0.20)$\times$$10^{-22}$ &  0.07$\pm$0.01 & 8.63($\pm$0.79)$\times$$10^{-22}$ &  0.47$\pm$0.04 &  $<$4.80$\times$$10^{-23}$ &  \nodata  &  \nodata  &  \nodata  \\ 
  MIPS8430 & 4.57($\pm$1.04)$\times$$10^{-22}$ &  1.03$\pm$0.23 &  $<$6.53$\times$$10^{-22}$ &  \nodata  & 1.30($\pm$0.21)$\times$$10^{-22}$ &  0.19$\pm$0.03 & 5.25($\pm$1.43)$\times$$10^{-23}$ &  0.08$\pm$0.02 \\ 
  MIPS8450 & 1.08($\pm$0.05)$\times$$10^{-21}$ &  3.93$\pm$0.18 & 7.25($\pm$0.59)$\times$$10^{-22}$ &  3.23$\pm$0.26 & 1.15($\pm$0.12)$\times$$10^{-22}$ &  0.55$\pm$0.06 &  \nodata  &  \nodata  \\ 
  MIPS8462 & 2.97($\pm$0.58)$\times$$10^{-22}$ &  0.33$\pm$0.06 & 3.11($\pm$0.72)$\times$$10^{-22}$ &  0.28$\pm$0.06 & 8.91($\pm$1.16)$\times$$10^{-23}$ &  0.07$\pm$0.01 & 5.97($\pm$1.48)$\times$$10^{-23}$ &  0.05$\pm$0.01 \\ 
  MIPS8477 & 3.04($\pm$0.52)$\times$$10^{-22}$ &  0.38$\pm$0.07 &  \nodata  &  \nodata  &  \nodata  &  \nodata  &  \nodata  &  \nodata  \\ 
  MIPS8479 &  $<$5.34$\times$$10^{-22}$ &  \nodata  & 4.47($\pm$0.64)$\times$$10^{-22}$ &  0.14$\pm$0.02 &  $<$8.37$\times$$10^{-23}$ &  \nodata  &  \nodata  &  \nodata  \\ 
  MIPS8495 &  $<$6.20$\times$$10^{-23}$ &  \nodata  &  $<$1.30$\times$$10^{-22}$ &  \nodata  & 5.40($\pm$1.11)$\times$$10^{-23}$ &  0.06$\pm$0.01 &  \nodata  &  \nodata  \\ 
  MIPS8499 & 1.49($\pm$0.09)$\times$$10^{-21}$ &  7.86$\pm$0.47 & 1.41($\pm$0.13)$\times$$10^{-21}$ &  5.55$\pm$0.51 & 1.94($\pm$0.33)$\times$$10^{-22}$ &  0.73$\pm$0.13 &  $<$2.95$\times$$10^{-23}$ &  \nodata  \\ 
  MIPS8507 & 7.46($\pm$0.98)$\times$$10^{-22}$ &  1.33$\pm$0.17 & 7.54($\pm$0.87)$\times$$10^{-22}$ &  1.26$\pm$0.15 &  6.25($\pm$1.83)$\times$$10^{-23}$ & 0.11$\pm$0.03 &  $<$4.89$\times$$10^{-23}$ &  \nodata  \\ 
  MIPS8521 & 3.55($\pm$0.42)$\times$$10^{-22}$ &  0.33$\pm$0.04 & 4.27($\pm$0.56)$\times$$10^{-22}$ &  0.24$\pm$0.03 &  $<$6.17$\times$$10^{-23}$ &  \nodata  &  $<$1.16$\times$$10^{-22}$ &  \nodata  \\ 
  MIPS8526 & 1.04($\pm$0.09)$\times$$10^{-21}$ &  2.37$\pm$0.21 & 4.00($\pm$0.83)$\times$$10^{-22}$ &  0.72$\pm$0.15 & 5.77($\pm$1.50)$\times$$10^{-23}$ &  0.10$\pm$0.03 & $<$6.38$\times$$10^{-23}$  &  \nodata  \\ 
  MIPS8532 & 4.61($\pm$0.83)$\times$$10^{-22}$ &  0.79$\pm$0.14 &  $<$1.25$\times$$10^{-22}$ &  \nodata  & 5.80($\pm$1.66)$\times$$10^{-23}$ &  0.08$\pm$0.02 & 6.35($\pm$1.13)$\times$$10^{-23}$ &  0.10$\pm$0.02 \\ 
  MIPS8543 & 2.73($\pm$0.09)$\times$$10^{-21}$ &  3.19$\pm$0.11 & 1.83($\pm$0.15)$\times$$10^{-21}$ &  2.01$\pm$0.16 & 1.83($\pm$0.35)$\times$$10^{-22}$ &  0.20$\pm$0.04 &  $<$4.81$\times$$10^{-23}$ &  \nodata  \\ 
  MIPS8550 &  $<$9.33$\times$$10^{-22}$ &  \nodata  & $<$5.21$\times$$10^{-22}$ &  \nodata &  $<$3.21$\times$$10^{-23}$ &  \nodata  &  $<$3.18$\times$$10^{-23}$ &  \nodata  \\ 
 MIPS15690 &  $<$3.30$\times$$10^{-22}$ &  \nodata  & 2.84($\pm$0.37)$\times$$10^{-22}$ &  0.05$\pm$0.01 &  $<$6.28$\times$$10^{-23}$ &  \nodata  & 7.45($\pm$1.21)$\times$$10^{-23}$ &  0.01$\pm$0.00$^*$ \\ 
 MIPS15755 & 2.58($\pm$0.12)$\times$$10^{-21}$ &  1.43$\pm$0.07 & 1.47($\pm$0.10)$\times$$10^{-21}$ &  0.83$\pm$0.06 & 2.48($\pm$0.23)$\times$$10^{-22}$ &  0.14$\pm$0.01 & 8.64($\pm$1.05)$\times$$10^{-23}$ &  0.05$\pm$0.01 \\ 
 MIPS15776 & 3.05($\pm$0.37)$\times$$10^{-22}$ &  0.10$\pm$0.01 & 1.43($\pm$0.22)$\times$$10^{-22}$ &  0.04$\pm$0.01 &  $<$3.72$\times$$10^{-23}$  &  \nodata  & \nodata & \nodata \\ 
 MIPS15999 & 6.63($\pm$0.95)$\times$$10^{-22}$ &  1.24$\pm$0.18 &  $<$5.59$\times$$10^{-22}$ &  \nodata  & 2.30($\pm$0.36)$\times$$10^{-22}$ &  0.26$\pm$0.04 & 1.02($\pm$0.17)$\times$$10^{-22}$ &  0.10$\pm$0.02 \\ 
 MIPS16037 &  $<$2.07$\times$$10^{-22}$ &  \nodata  & 1.43($\pm$0.34)$\times$$10^{-22}$ &  0.06$\pm$0.01 &   $<$5.99$\times$$10^{-23}$  &  \nodata  &  \nodata  &  \nodata  \\ 
 MIPS16047 & 3.58($\pm$0.77)$\times$$10^{-22}$ &  0.34$\pm$0.07 &  $<$5.29$\times$$10^{-22}$ &  \nodata  & 1.01($\pm$0.33)$\times$$10^{-22}$ &  0.07$\pm$0.02 & 8.38($\pm$1.28)$\times$$10^{-23}$ &  0.07$\pm$0.01 \\ 
 MIPS16066 & 4.68($\pm$0.53)$\times$$10^{-22}$ &  0.43$\pm$0.05 & 3.43($\pm$0.50)$\times$$10^{-22}$ &  0.23$\pm$0.03 &  $<$4.19$\times$$10^{-23}$ &  \nodata  & 4.37($\pm$1.12)$\times$$10^{-23}$ &  0.03$\pm$0.01 \\ 
 MIPS16118 & \nodata  &  \nodata  &  \nodata  &  \nodata  &  \nodata  &  \nodata  &  \nodata  &  \nodata  \\ 
 MIPS16152 & 2.20($\pm$0.52)$\times$$10^{-22}$ &  0.40$\pm$0.09 &  \nodata  &  \nodata  &  \nodata  &  \nodata  &  \nodata  &  \nodata  \\ 
 MIPS16156 & 4.86($\pm$0.75)$\times$$10^{-22}$ &  0.94$\pm$0.15 & 1.67($\pm$0.37)$\times$$10^{-22}$ &  0.23$\pm$0.05 &  $<$5.39$\times$$10^{-23}$ &  \nodata  & 8.00($\pm$1.25)$\times$$10^{-23}$ &  0.10$\pm$0.02 \\ 
 MIPS16170 & 1.11($\pm$0.09)$\times$$10^{-21}$ &  2.92$\pm$0.24 & 5.27($\pm$1.26)$\times$$10^{-22}$ &  1.00$\pm$0.24 & 1.56($\pm$0.23)$\times$$10^{-22}$ &  0.28$\pm$0.04 &  $<$1.00$\times$$10^{-22}$ &  \nodata  \\ 
 MIPS16206 & 9.59($\pm$0.81)$\times$$10^{-22}$ &  1.07$\pm$0.09 & 3.95($\pm$0.83)$\times$$10^{-22}$ &  0.48$\pm$0.10 & 7.63($\pm$1.38)$\times$$10^{-23}$ &  0.09$\pm$0.02 & 5.60($\pm$1.45)$\times$$10^{-23}$ &  0.11$\pm$0.03 \\ 
 MIPS16249 & 1.04($\pm$0.08)$\times$$10^{-21}$ &  2.00$\pm$0.15 & 5.48($\pm$1.25)$\times$$10^{-22}$ &  0.86$\pm$0.20 & 1.18($\pm$0.32)$\times$$10^{-22}$ &  0.17$\pm$0.05 & 7.77($\pm$1.33)$\times$$10^{-23}$ &  0.10$\pm$0.02 \\ 
 MIPS16267 &  $<$1.78$\times$$10^{-22}$ &  \nodata  &  $<$2.61$\times$$10^{-22}$ &  \nodata  &  $<$3.95$\times$$10^{-23}$ &  \nodata  &  \nodata  &  \nodata  \\ 
 MIPS22235 & 2.64($\pm$0.11)$\times$$10^{-21}$ &  1.34$\pm$0.06 & 2.32($\pm$0.14)$\times$$10^{-21}$ &  0.87$\pm$0.05 & 4.43($\pm$0.34)$\times$$10^{-22}$ &  0.15$\pm$0.01 &  $<$7.33$\times$$10^{-23}$ &  \nodata  \\ 
 MIPS22307 & 1.07($\pm$0.10)$\times$$10^{-21}$ &  0.69$\pm$0.06 & 7.92($\pm$0.99)$\times$$10^{-22}$ &  0.51$\pm$0.06 & 9.35($\pm$1.91)$\times$$10^{-23}$ &  0.06$\pm$0.01 & 8.36($\pm$1.17)$\times$$10^{-23}$ &  0.05$\pm$0.01 \\ 
 MIPS22323 & 5.59($\pm$0.51)$\times$$10^{-22}$ &  0.27$\pm$0.02 & 2.30($\pm$0.65)$\times$$10^{-22}$ &  0.12$\pm$0.03 & 1.03($\pm$0.14)$\times$$10^{-22}$ &  0.06$\pm$0.01 &  \nodata  &  \nodata  \\ 
 MIPS22352 & 2.07($\pm$0.10)$\times$$10^{-21}$ &  2.83$\pm$0.14 & 1.23($\pm$0.15)$\times$$10^{-21}$ &  1.47$\pm$0.18 & 1.76($\pm$0.21)$\times$$10^{-22}$ &  0.21$\pm$0.03 & 6.19($\pm$1.23)$\times$$10^{-23}$ &  0.06$\pm$0.01 \\ 
 MIPS22356 & 3.92($\pm$0.44)$\times$$10^{-22}$ &  0.20$\pm$0.02 & 3.34($\pm$0.64)$\times$$10^{-22}$ &  0.15$\pm$0.03 & 5.96($\pm$1.25)$\times$$10^{-23}$ &  0.02$\pm$0.00$^*$ & 7.05($\pm$1.70)$\times$$10^{-23}$ &  0.03$\pm$0.01 \\ 
 MIPS22371 & 4.66($\pm$0.36)$\times$$10^{-22}$ &  0.26$\pm$0.02 & 1.94($\pm$0.38)$\times$$10^{-22}$ &  0.11$\pm$0.02 &  $<$5.94$\times$$10^{-23}$ &  \nodata  &  \nodata  &  \nodata  \\ 
 MIPS22379 & 5.00($\pm$0.77)$\times$$10^{-22}$ &  0.44$\pm$0.07 & 1.50($\pm$0.46)$\times$$10^{-22}$ & 0.12$\pm$0.04  &  $<$6.03$\times$$10^{-23}$ &  \nodata  & 4.40($\pm$1.10)$\times$$10^{-23}$ &  0.03$\pm$0.01 \\ 
 MIPS22417 & \nodata &  \nodata &  \nodata  &  \nodata  &  \nodata  &  \nodata  &  \nodata  &  \nodata  \\ 
 MIPS22432 & 5.37($\pm$0.33)$\times$$10^{-22}$ &  0.54$\pm$0.03 &  $<$4.25$\times$$10^{-22}$ &  \nodata  & 1.69($\pm$0.19)$\times$$10^{-22}$ &  0.08$\pm$0.01 &  \nodata  &  \nodata  \\ 
 MIPS22536 & 4.09($\pm$0.58)$\times$$10^{-22}$ &  0.27$\pm$0.04 & 7.58($\pm$0.87)$\times$$10^{-22}$ &  0.34$\pm$0.04 &  $<$5.42$\times$$10^{-23}$ &  \nodata  &  \nodata  &  \nodata  \\ 
 MIPS22549 & 3.53($\pm$0.59)$\times$$10^{-22}$ &  0.38$\pm$0.06 & 3.00($\pm$0.79)$\times$$10^{-22}$ &  0.32$\pm$0.08 &  $<$2.99$\times$$10^{-23}$ &  \nodata  & \nodata &  \nodata  \\ 
 MIPS22555 &  $<$2.38$\times$$10^{-22}$ &  \nodata  &  \nodata  &  \nodata  &  \nodata  &  \nodata  &  \nodata  &  \nodata  \\ 
 MIPS22557 & 6.40($\pm$1.58)$\times$$10^{-22}$ &  1.13$\pm$0.28 & 7.66($\pm$1.34)$\times$$10^{-22}$ &  1.24$\pm$0.22 &  $<$7.04$\times$$10^{-23}$ &  \nodata  & 6.57($\pm$1.67)$\times$$10^{-23}$ &  0.08$\pm$0.02 \\ 
 MIPS22635 & 6.51($\pm$0.99)$\times$$10^{-22}$ &  1.58$\pm$0.24 & 5.69($\pm$0.96)$\times$$10^{-22}$ &  0.93$\pm$0.16 & 7.41($\pm$1.54)$\times$$10^{-23}$ &  0.11$\pm$0.02 & 4.35($\pm$1.13)$\times$$10^{-23}$ &  0.06$\pm$0.02 \\ 
 MIPS22638 & 2.96($\pm$0.57)$\times$$10^{-22}$ &  0.38$\pm$0.07 & 3.18($\pm$0.63)$\times$$10^{-22}$ &  0.36$\pm$0.07 &  $<$3.93$\times$$10^{-23}$ &  \nodata  & 5.28($\pm$1.27)$\times$$10^{-23}$ &  0.05$\pm$0.01 \\ 
 MIPS22690 & \nodata & \nodata &  \nodata  &  \nodata  &  \nodata  &  \nodata  &  \nodata  &  \nodata  \\ 
 MIPS22722 & \nodata & \nodata &  \nodata  &  \nodata  &  \nodata &  \nodata  &  \nodata  &  \nodata  \\ 
\enddata
\tablecomments{ ~~ Fluxes may not be given for lines that are in the IRS wavelength range when their neighboring continuum cannot be well determined. \\
~~*~Error bars of 0 correspond to EW uncertainties that are lower than the selected measurement accuracy.}
\end{deluxetable}
\end{centering}